\documentclass[a4paper, onecolumn, usenatbib]{mn2e}

\usepackage{amssymb,amsmath,hyperref}
\usepackage{graphicx}
\usepackage{float}
\usepackage{cleveref}
\usepackage{color}
\usepackage{soul}
\usepackage{tensor}
\usepackage{subfigure}
\usepackage{natbib}

\usepackage[T1]{fontenc}
\usepackage{aecompl}

\newcommand{\vep}{\varepsilon}
  
\newcommand{\msun}{M_{\odot}} % solar mass
\newcommand{\dv}[2]{\frac{d #1}{d #2}}

\newcommand{\ppv}[2]{\frac{\partial^2 #1}{\partial {#2}^2}}

\newcommand{\babla}{\boldsymbol{\nabla}}
\newcommand{\boxi}{\boldsymbol{\xi}}
\newcommand{\boeta}{\boldsymbol{\eta}}

\title{Developing a model for neutron star oscillations following starquakes}
\author[L. Keer et al.]
{L.~Keer$^1$\thanks{Email: lucy.keer@gmail.com}
and D.I.~Jones$^1$ 
\\ $^1$Mathematical Sciences, University of Southampton, University
Road, Southampton SO17 1BJ}

\begin{document}
\maketitle
\begin{abstract}
Glitches -- sudden increases in spin rate -- are observed in many pulsars. One mechanism
advanced to explain glitches in the youngest pulsars 
is that they are caused by a starquake, a sudden rearrangement of the crust of the neutron star.
Starquakes have the potential to excite some of the oscillation modes of the
neutron star, which means that they are of interest as a source of 
gravitational waves. These oscillations could also have an impact on radio emission.
In this paper we make upper estimates of the amplitude of the oscillations produced
by a starquake, and the corresponding gravitational wave emission. We then develop
a more detailed framework for calculating the oscillations excited by the starquake,
using a toy model of a solid, incompressible star where all strain is lost 
instantaneously from the star at the glitch.
For this toy model, we give plots of the amplitudes of the modes excited, as the
shear modulus and rotation rate of the star are varied. We find that for our specific model,
the largest excitation is generally of a mode similar to the $f$-mode of an incompressible fluid star, 
but that other modes of the star are excited to a significant degree over small parameter ranges 
of the rotation rate.    
\end{abstract}

%%%%%%%%%%%%%%%%%%%%%%%%%%%%%%%%%%%%%%%%%%%%%%%%%%%%%%%%%%%%%%%%%%
% INTRODUCTION
%%%%%%%%%%%%%%%%%%%%%%%%%%%%%%%%%%%%%%%%%%%%%%%%%%%%%%%%%%%%%%%%%%

\section{Introduction}
\label{sec:intro}

Neutron stars normally rotate at an extremely regular rate.
Once the steady slowdown over time from magnetic dipole radiation
is accounted for, pulsars can keep time to an accuracy
of one part in $10^{11}$ or more (\cite{PulsarAstronomy}).
However, younger pulsars often display more erratic behaviour. 
Some of this variation comes from timing noise, small unmodelled deviations in the pulse period.
More dramatically, many have been observed to undergo a sudden speedup in rotation
rate, known as a glitch. 

Glitch observations are discussed in detail
in \cite{2011Espinoza}, where glitching pulsars are
shown to exhibit a diverse range of behaviours.
The size of a glitch can be characterised by the fractional change in spin frequency
$\frac{\Delta\Omega}{\Omega}$ of the pulsar. Glitches have been observed
spanning the range $\frac{\Delta\Omega}{\Omega}\sim 10^{-11}~-~10^{-5}$,
and the distribution of glitch sizes experienced by
an individual pulsar also varies widely. 
Some pulsars, such as the Vela, undergo `giant' glitches of approximately regular size
with $\frac{\Delta\Omega}{\Omega}\sim 10^{-6}$, 
whereas some of the youngest pulsars such as the Crab
have smaller glitches with $\frac{\Delta \Omega}{\Omega} \lesssim 10^{-8}$, and
glitch sizes are distributed over a wide range (\cite{2001Wong}).

Glitches are likely to excite oscillation modes
of the star, making them an interesting candidate source for gravitational wave emission. 
These oscillations could also be expected to shake the star's magnetosphere, 
leading to the possibility of radio emission observable by next generation radio
telescopes. The increased sensitivity and time resolution of the Square Kilometre Array may allow
the glitch event to be detected directly.

However, the nature and amplitude of the oscillation modes excited will depend strongly
on the mechanism producing the glitches, which is as yet unknown. The leading explanation
for the larger glitches is that they are caused by a rapid transfer of angular momentum from the superfluid 
component of the star to its crust (\cite{1975Anderson}). This superfluid component
may ordinarily be only weakly coupled to the crust, which gradually slows down 
through magnetic braking. This is because the superfluid can only slow down
through the outward migration of its quantised vortices.
As they reach the solid crust they may become pinned to the nuclei in the crystal lattice,
fixing the angular momentum of this part of the superfluid.
A glitch would then be triggered by an event that could increase the coupling between the two components by producing a
sudden collective unpinning of these vortices. 
Various mechanisms for this unpinning have been proposed (\cite{1984Alpar},
\cite{2009Glampedakis}, \cite{2008Melatos}).

Another idea is that glitches are produced by a starquake -- a sudden rearrangement of the crust of the
star (\cite{1971Baym}). This model uses the fact that while a fluid
star would be able to freely adjust its shape as the neutron star slows down over time,
the solid crust of a neutron star will resist the deformation from 
its relaxed state; consequently, it will stay more oblate than a fluid
star would. The level of strain in the crust builds up, and this may reach a critical value 
at which the crust can no longer withstand the strain and breaks.
The release of strain means that the star loses some of
its residual oblateness. The moment of inertia decreases, and so by 
conservation of angular momentum there must be a corresponding increase
in the star's angular velocity, producing the observed glitch. This model
is not able to produce a large enough energy release to account for the largest 
glitches, such as those of Vela (\cite{1996Alpar}), 
but is still promising for describing smaller glitches such as those of the Crab.

The prospects for gravitational wave detection from the superfluid model
have been discussed in papers by \cite{2010Sidery} and \cite{2008vanEysden}.
In this paper, we will concentrate on the starquake model, 
starting by making some order-of-magnitude estimates for the maximum energy released
in a starquake, and the amplitudes of the oscillations produced, using a formalism originally
developed by \cite{1971Baym}. We then develop a more detailed framework for
calculating the oscillations excited by a glitch, based on calculating initial data
for the glitch and projecting it against a basis of normal modes of the star. 
We carry this out for a toy model in which the star is completely solid and incompressible,
and the glitch is modelled by a instantaneous loss of all strain.
Although this acausal glitch mechanism is not realistic, it is a sensible starting point
which allows us to develop the model, and even this simple model produces some
interesting results.

We will start in Section \ref{sec:model} by giving an overview of our model for a starquake
and fixing notation for the rest of the paper.
In Section \ref{sec:estimates} we make order-of-magnitude estimates of the amplitude
of oscillations produced by the quake, 
both for our specific model and for an optimistic scenario in which all energy released
in the quake goes into oscillations. For this scenario, we also calculate
the gravitational wave emission produced by the oscillation,
both for an ordinary neutron star with a fluid interior and a solid quark star.
We then move on to the more detailed model, describing the calculation of the
initial data for the glitch in Section \ref{sec:initialdata} and the normal
modes of the rotating star in Section \ref{sec:modes}. In Section 
\ref{sec:projection} we develop a projection scheme that allows us to calculate
the amplitudes of the oscillation modes excited, and discuss the results of this. Section
\ref{sec:conclusions} then provides a summary of the results of the paper.

%%%%%%%%%%%%%%%%%%%%%%%%%%%%%%%%%%%%%%%%%%%%%%%%%%%%%%%%%%%%%%%%%%
% STARQUAKE MODEL
%%%%%%%%%%%%%%%%%%%%%%%%%%%%%%%%%%%%%%%%%%%%%%%%%%%%%%%%%%%%%%%%%%

\section{The starquake toy model}
\label{sec:model}

%figure: glitch model%%%%%%%%%%%%%%%%%%%%%%%%%%%%%%%%%%%%%%%%%%
\begin{figure}
\centering
\includegraphics[scale=0.5]{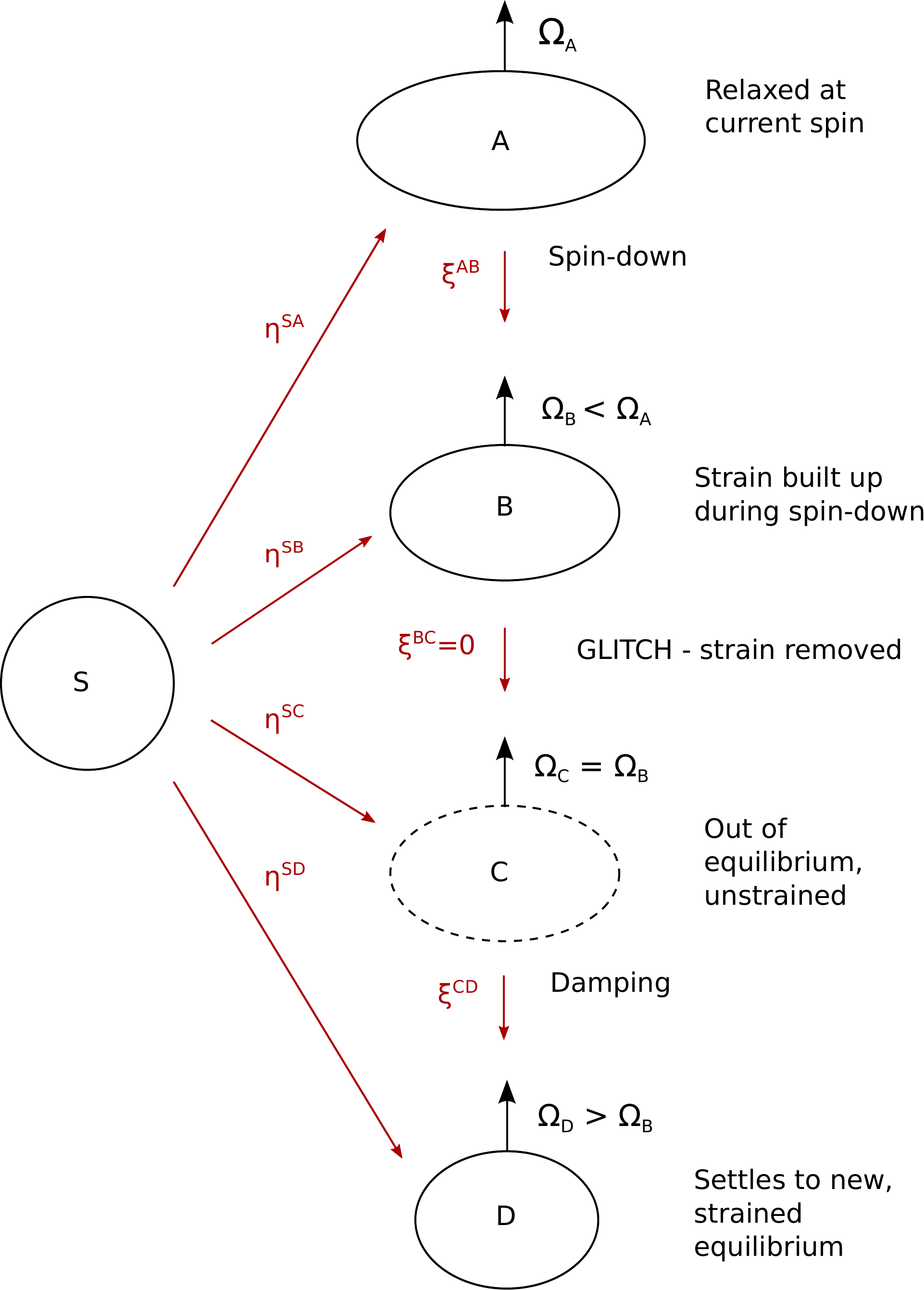}
\caption[The starquake model]{Diagram of our starquake model. As the star
spins down from an initial, relaxed configuration (Star A) spinning at angular
velocity $\Omega_A$, strain builds up in the crust. When this reaches a critical level (Star B)
at some angular velocity $\Omega_B$, the crust cracks and strain is removed. Immediately after the starquake
the star is out of equilibrium (Star C), and oscillates briefly 
before settling down to a new equilbrium state (Star D) spinning with 
angular velocity $\Omega_D$. We will treat the equilibrium states as perturbations
about a spherical background configuration, Star S. The maps $\eta$ track the displacements of 
particles in S to their new positions in Stars A -- D, while the 
$\xi$ displacement fields map between A -- D.}
\label{fig:potatoes}
\end{figure}
%%%%%%%%%%%%%%%%%%%%%%%%%%%%%%%%%%%%%%%%%%%%%%%%%%%%%%%%%%%%%%%%

In this section, we describe our simplified toy model for a glitch. 
This is illustrated in Figure \ref{fig:potatoes}, where the main stages of 
the glitch are labelled as Stars A--D. We will work in Newtonian gravity
throughout the paper. We will also only consider the case of slow rotation,
in which rotation is treated as a perturbation about a spherical background
star (Star S of the figure).

In this toy model we take our star to be completely solid and incompressible:
this will allow us to use analytic results
for the equilibrium configurations of the rotating star, and for the normal modes 
of Star S. The star is also modelled as an elastic solid, so that it has some relaxed state of zero strain,
and deformation from this state will induce a strain field in the star. 
This strain field is built from the displacement vector field $\boxi$ connecting points of the star
in its relaxed configuration to their new positions in its current configuration.

In our model, the star starts in an unstrained state, \textbf{Star A},
spinning with angular velocity $\Omega_A$. 
In this relaxed state, the star will have the same shape as a completely fluid star.
This star will then spin down as it loses energy. This will cause
it to become less oblate, inducing a strain field in the star 
as it is deformed from its relaxed configuration. 
The starquake then occurs at the point where the strain has built
up to some critical level where the crust can no longer support it -- we 
label this as \textbf{Star B}, which is spinning at some angular velocity
$\Omega_B$. The strain field at this point can 
be calculated from the displacement field that 
connects particles in the unstrained Star A to their new positions
in Star B, which we will label $\boxi^{\text{AB}}$.

We now need to specify our model of the glitch itself. For our toy model,
we will take the extreme case where \textit{all} of the strain energy of Star B is lost from 
the star. More precisely:

\begin{quotation}
\noindent 
We assume that all the strain is lost from the 
star at the glitch, so that the new, unstrained reference state of the
star is that of Star B. Furthermore, we assume that this energy
is just lost from the star as heat, rather than going into kinetic or 
gravitational energy. This means that the mass distribution 
of the star will not be changed at the glitch: particles in Star
C immediately after the glitch are not displaced from their positions in Star B:
using the same labelling method for the displacement field between
Stars B and C, we have $\boxi^{\text{BC}} = 0$.
\end{quotation}
After the glitch, the star will now be out of equilibrium (\textbf{Star C}),
and so it will start to oscillate.
These oscillations will be damped until the star finally reaches
a new equilibrium configuration, \textbf{Star D}, with angular velocity $\Omega_D$. 
This equilibrium state is
deformed from its unstrained state, Star C, and so has
some residual strain constructed from the displacement field
$\boxi^{\text{CD}}$.

%%%%%%%%%%%%%%%%%%%%%%%%%%%%%%%%%%%%%%%%%%%%%%%%%
% Plan of the calculation 
 
\subsection{Plan of the calculation}
\label{subsec: plan}

The basic idea of our starquake model is to project initial data 
describing the starquake against the set of oscillation modes of the star after the 
glitch, in order to find the amplitudes of the excited oscillation modes. 
This process has three stages:

\begin{enumerate}
\item First we construct the initial data, which will take the
form of a displacement field $\boxi^{\text{DC}}$ and velocity field
$\dot{\boxi}^{\text{DC}}$ describing how particles in
the star are perturbed from equilibrium after the glitch. Note that
the initial data links the final state, Star D, to the state of the
star before the glitch, so the displacement field initial data is in
the opposite direction to the strain field $\boxi^{\text{CD}}$.
This construction will involve finding the new equilibrium
state of the star after the starquake, Star D, which is rotating
with angular velocity $\Omega_D$. We do this in the case of slow
rotation, where the change in the star's equilibrium configuration due to rotation is
treated as a perturbation about the spherical background star, Star S. 
\item Next, we calculate the spectrum of oscillation modes of Star D. We will start
by finding analytic results for the eigenvalues and eigenfunctions of the 
spherical Star S. We then introduce rotation as a perturbation and 
calculate the first order corrections to these modes.
\item Finally, we can project our initial data, $\boxi^{\text{DC}}$ and $\dot{\boxi}^{\text{DC}}$, 
against this set of modes of Star D, to see which ones are excited by it. 
The eigenfunctions of Star S are orthogonal, which we 
show in Appendix \ref{appx:orthogonal}. However, once we include rotation, this is no longer the 
case. We describe a scheme that will nevertheless allow us to 
carry out the projection.
\end{enumerate}

%%%%%%%%%%%%%%%%%%%%%%%%%%%%%%%%%%%%%%%%%%%%%%%%%%%%%%%%%%%%%%%%%%
% ENERGY ESTIMATES
%%%%%%%%%%%%%%%%%%%%%%%%%%%%%%%%%%%%%%%%%%%%%%%%%%%%%%%%%%%%%%%%%%

\section{Energy estimates for the model}
\label{sec:estimates}

Before making any detailed calculations, we can get more 
insight into the starquake mechanism for glitches by making some 
estimates for the energy released by the quake. 
An estimate that has appeared in the literature (\cite{LSCVela}) takes
the rotational kinetic energy available to the star and multiplies it by the fractional
change in velocity characterising the glitch, to obtain 
%$\Delta E \sim I\Omega \left(\frac{\Delta\Omega}{\Omega}\right).$
%
\begin{equation}
\label{literature}
\Delta E \sim I\Omega^2 \left(\frac{\Delta\Omega}{\Omega}\right).
\end{equation}
However, the
actual value of the change in energy will depend strongly on the detailed physics of the glitch.
This in turn depends on the mechanism that triggers the 
glitch. Here we will concentrate on the starquake model, and make some 
estimates for the energy released in the glitch in this case, and
the amplitude of the oscillations produced. 
We will use the notation of the previous section, where the stages
of the starquake model are labelled by Stars A -- D.

In the specific model outlined in the previous section, the strain energy
is just lost from the star at the glitch. 
In this case, the energy available to go into oscillations comes only 
from the change in the star's gravitational and kinetic energy
as it settles to a new equilibrium state. This is the energy difference
between Stars C and D, $\Delta E_{CD}=E_C - E_D$.

Although this is perhaps the simplest option to model, it is
not necessarily the most realistic. We can also imagine
taking the strain energy of the star before the glitch
and putting some of it into, for example, the kinetic energy of the star after the quake (if the crust cracks
or otherwise moves about). 

The upper bound on this would be to put \textit{all} the strain energy made available into 
gravitational or kinetic energy. This would correspond to the total energy
immediately after the glitch being that of Star B. This gives
us an upper limit on the energy available to go into oscillations at the
glitch, $\Delta E_{BD}=E_B - E_D$. 

In this section, we will make an estimate for both the upper limit on energy released, $\Delta E_{BD}$,
and the energy released in our specific model, $\Delta E_{CD}$. 
To do this, we will start by building on the starquake model of \cite{1971Baym}. 
This model characterises the distortion of the rotating neutron star
using a single parameter, the oblateness parameter $\vep$. 
This is defined by

\begin{equation}
\label{glitch:oblate}
I = I_{\text{S}} (1+\vep),
\end{equation}
where $I$ is the moment of inertia of the distorted star and $I_{\text{S}}$ is that of the star while nonrotating
and spherical. For the incompressible stellar models we will consider in our toy model,
$I_{\text{S}}$ would be that of a spherical star of the same volume. 
%For a neutron star of radius 10km and mass 1.4$\msun$, we have $I_{\text{S}} \sim 1 \times 10^{45}$ g cm$^2$.

The value of $\vep$ is then found by minimising the total energy of the star.
For each equilibrium configuration A, B and D, this can be written in the form

\begin{equation}
\label{energies:energy recap}
E = E_{\text{S}} + \frac{1}{2}I_S(1+\vep)\Omega^2 + A\vep^2 + B(\vep - \vep_{\text{ref}})^2,
\end{equation}
where $E_{\text{S}}$ is the energy of the spherical star, $A$ and $B$ are constants, and the 
three corrections are kinetic, gravitational and strain energy terms
respectively. The strain energy depends on both the current oblateness $\vep$ of
the star, and the `reference' oblateness $\vep_{\text{ref}}$ at which the star is
unstrained. By minimising with respect to the ellipticity
$\vep$ at fixed angular momentum, the ellipticity can be written as

\begin{equation}
\label{energies:epsilon recap}
\vep = \frac{I_S\Omega^2}{4(A+B)}  + \frac{B}{A+B} \vep_{\text{ref}}.
\end{equation}
It will be useful to get an idea of the relative
sizes of these energies by putting in some typical numerical
values for a neutron star. 
\cite{1971Baym} make an estimate for $A$ using the exact value for a homogeneous
incompressible star,

\begin{equation}
\label{A estimate}
A = \frac{3}{25}\frac{M^2G}{R}.
\end{equation}
%\begin{align}
%\label{B estimate}
%B = \frac{38}{25}\mu R^3, \\
%\label{IS estimate}
%I_S = \frac{2}{5}MR^2, \\
%\label{T estimate}
%T = \frac{1}{2}I_S \Omega^2.
%\end{equation}
%For a 10 km radius star with a mass of 1.4$\msun$, we obtain $A \sim 6\times 10^{52}$ erg.
As a rough estimate for the value of $B$, 
the form $E_{\text{strain}}\equiv 
B\left(\vep - \vep_{\text{ref}}\right)^2$ for the strain energy can be used to find the mean
stress in the crust,
 
\begin{equation}
\label{glitch:stress}
\sigma \equiv \left| \frac{1}{V_{\text{crust}}} \frac{\partial E_{\text{strain}}}{\partial \vep} \right|,
\end{equation}
where $V_{\text{crust}}$ is the volume of the crust. This gives 

\begin{equation}
\label{glitch:stress2}
\sigma = \mu(\vep - \vep_{\text{ref}}),
\end{equation}
with $\mu =\frac{2B}{V_{\text{crust}}}$ as the mean shear modulus of the crust,
which can then be rearranged to find $B$. 
For a a 10 km radius star with a mass of 1.4$\msun$, 
we obtain $I_{\text{S}} \sim 1 \times 10^{45}$ g cm$^2$ and $A \sim 6\times 10^{52}$ erg.
Using the estimate of \cite{1991Strohmayer} of $10^{30}$ erg cm$^{-3}$
for the shear modulus of the crust, we have
$B \sim 6 \times 10^{47}$ erg, assuming the crust is 1 km thick. 

The kinetic energy is of order $T= I_{\text{S}}\Omega_B^2$;
again, this is generally small compared to the gravitational binding energy.
%($T \sim 4 \times 10^{49}$ erg for the Crab pulsar, for example).
As these energies will recur frequently throughout, we will define

\begin{align}
\label{glitch:b}
b &\equiv \frac{B}{A} \sim 9 \times 10^{-6}, \\
\label{glitch:t}
t &\equiv \frac{T}{A} \sim 7 \times 10^{-7} \left(\frac{\Omega_B /2\pi}{\text{1\,Hz}}\right)^2.
\end{align}
Rewriting in terms of $t$ and $b$ and using the assumption that $b$ is small, we can approximate $\vep$
\eqref{energies:epsilon recap} as  

\begin{equation}
\label{energies:eps blla}
\vep = \frac{t}{4}\left(\frac{\Omega}{\Omega_B}\right)^2(1-b) + b\,\vep_{\text{ref}}
+ O(b^2).
\end{equation}

\subsection{Energies released in the starquake}
We can now apply these results to the stages of our starquake model.
Star A is unstrained, so that $\vep_{\text{ref}}=\vep_A$.
This yields an ellipticity of

\begin{equation}
\label{energies:epsA}
\vep_A =  \frac{I_{\text{S}}\Omega_A^2}{4A}.
\end{equation}
%and the corresponding energy is 
%
%\begin{equation}
%\label{glitch:EA}
%E_A = E_S + \frac{1}{2}I_S(1+\vep_A)\Omega^2_A + A\vep_A^2.
%\end{equation}
The star then spins down to angular velocity $\Omega_B$, still
with $\vep_{\text{ref}}=\vep_A$, so that

\begin{equation}
\label{energies:epsB}
\vep_B = \frac{I_S\Omega_B^2}{4A}(1-b) + b\,\vep_A
\end{equation}
to first order. 
We will now introduce the variable

\begin{equation}
\label{energies:x}
X= \frac{\Omega_A^2-\Omega_B^2}{\Omega_B^2}
\end{equation}
and rewrite in terms of $t$ \eqref{glitch:t}, so that the ellipticities of Stars A and B become

\begin{align}
\label{energies:epsA bxt}
\vep_A &=  \frac{t}{4}(1+X),  \\
\label{energies:epsB bxt}
\vep_B &= \frac{t}{4}\left(1+bX\right). 
\end{align} 
We would expect $X$, which is related to the difference
in angular velocity between Star A and Star B, to also be a small quantity.
This is certainly the case for the glitches seen in the observed pulsar population.
Here $t$ is defined in terms of $\Omega_B$, so that although $\vep_A$
appears to depend on $X$, the $\Omega_B$ terms actually cancel. 

Next we consider what happens after the starquake. 
At the glitch, all strain is lost,
so that the reference ellipticity is set to that of Star B.
This removal of strain will cause the star to be temporarily out
of equilibrium (Star C). It will then oscillate before damping down to the new
equilibrium state Star D. 
We can calculate the oblateness $\vep_D$ of this star again by using our
general expression for the ellipticity \eqref{energies:eps blla}. Star
D is rotating with angular velocity $\Omega_D$ and unstrained
at a reference ellipticity of $\vep_B$, so that 

\begin{equation}
\label{energies:eps blla vepD}
\vep_D = \frac{t}{4}\left(\frac{\Omega_D}{\Omega_B}\right)^2(1-b) + b\,\vep_B
\end{equation}
to first order in $b$. Rewriting in terms of $\vep_B$ \eqref{energies:epsB bxt}, we obtain

\begin{equation}
\label{energies:epsD}
\vep_D = \frac{t}{4}\left[
\left(\frac{\Omega_D}{\Omega_B}\right)^2 
+ b\left(1-\left(\frac{\Omega_D}{\Omega_B}\right)^2\right)
\right],
\end{equation}
where we have written this in terms of the still unknown
angular velocity $\Omega_D$ of Star D. 
We can find $\Omega_D$, given the fact that angular momentum is conserved
at the glitch, so that

\begin{equation}
\label{energies:ang m}
\Omega_B (1+\vep_B) = \Omega_D (1+\vep_D).
\end{equation}
Substituting in the expressions for $\vep_B$ \eqref{energies:epsB bxt}
and $\vep_D$ \eqref{energies:epsD}, we obtain

\begin{equation}
\label{energies:ang m bxt}
1+ \frac{t}{4}\left(1+bX\right) = \frac{\Omega_D}{\Omega_B}\left[
1+\frac{t}{4}\left(\left(\frac{\Omega_D}{\Omega_B}\right)^2 + b\left(1-\left(\frac{\Omega_D}{\Omega_B}\right)^2\right)\right)
\right].
\end{equation}
The right hand side has a term of order $\Omega_D^3$, but 
we can simplify this by making one final approximation, which
is that we know the fractional change in angular velocity at the glitch is small. Defining this as

\begin{equation}
\label{energies:delta OmegaBD}
z \equiv \frac{\Omega_D-\Omega_B}{\Omega_B}
\end{equation}
we can solve \eqref{energies:ang m bxt} to first order in $z$ as

\begin{equation}
\label{02}
z = \frac{btX}{4+3t(1-b)}.
\end{equation}
As a check, we see that if
$b=0$ (zero strain energy) this reduces to $z=0$, as expected.
We will keep only terms up to second order in either $b$ or $t$, so 
that

\begin{equation}
\label{energies:z b firstorder}
z= \frac{1}{4}btX.
\end{equation}
We would expect the change in angular velocity between the unstrained
state, Star A, and the state at the glitch, Star B, to be relatively 
small, given the observed time between glitches, so that $X= \left( \frac{\Omega_A^2}{\Omega_B^2}-1\right)$ is small. 
This gives us a consistency
constraint for the model: we can only use it for a change in spin
at the glitch $z=\frac{\Delta\Omega_{BD}}{\Omega_B}$ that gives a sensible
result for $X$. Putting in typical values of $b$ and $t$, we have 

\begin{equation}
\label{energies:small glitch}
\frac{\Delta\Omega_{BD}}{\Omega_B} \sim 10^{-8} 
\left(\frac{b}{10^{-5}}\right)\left(\frac{t}{10^{-3}}\right)\left(\frac{X}{1}\right).
\end{equation}
This means that the model only works for small glitches. For example, we can make an estimate of
the predicted size of glitches our formula gives for the Crab. We have
$t\sim 6\times 10^{-4}$, and we can make an estimate of $X \sim 10^{-3}$ based on approximately 
one glitch observed per year and a spindown timescale of $\sim 10^3$ years.
Assuming a value of $b\sim 10^{-5}$, we then find $\frac{\Delta\Omega_{BD}}{\Omega_B} \sim 6\times 10^{-12}$, 
much smaller than the largest observed Crab glitches of size $\sim 10^{-8}$.

We can then substitute our result for $z$ back into 
the ellipticity $\vep_D$ \eqref{energies:epsD}.
If we keep only terms up to first order in either $b$ or $t$,
no terms depending on $z$ remain and we are left with just

\begin{equation}
\label{energies:epsD bt}
\vep_D = \frac{t}{4},
\end{equation}
i.e. we find that to leading order, Star D is the same shape as a purely
fluid star rotating with angular velocity $\Omega_B$.
These expressions for $\vep_D$, $\vep_B$ \eqref{energies:epsB bxt} and $\vep_A$ 
\eqref{energies:epsA bxt} can now
be used to calculate the energy released at the glitch, both for our upper
estimate $\Delta E_{BD}$ and for our specific toy model, $\Delta E_{CD}$.

The energy of Star B can be written as

\begin{equation}
\label{energies:EB bt}
E_B = E_{\text{S}} + A\left[
\frac{1}{2}t(1+\vep_B)+ \vep_B^2 + b(\vep_B - \vep_A)^2
\right].
\end{equation}
To be consistent with our approximation for $\vep_D$, we will
keep terms up to fourth order in either $b$ or $t$, so that

\begin{equation}
\label{energies:EB 4order}
E_B=E_S+
A\left[\frac{t}{2}+\frac{3t^2}{16}
+\left(\left(\frac{X}{4}+\frac{X^2}{16}\right) t^2\right) b
-\frac{1}{16}X^2 t^2b^2
\right].
\end{equation}
After the glitch, Star D has energy

\begin{equation}
\label{energies:ED bt}
E_D = E_{\text{S}} + A\left[
\frac{1}{2}t(1+\vep_D)
+ \vep_D^2 + b(\vep_D - \vep_B)^2
\right],
\end{equation}
or, again keeping terms up to fourth order, 

\begin{equation}
\label{energies:ED 4order}
E_D=E_S+
A\left[
\frac{t}{2}+\frac{3t^2}{16}+ \left(\left(\frac{X}{4} +\frac{Xt}{16} \right)t^2\right)b
\right].
\end{equation}
We can now calculate the maximum energy available to go into
oscillations, $\Delta E_{BD} = E_B - E_D$. The lowest order
term of this is

\begin{equation}
\label{energies:Delta EBD}
\Delta E_{BD} = \frac{1}{16} A X^2 t^2 b.
\end{equation}
We can use our expression for $z$ \eqref{energies:z b firstorder} to rewrite this as
\begin{equation}
\label{energies:Delta EBD rewrite}
\Delta E_{BD} \sim AXtz \sim X I \Omega_B^2  \left(\frac{\Delta \Omega_{DB}}{\Omega_B}\right).
\end{equation}
Comparing this with the form \eqref{literature} that has appeared in the literature
as an upper bound on the energy made available at a glitch, we see that in our case the
energy is suppressed by a factor of $X$. \color{black}

To find the scaling for the amplitudes of the oscillations excited, we can equate this energy to 
the kinetic energy of the oscillations $\left(E_{\text{mode}}\right)_{BD}$,
which has the approximate form

\begin{equation}
\label{energies:Emode}
\left(E_{\text{mode}}\right)_{BD} \sim I_S \omega^2 \alpha_{BD}^2
\end{equation}
where $\omega$ is the frequency of the oscillations and
$\alpha_{BD}$ is a dimensionless number characterising
their amplitude.
Using this, 

\begin{equation}
\label{energies:alpha bxt}
\alpha_{BD} \sim\left(\frac{1}{\omega}\right) \, \sqrt{\frac{A}{I_S}} \sqrt{b} \, tX.
\end{equation}
%We can rewrite this in terms of $z$ \eqref{energies:z b firstorder} as 
%
%\begin{equation}
%\label{energies:alpha z}
%\alpha_{BD}\sim \frac{4}{\sqrt{b}} \, \sqrt{\frac{A}{I_S}}
%\left(\frac{1}{\omega}\right) z.
%\end{equation}
%Note that this is independent of the rotation rate $\Omega_B$.
%
Again using our previous estimates of
$A$ and $I_S$, we have

\begin{equation}
\label{energies:numerical estimate}
\alpha_{BD}\sim 2\times 10^{-6} \,
\biggl(\frac{\omega/2\pi}{2000 \,\text{Hz}}\biggr)^{-1}
\biggl(\frac{b}{10^{-5}}\biggr)^{\frac{1}{2}}
%\biggl(\frac{A}{10^{-8}}\biggr)^{\frac{1}{2}}
%\biggl(\frac{I_S}{10^{-8}}\biggr)^{-\frac{1}{2}}
\biggl(\frac{t}{10^{-3}}\biggr)
\biggl(\frac{X}{1}\biggr).
\end{equation}
Here we have used a typical $f$-mode frequency of 2000 Hz for the mode excited.
This amplitude estimate corresponds to 
displacements at the surface of the order $\delta r\sim\alpha_{BD} R$, i.e.
surface oscillations of around 1 cm, in the maximally optimistic case of large spin
down between glitches ( $X\sim 1$). 
%This could be of interest in terms of
%electromagnetic radiation from the shaking of field lines at the surface.

To calculate the energy $\Delta E_{CD}$ released in our specific model,
we also need the energy of Star C. Star C has the same oblateness $\vep_B$ as
Star B, but its energy differs from $E_B$ \eqref{energies:EB bt}
in that the strain energy has been removed, so that it has energy 

\begin{equation}
\label{arb:EC bt}
E_C = E_{\text{S}} + A\left[
\frac{1}{2}t(1+\vep_B)+ \vep_B^2 \right].
\end{equation}
Subtracting this from $E_D$ \eqref{energies:ED 4order} and again keeping terms
only up to fourth order in $b$ and $t$, we find that

\begin{equation}
\label{03}
\Delta E_{CD} = \frac{1}{16} A\, b^2 X^2 t^2.
\end{equation}
Note the extra factor of $b$ compared to the energy change $\Delta E_{BD}$
\eqref{energies:Delta EBD} above; this means that the energy released
in our model will be considerably less than that in our earlier
upper estimate. The corresponding amplitude excited in this case will be

\begin{equation}
\label{energies:alpha z}
\alpha_{CD}\sim 4
\, \sqrt{\frac{A}{I_S}}
\left(\frac{1}{\omega}\right) z,
\end{equation}
a factor of $\sqrt{b}$ smaller than the upper limit $\alpha_{BD}$.

We can also make estimates for the gravitational wave emission from a 
starquake. For this, we will look at the case where an $l=2$, $m=0$ 
spheroidal oscillation mode has been excited, with the amplitude 
$\alpha_{BD}$ just calculated. For these modes, the pulsating stellar surface can be 
described by the polar equation

\begin{equation}
\label{gws:stellar surface}
r(\theta,t) = R(1  + \alpha_{BD}e^{-i\omega t}P_2 (\cos\theta)),
\end{equation}
where $\omega$ is the frequency of the oscillation. 
This is a reasonable choice of oscillation mode, given that the rotating star also has this shape 
and that the starquake produces a change in ellipticity of the star. In particular, the toy model
we will construct largely excites modes of this type.
It can be shown that for a constant density star, oscillations of this type will produce
a gravitational wavefield that can be chosen to be purely in the `plus' polarisation, with

\begin{equation}
\label{gws:h+}
h_+ = -\frac{3G}{5c^4}\omega^2 \alpha_{BD}\sin^2\theta MR^2 \frac{e^{i\omega(r-t)}}{r}.
\end{equation}
The rate of energy loss from this mode is 

\begin{equation}
\label{gws:luminosity}
\dv{E}{t} = -\frac{3G}{125c^5} \,\alpha_{BD}^2 \,\omega^6 M^2 R^4,
\end{equation}
a result first shown by \cite{1967Chau}.
Comparing this with our expression $\left(E_{\text{mode}}\right)_{BD}$ \eqref{energies:Emode} 
for the initial energy in the mode, we find that the energy as a function of time satisfies

\begin{equation}
\label{gws:E t}
E_{BD}(t) \sim  \left(E_{\text{mode}}\right)_{BD} e^{-\frac{t}{2\tau}},
\end{equation}
where $\tau$ is the damping timescale,

\begin{equation}
\label{gws:tau}
\tau = \frac{100c^5}{3GMR^2\omega^4}\,.
\end{equation}
The energy $E_{BD} \propto h_+^2$, and so the wavefield decays as

\begin{equation}
\label{gws:h plus full}
h_+(t) = -\frac{3G}{5c^4}\omega^2 \alpha_{BD}\sin^2\theta MR^2\frac{e^{i\omega(r-t)}}{r} e^{-\frac{t}{\tau}}.
\end{equation}
We now make estimates for a homogeneous fluid star with
a mass of 1.4$\msun$ and a radius of 10 km. We will again 
put in an estimate of $2000$ Hz for the excited mode, appropriate
for an $f$-mode.
The damping timescale \eqref{gws:tau} is then

\begin{equation}
\label{gws:tau numerical}
\tau \sim 0.07 \; \text{s}.
\end{equation}
Taking the Crab's rotation rate of 30Hz as a typical example, we
can put in our value of $\alpha_{BD}\sim 2\times 10^{-6}$. For a source at a 
distance of 1 kiloparsec, the strain $h_+$ is then

\begin{equation}
\label{gws:hplus numerical}
h_+ \sim 1 \times 10^{-23}\,
\biggl(\frac{\omega/2\pi}{2000\, \text{Hz}}\biggr)
\biggl(\frac{r}{1 \,\text{kpc}}\biggr)^{-1}
\biggl(\frac{b}{10^{-5}}\biggr)^{\frac{1}{2}}
\biggl(\frac{t}{10^{-3}}\biggr)
\biggl(\frac{X}{1}\biggr)
%\biggl(\frac{A}{10^{-8}}\biggr)^{\frac{1}{2}}
%\biggl(\frac{I_S}{10^{-8}}\biggr)^{-\frac{1}{2}}
%\biggl(\frac{z}{10^{-8}}\biggr),
\end{equation}
where we have used our value for the amplitude $\alpha_{BD}$
\eqref{energies:numerical estimate} (note that our estimate for $z$ used the
high value of $X=1$, i.e. the star spins down considerably before the quake occurs). 
We can compare this to the upper limits on the strain from the Vela pulsar, of
$h = 1.4 \times 10^{-20}$ for $l=2$, $m=0$ modes \cite{LSCVela}; this is a meaningful 
comparison, as we are looking at oscillations of similar frequency and duration.
Our estimate for the amplitude of $h_+$ produced by a starquake is around three orders 
of magnitude smaller, meaning that detection is unlikely except in the improbable case of 
a very close and rapidly spinning star.

%For the purposes
%of comparing to a detector sensitivity curve, we are more interested in 
%the characteristic strain, 
%
%\begin{equation}
%\label{gws:hplus tau}
%h_+\sqrt{\tau} \sim 4 \times 10^{-24}\,
%\biggl(\frac{\omega/2\pi}{2000\, \text{Hz}}\biggr)
%\biggl(\frac{r}{1 \,\text{kpc}}\biggr)^{-1}
%\biggl(\frac{b}{10^{-5}}\biggr)^{-\frac{1}{2}}
%%\biggl(\frac{A}{10^{-8}}\biggr)^{\frac{1}{2}}
%%\biggl(\frac{I_S}{10^{-8}}\biggr)^{-\frac{1}{2}}
%\biggl(\frac{z}{10^{-8}}\biggr) \; \text{Hz}^{-\frac{1}{2}}.
%\end{equation} 
%This level of strain should be detectable with third generation
%detectors. Figure \ref{fig:et} shows a preliminary sensitivity 
%curve for the Einstein Telescope (\cite{2009Punturo}). Our upper
%estimate is plotted for a star
%oscillating at an $\textit{f}$-mode frequency 2000Hz and lies above
%the curve. The figure also shows the planned Advanced LIGO sensitivity curve for 
%a representative choice of detector configuration (the Zero Det, High
%Power configuration) (\cite{2009Shoemaker}). Our upper estimate on the 
%strain level is below this curve.

\subsection{Estimates for solid quark stars}
The estimates given so far have been for a normal neutron star
with a crust around 1 km thick. 
More exotic models have however been proposed, including quark stars
with a large solid core (\cite{Glendenning,2003Xu}). These models are of interest 
in terms of gravitational wave emission, as they are expected to 
be able to sustain a larger maximum ellipticity (\cite{2005Owen,2007Haskell,2007Lin}).
More relevant to us here is the possibility that a solid star of this type 
could be expected to undergo a much larger quake. 

In this section we will make some estimates for this scenario, based 
on a solid star with a shear modulus of 
$\mu_{\text{solid}} \sim 4 \times 10^{32}$ erg cm$^{-3}$ \cite{2003Xu}.
For a completely solid star of radius 10 km, we have $B_{\text{solid}} \sim 8\times 10^{50}$ erg,
i.e. $b_{\text{solid}} \equiv \frac{B_{\text{solid}}}{A}
\sim 1\times 10^{-2}$, a much larger value than our previous estimate of $b\sim 10^{-5}$ 
for a star with a fluid core. This means that the star in our model can sustain larger glitches:
our criterion \eqref{energies:small glitch} that the spin down between Stars A and B is
of order unity yields a fractional change in spin rate $z_{\text{solid}}$ of

\begin{equation}
\label{04}
z_{\text{solid}}\sim 
2\times 10^{-6} \left(\frac{b_{\text{solid}}}{10^{-2}}\right)\left(\frac{t}{10^{-3}}\right)\left(\frac{X}{1}\right).
\end{equation}
We are thus able to use this model for glitches of around a factor of
100 larger than for a normal neutron star model.
Using our new value of $b_{\text{solid}}$,
the amplitude of the oscillations in our model \eqref{energies:alpha z} is

\begin{equation}
\label{05}
\left(\alpha_{BD}\right)_{\text{solid}}  \sim 1\times 10^{-2} \,
\biggl(\frac{b_{\text{solid}}}{10^{-2}}\biggr)^{\frac{1}{2}}
\biggl(\frac{t}{10^{-3}}\biggr)
\biggl(\frac{X}{1}\biggr)
\biggl(\frac{\omega/2\pi}{2000 \,\text{Hz}}\biggr)^{-1},
%\biggl(\frac{z_{\text{solid}}}{10^{-6}}\biggr),
\end{equation}
and the corresponding value of the strain is

\begin{equation}
\label{gws:hplus numerical quark}
h_+ \sim 3 \times 10^{-22}\,
\biggl(\frac{\omega/2\pi}{2000\, \text{Hz}}\biggr)
\biggl(\frac{r}{1 \,\text{kpc}}\biggr)^{-1}
\biggl(\frac{b}{10^{-2}}\biggr)^{\frac{1}{2}}
\biggl(\frac{t}{10^{-3}}\biggr)
\biggl(\frac{X}{1}\biggr).
\end{equation}
This is still two orders of magnitude smaller that the Vela upper limit quoted previously, so 
would again require a rapidly spinning star with large spin down at the glitch to be 
detectable.

%
%\begin{equation}
%\label{gws:hplus tau}
%\left(h_+\sqrt{\tau}\right)
%\sim 5 \times 10^{-22} \; \text{Hz}^{-\frac{1}{2}},
%\end{equation} 
%a factor of 100 larger. 
%
%This is plotted on the sensitivity curve
%in Figure \ref{fig:et}, again for an oscillation at 2000 Hz, and this
%time is above the projected sensitivity curves for both ET and Advanced LIGO. 
%%figure: sensitivity curve%%%%%%%%%%%%%%%%%%%%%%%%%%%%%%%%%%%%%%%%%%
%\begin{figure}[H]
%\centering
%\includegraphics[scale=1]{figures/etdata.pdf}
%\caption[Einstein Telescope sensitivity curve]{A plot of the proposed Advanced LIGO (\textit{blue})
%and Einstein Telescope (\textit{red}) sensitivity curves, with our upper estimates for the
%characteristic strain produced by a normal neutron star with a fluid
%core (marked by a cross) and by a solid quark star (marked by a diamond).
%Both of these are plotted for a typical \textit{f}-mode frequency of 
%2000 Hz.}
%\label{fig:et}
%\end{figure}
%%%%%%%%%%%%%%%%%%%%%%%%%%%%%%%%%%%%%%%%%%%%%%%%%%%%%%%%%%%%%%%%%%%%%%
%\newline\newline
To summarise, in this section we have made estimates of the amplitudes of the 
oscillation produced in a
starquake, both for our toy model and a more optimistic scenario where all the
energy released in the glitch goes into mode oscillations. We have found
that in this optimistic case, amplitudes of $10^{-6} \delta R / R$ can be excited at the
surface of the star, for the case where the star spins down by a large amount at the glitch ($X \sim 1$).
For our specific toy model where the strain energy is lost at the glitch, we 
find that these amplitudes are suppressed by a factor of $\sqrt{b}$.
We then briefly considered how these estimates were altered in the case of a solid quark star, 
where $b_{\text{solid}}$ could be as large as $10^{-2}$, allowing for much larger 
glitches. In this case, the upper bound from the optimistic scenario gives
amplitudes of $10^{-2} \delta R / R$.

%%%%%%%%%%%%%%%%%%%%%%%%%%%%%%%%%%%%%%%%%%%%%%%%%%%%%%%%%%%%%%%%%%
% INITIAL DATA
%%%%%%%%%%%%%%%%%%%%%%%%%%%%%%%%%%%%%%%%%%%%%%%%%%%%%%%%%%%%%%%%%%

\section{Construction of the initial data}
\label{sec:initialdata}
We will next carry out the first stage of the more detailed glitch model
discussed in Section \ref{sec:model}, by constructing the initial
displacement and velocity fields $\boxi^{\text{DC}}$ and $\dot{\boxi}^{\text{DC}}$.
We describe Stars A--D in terms of their deformation from the spherical 
background, Star S. This star has constant density $\rho$, pressure $p$, 
and a gravitational potential $\Phi$ obeying Poisson's equation

\begin{equation}
\label{poisson}
\nabla^2 \Phi = 4\pi G \rho,
\end{equation}
and is in hydrostatic equilibrium

\begin{equation}
\label{zero:hydrostatic}
\nabla_i \tau\indices{^i_j} - \rho\nabla_j\Phi =0,
\end{equation}
where $\tau_{ij}$ is the isotropic stress tensor of the background star,
$\tau_{ij}\equiv -p\,\delta_{ij}$.
The spherically symmetric solutions of these equations have pressure

\begin{equation}
\label{zero:p}
p(r) = \frac{2\pi}{3} G\rho^2 (R^2-r^2),
\end{equation}
where $R$ is the radius of the star.
We work in terms of Lagrangian displacements $\boxi$ between configurations,
using the definition of e.g. \cite{Shapiro}, in 
which the Lagrangian perturbation of a quantity $Q$ is related to the Eulerian perturbation by

\begin{equation}
\label{EL}
\Delta Q \equiv \delta Q + \xi^i\nabla_i Q.
\end{equation}
To keep track of Lagrangian perturbations
between different configurations, we label maps from the spherical background
to Stars A-D with displacement fields $\boldsymbol{\eta}$, and maps between Stars A-D
with displacement fields $\boldsymbol{\xi}$ (see Figure \ref{fig:potatoes}). 

The velocity initial data is just generated by the change in angular velocity from 
$\Omega_B = \Omega_C$ to $\Omega_D$ at the glitch, so that we have

\begin{equation}
\label{arb:vel}
\boldsymbol{\dot{\boxi}}^{\text{DC}} = r\Delta\Omega\,\boldsymbol{\hat{\phi}},
\end{equation}
where $\Delta \Omega \equiv \left(\Omega_D-\Omega_B\right)$.
To find the displacement field $\boldsymbol{\xi}^{\text{DC}}$, we have to solve the equations of
motion for Star D, regarded as a perturbation about Star S. 
Star D is in equilibrium and rotating at angular velocity $\Omega_D$, with a centrifugal 
force governed by the potential $\phi^c_D = \Omega_D^2 r^2 \sin^2 \theta$. We will
decompose this as

\begin{equation}
\label{centrifugal}
\phi^c_D = \frac{\Omega_D^2r^2}{3}\left(1 - P_2(\cos\theta) \right)
\end{equation}
where $P_2(\cos\theta)$ is the second Legendre polynomial, and look for solutions with
this same symmetry.

The rotation will introduce corresponding Lagrangian perturbations $\Delta p^{\text{SD}}$
and $\Delta \Phi^{\text{SD}}$ in the star.
The star is also a strained configuration, and its unstrained state is that of Star C. 
Star C is out of equilibrium and cannot be described so easily as a perturbation about S, but 
for our model it will be enough to specify its surface shape.
Star D then obeys the equation of motion

\begin{equation}
\label{forces D}
-\rho\nabla_i\phi^c_D = \nabla_j (\Delta \tau^{\text{SD}})\indices{_i^j}  
- \rho\nabla_i\Delta\Phi^{\text{SD}},
\end{equation}
where the perturbed stress tensor $(\Delta \tau^{\text{SD}})_{ij}$ has the form

\begin{equation}
\label{Delta T D 2}
(\Delta \tau^{\text{SD}})_{ij} = -(\Delta p^{\text{SD}})\delta_{ij} + 2\mu u^{\text{CD}}_{ij},
\end{equation}
Notice that the shear strain term $u^{\text{CD}}_{ij}$ is built from the 
displacement field $\boldsymbol{\xi}^{\text{CD}}$ connecting Stars C and D. 
This is because all strain is removed in the glitch, so that Star C is the new
reference state of zero strain. 
We also have Poisson's equation for the gravitational perturbation,
and the incompressibility condition

\begin{align}
\label{poisson D}
\nabla^2 \delta\Phi^{\text{SD}} &= 4\pi G\delta \rho^{\text{SD}}, \\
\label{cont D}
\nabla_i \left(\eta^{\text{SD}}\right)^i &= 0.
\end{align}
Incompressibility means that the Lagrangian perturbation of the density $\Delta \rho$ is zero.
This gives us a formula for the Eulerian perturbation,

\begin{equation}
\label{eqm:delta rho}
\delta\rho = \rho \delta(r-R)\eta^r,
\end{equation}
where $\eta^r$ is the radial component of $\eta^i$ and $\delta(r-R)$ is the Dirac delta function. We
can see that this vanishes everywhere except at the surface.  
By substituting this into Poisson's equation \eqref{poisson D}, 
the gravitational perturbation $\delta\Phi^{\text{SD}}$ can be shown to satisfy the jump conditions 

\begin{align}
\label{phi cont D}
\left[\delta \Phi^{\text{SD}}\right]^{R+\epsilon}_{R-\epsilon} =0,\\
\label{phi discont D}
 \left[\frac{d}{dr}(\delta \Phi^{\text{SD}})\right]^{R+\epsilon}_{R-\epsilon} 
 =4\pi G\rho \eta^{\text{SD}}_r(R,\theta,\phi)
\end{align}
at the surface. By matching solutions finite inside and outside the star, we find that

\begin{equation}
\label{zero:delta Phi SD UCD}
\Delta\Phi^{\text{SD}}= \frac{8\pi G\rho}{15} r \left(\eta'^{\text{SC}}_r + U^{\text{CD}}(r)\right)P_2(\cos\theta).
\end{equation}
We also have boundary conditions on the stress tensor at the surface.
Particles on the surface of Star S, where $\tau_{ir}=0$, should still 
satisfy the same condition (`zero traction') after the perturbation, leading to the conditions

\begin{equation}
\label{traction D}
\Delta\tau_{ir}^{\text{SD}}(R)=0.
\end{equation}
This gives two independent boundary conditions for the $(rr)$ and $(r\theta)$ components.
To solve these equations, we will need to make use of the equilibrium configuration
of Star B, which was unstrained at angular velocity $\Omega_A$ and is now spinning at angular velocity $\Omega_B$.
The displacement field $\boxi^{\text{AB}}$ is found in \cite{1971Baym}
to be of the form

\begin{equation}
\label{xiAB}
\boldsymbol{\xi}^{\text{AB}} = U^{\text{AB}}(r)P_2(\cos\theta) \boldsymbol{\hat{r}} 
+ V^{\text{AB}}(r) \babla P_2,
\end{equation}
where $P_2(\cos\theta)$ is the second Legendre polynomial and 
the radial functions $U^{\text{AB}}$ and $V^{\text{AB}}$ are given by

\begin{align}
\label{eqm:collect UAB}
U^{\text{AB}}(r)&=\frac{1}{20R^2}\frac{tX}{1+b}
\left(8R^2r-3r^3\right), \\
\label{eqm:collect VAB}
 V^{\text{AB}}(r)&=\frac{1}{20R^2}\frac{tX}{1+b}
\left(4R^2r^2-\frac{5}{2}r^4\right).
\end{align}
These have been written in terms of our small parameters $b$, $t$
and $X$ \eqref{energies:x}, where for a homogeneous elastic star $b$ and $t$ have the exact values

\begin{align}
\label{b exact}
b &= \frac{57\mu}{8\pi G \rho^2R^2},
\\
\label{t exact}
t &= \frac{5}{2\pi G\rho}\Omega_B^2.
\end{align}
We can also find the surface shape of the Star B (and therefore Star C)
by calculating a map from the surface of Star S to that of Star B,

\begin{equation}
\label{eqm:eta SB R}
\left(\eta^r\right)^{\text{SB}}(R) 
= -\frac{tR}{4}\frac{1}{1+b}\left(1+b\,\frac{\Omega_A^2}{\Omega_B^2}\right)P_2(\cos\theta).
\end{equation}
For the rest of the calculation, we follow a similar method to that of \cite{2000Franco}, defining

\begin{equation}
\label{arb:hSD}
\mu h^{\text{SD}}= -\Delta p^{\text{SD}} -\rho\Delta\Phi^{\text{SD}}+\rho\nabla_i(\Omega_D^2r^2\cos^2 \theta).
\end{equation}
so that $\nabla^2 h^{\text{SD}} = 0$. The solutions with the same symmetries as the 
centrifugal potential satisfy

\begin{equation}
\label{arb:h2 H2H0}
h^{\text{SD}} = H_2r^2P_2(\cos\theta) + H_0,
\end{equation}
for $H_2$ and $H_0$ constant.
We decompose the displacement vector $\xi^{\text{CD}}_i$ as

\begin{equation}
\label{arb:xiCD decomposition}
\boldsymbol{\xi}^{\text{CD}} = U^{\text{CD}}(r)P_2(\cos\theta)\boldsymbol{e}_r + V^{\text{CD}}(r)\boldsymbol{\nabla}P_{2}(\cos\theta).
\end{equation}
Substituting this back into the force equation \eqref{forces D} and
using incompressibility, we find that

\begin{align}
\label{UCD CH2}
U^{\text{CD}}(r) &= Cr - \frac{1}{7}H_2r^3, \\
\label{VCD CH2}
V^{\text{CD}}(r)&=\frac{1}{2}Cr^2 - \frac{5}{42}H_2r^4,
\end{align}
where $C$ is another constant. We can then use our surface
boundary conditions to fix $C$ and $H_2$. The $(r\theta)$ component
of the surface boundary condition \eqref{traction D} 
gives $C=\frac{8}{21}H_2R^2$.
For the $(rr)$ component, we rewrite $\Delta p^{\text{SD}}$ in terms
of $h^{\text{SD}}$, and substitute this in along
with our expression for the gravitational potential
$\Delta\Phi^{\text{SD}}$ \eqref{zero:delta Phi SD UCD}, obtaining

\begin{align}
\label{arb:H0}
H_0 &= \frac{\rho}{3\mu}\Omega_D^2R^2,
\\
\label{arb:H2 initial}
H_2 &= - \frac{21}{57\mu R^2 + 8\pi G\rho^2 R^4}\cdot \left(
\rho\Omega_D^2R^2 + \frac{8\pi G\rho^2R}{5}\eta'^{\text{SC}}_r (R)
\right).
\end{align}
It remains to include the shape of Star C at the surface, $\eta'^{\text{SC}}_r (R)$ \eqref{eqm:eta SB R}.
Inserting this and rewriting in terms of $b$ and $t$, we have

\begin{equation}
\label{arb:H2}
H_2= -\frac{21}{20 R^2} \frac{t}{1+b}\left(
\frac{\Omega_D^2}{\Omega_B^2} -  \frac{1}{1+b}\left(1+b\frac{\Omega_A^2}{\Omega_B^2}\right)\right).
\end{equation}
This is still written in terms of $\Omega_D$, which we can eliminate
using conservation of momentum \eqref{energies:ang m}, as in Section \ref{sec:estimates}.
Keeping terms only up to second order in either $b$ or $t$, we find that

\begin{equation}
\label{H2 approx}
H_2 = \frac{21}{20 R^2} btX.
\end{equation}
Rewriting in terms of the fractional change in angular velocity at
the glitch $z$ \eqref{energies:z b firstorder} and substituting $H_2$ into 
our form for the radial functions $U^{\text{CD}}$ \eqref{UCD CH2}
and $V^{\text{CD}}$ \eqref{VCD CH2}, we finally obtain 

\begin{equation}
\label{xiCD 1}
\boxi^{\text{DC}} = 
\left(\frac{5z}{R^2} \left(3r^3-8R^2r\right)\right)P_2\,\boldsymbol{\hat{r}} 
+ \left(\frac{5z}{R^2} \left(\frac{5}{2} r^4-4R^2r^2\right)\right)\babla P_2
\end{equation}
as our displacement field initial data.

We can now compare the results of this calculation with the estimate we made in Section \ref{sec:estimates} for
the size of the amplitude $\alpha_{CD}$ \eqref{energies:alpha z}. 
As a measure of the amplitude of oscillations we expect to produce
with our initial data, we can take 

\begin{equation}
\label{U scaling}
\frac{U^{\text{CD}}(R)}{R}= -25z. 
\end{equation}
To compare this with $\alpha_{CD}$, we can substitute the exact values of $A$ and $I_S$ 
\eqref{A estimate} for an incompressible star,
and assume excitation of an $f$-mode with scaling $\omega\sim \sqrt{G\rho}$
(we shall see that when we carry out the projection, we do preferentially 
excite a mode with this scaling). 
Given this, we have $\sqrt{\frac{A}{I_S}}
\left(\frac{1}{\omega}\right)\sim O(1)$ and so $\alpha_{CD}\sim z$, 
consistent with the scaling of our initial data \eqref{U scaling}.

%%%%%%%%%%%%%%%%%%%%%%%%%%%%%%%%%%%%%%%%%%%%%%%%%%%%%%%%%%%%%%%%%%
% NORMAL MODES
%%%%%%%%%%%%%%%%%%%%%%%%%%%%%%%%%%%%%%%%%%%%%%%%%%%%%%%%%%%%%%%%%%

\section{Oscillation modes of the rotating star}
\label{sec:modes}
In this section we will find the spectrum of eigenvalues and associated 
eigenfunctions for Star D; we will later project the initial 
data of Section \ref{sec:initialdata} against these eigenfunctions to find which modes are excited.
We will start by calculating the oscillation modes of the spherical star, Star S.
This problem was first studied for the constant density case we are interested
in by \cite{Bromwich}. We will first restate the problem and
summarise his analytic results
for the eigenvalues and eigenfunctions, and then carry out a further numerical
investigation of these results. Next, we will make the slow rotation approximation
and consider rotation as a perturbation about this background model.

An alternative approach would be to start by computing the oscillation modes of a rotating
\textit{fluid} star, and then adding elasticity as a small perturbation. We instead chose to 
use the exact results of Bromwich we had available for the non-rotating elastic star. \color{black}

%%%%%%%%%%%%%%%%%%%%%%%%%%%%%%%%%%%%%%%%%%%%%%%%%%
% Modes of a spherical star

\subsection{Oscillation modes of the spherical background star}
\label{subsec:spherical}
We are interested in normal mode solutions $\xi^i(x,t) = e^{i\omega t}\xi^i(x)$
of the equation

\begin{equation}
\label{modes:elastic force modes}
\rho \ddot{\xi}_i =-\nabla_j (\Delta \tau)\indices{_i^j} - \rho\nabla_i(\Delta\Phi),
\end{equation}
with stress tensor

\begin{equation}
\label{Delta T D}
\Delta \tau_{ij} = -(\Delta p)\delta_{ij} + 2\mu u_{ij},
\end{equation}
and surface boundary conditions

\begin{align}
\label{traction mode}
\Delta\tau_{ir}&=0,
\\
\label{phi cont mode}
\left[\delta \Phi\right]^{R+\epsilon}_{R-\epsilon} &=0,\\
\label{phi discont mode}
 \left[\frac{d}{dr}(\delta \Phi)\right]^{R+\epsilon}_{R-\epsilon} 
 &=4\pi G\rho \xi_r(R,\theta,\phi).
\end{align}
The eigenfunctions $\boldsymbol{\xi}$ can be classified into spheroidal and toroidal types, which
we will label $\boldsymbol{S}$ and $\boldsymbol{T}$. 
For each type, the eigenfunctions can be labelled by radial, azimuthal and axial eigenvalue numbers
$n$, $l$ and $m$.
We will specialise to axisymmetric $m=0$ solutions, as the star is axisymmetric in all stages
of our starquake model. The spheroidal eigenfunctions then have the form

\begin{equation}
\label{Sln}
\mathbf{S}^{ln} = U^{ln}(r)\boldsymbol{\hat{r}} + V^{ln}(r)\babla P_{l},
\end{equation}
with corresponding eigenvalues $\left(\omega_S\right)^{ln}$. Bromwich shows that these
eigenvalues can be found from the roots
$\left(k_S\right)^{ln}\equiv \sqrt{\frac{\rho}{\mu}}\left(\omega_S\right)^{ln}$ of the equation

\begin{equation}
\label{fs}
\begin{split}
f_S\left(k_S,l, \frac{\rho}{\mu}\right)\equiv\left(l+\frac{lgR}{(2l+1)}\left(\frac{\rho}{\mu}\right)
-\frac{k_S^2}{2(l-1)}\right)\left(k_S+2\frac{\psi'_l(k_S)}{\psi_l(k_S)}\right) \\
+(l+1)\left(k_S+2l\frac{\psi'_l(k_S)}{\psi_l(k_S)}\right),
\end{split}
\end{equation}
where the functions $\psi_l$ are defined in terms of the spherical Bessel functions 
$j_l$ as $\psi_l(x) \equiv x^{-l}j_l(x)$, and $g = \frac{4\pi G \rho R}{3}$.
The radial functions $U^{ln}(r)$ and $V^{ln}(r)$ appearing in the eigenfunctions \eqref{Sln}
are 

\begin{align}
\label{Uln}
U^{ln}(x)&= C^{ln}\left[
		lq^{ln}r^{l-1} +\frac{l(l+1)}{r}\frac{1}{k_S^{ln}} j_l((k_S)^{ln}r)
	\right]
\\
\label{Vln}
V^{ln}(x) &= C^{ln}\left[
		q^{ln}r^{l-1} + \frac{1}{kr}j_l((k_S)^{ln}r) + j_l'((k_S)^{ln}r) 
	\right]
\end{align}
where the $C^{ln}$ are arbitrary constants and $q^{ln}$ is the constant

\begin{equation}
\label{qln}
q^{ln} = \frac{1}{2(l-1)R^{l-3}}\left[2j_l'((k_S)^{ln}R) +  
             \left(\frac{2(l^2+l-1)}{(k_S)^{ln}R} - (k_S)^{ln}R \right)\right]j_l((k_S)^{ln}R).
\end{equation}
The axisymmetric toroidal eigenfunctions have the form

\begin{equation}
\label{Tln}
\mathbf{T}^{ln} =W^{ln}(r)(\boldsymbol{\hat{r}}\times\babla P_{l}).
\end{equation}
These eigenfunctions are simpler to analyse because they have no radial 
component and so are unaffected by gravity. The toroidal eigenvalues 
$\left(\omega_T\right)^{ln}$ can be found from the roots
$\left(k_T\right)^{ln}\equiv \sqrt{\frac{\rho}{\mu}}\left(\omega_T\right)^{ln}$ of the equation

\begin{equation}
\label{ft}
f_T(k_T,l) = k_Tj'_l(k_T)-j_l(k_T)=0.
\end{equation}
The radial functions $W^{ln}$ have the form

\begin{equation}
\label{Wln}
W^{ln} = -D^{ln}j_l \left(\left(k_T\right)^{ln}r\right),
\end{equation}
where the $D^{ln}$ are arbitrary constants.

To further investigate the eigenvalue spectrum and associated eigenfunctions,
we will need to carry out some of the work numerically. We will find 
spheroidal and toroidal eigenvalues by finding roots of the nonlinear functions
$f_S$ \eqref{fs} and $f_T$ \eqref{ft}.
We can then plot the corresponding eigenfunctions by using these eigenvalues
in conjunction with the analytic forms of the radial functions $U$ \eqref{Uln}, $V$ \eqref{Vln}
and $W$ \eqref{Wln}.

To illustrate this, in this section we will specialise to the case of $l=2$, $m=0$ modes.
These are the most important for our glitch model, as our initial data for the
displacement field \eqref{xiCD 1} has a $P_2(\cos\theta)$-dependence which matches that of the 
$l=2$ eigenfunctions.

In general, we can parametrise our background stars by their
radius, density and shear modulus $(R,\rho,\mu)$. We can make
an arbitrary rescaling of the first two, as for given $l$, the only free parameter in 
the function $f_S$ \eqref{fs} used to determine the eigenvalues 
is the ratio $\frac{\mu}{\rho}$.

For our numerical work we first scale to unit radius, $R=1$. Instead
of rescaling the density $\rho$ directly, we make the more useful
physical choice of rescaling the frequencies against a typical mode 
frequency. For this, we will scale by the fundamental mode of a completely
fluid incompressible star. 
For this model, the oscillation frequencies were found by \cite{1863Thomson} (Lord Kelvin) to satisfy
\begin{equation}
\label{modes:kelvin}
\omega^2_l = \frac{8\pi G\rho}{3}\,\frac{l(l-1)}{2l+1},
\end{equation}
and are often called Kelvin modes,
with corresponding eigenfunctions
\begin{equation}
\label{modes:eigenfunction}
\left(\boxi_{\text{fluid}}\right)^{lm} = \frac{2(l-1)}{2l+1}\frac{A_{lm}}{\rho\omega_l^2}\babla_i(r^lY_{lm})
\end{equation}
where the $A_{lm}$ are arbitrary constants. We will choose to scale our results so that the $l=2$ Kelvin mode,
$\omega_K = \sqrt{\frac{16\pi}{15}}\sqrt{G\rho}$, becomes $\omega_K = 1$.
Finally, for comparison with our earlier results, we switch from 
the ratio of shear modulus to density $\frac{\mu}{\rho}$ to the ratio of
strain energy to gravitational energy $b$ \eqref{b exact}.
Combining this with the scaling choices above, we find that the parameter $\frac{\rho}{\mu}$
in our function $f_S$ for finding the spheroidal modes \eqref{fs} becomes
$\frac{\rho}{\mu}= \frac{38}{5b}$.

We find that the quantity  $\frac{\psi_l'(x)}{\psi_l(x)}$ occuring in $f_S$
can be expressed in terms of cylindrical Bessel functions $J_l(x)$ as

\begin{equation}
\label{modes:bessels}
\frac{\psi_l'(x)}{\psi_l(x)} = -\frac{J_{l+\frac{3}{2}}(x)}{J_{l+\frac{1}{2}}(x)},
\end{equation}
and so $f$ diverges when $x$ is a root of $J_{l+\frac{1}{2}}$. It is therefore
easier numerically to find roots of

\begin{equation}
\label{modes:g in terms f}
g_S\left(l,k_S,b\right) = J_{l+\frac{1}{2}}(k_S)\, f_S\left(l,k_S,b\right).
\end{equation}
We carry this out in Mathematica using the external package RootSearch \cite{Ersek}.
To interpret the results, we need to  convert back from 
the roots $(k_S)^{ln}$ to the frequencies $(\omega_S)^{ln}$.
%figure:eigenvalues%%%%%%%%%%%%%%%%%%%%%%%%%%%%%%%%%%%%%%%%%%%
\begin{figure}
\centering
\includegraphics[scale=1.5]{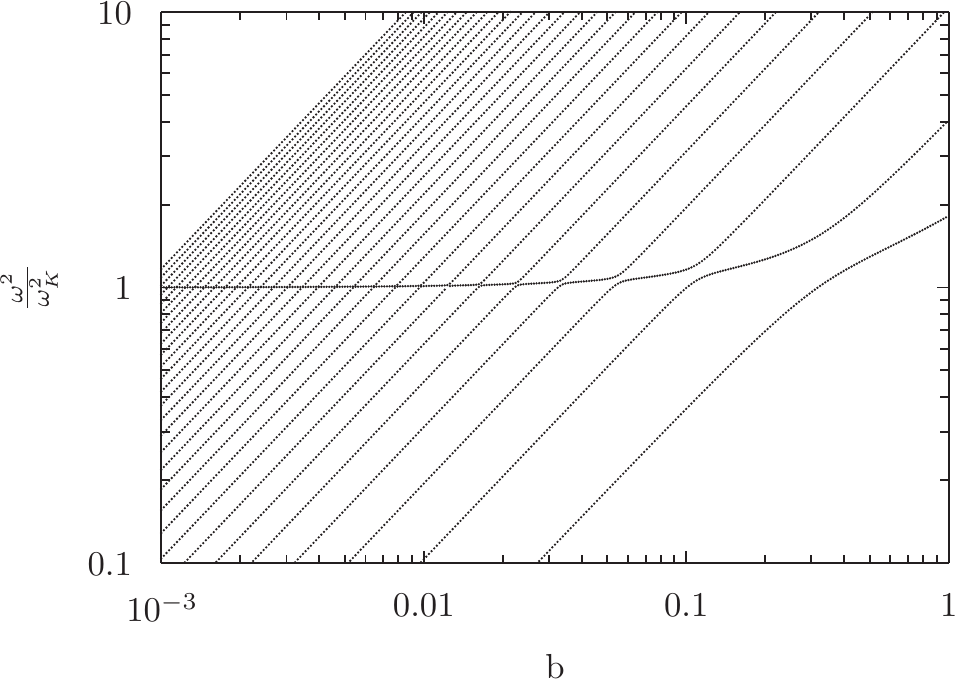}
\caption[Plot showing how the first 30 $l=2$ spheroidal mode
frequencies of an incompressible
elastic star vary with the ratio of strain to gravitational energy $b$.]{Plot showing how the $l=2$ mode
frequencies of an incompressible
elastic star vary with the ratio of strain to gravitational energy $b$.
On the $y$-axis, frequencies are scaled by the fundamental $l=2$ Kelvin mode
of an incompressible fluid star, $\omega_K$. For typical neutron star parameters,
$\omega_K\sim 2000$ Hz. For a given value of $b$, the lowest order modes  
have a similar character to those of an elastic solid, scaling as $\sim \sqrt{\frac{\mu}{\rho}}$.
There is one mode around $\frac{\omega}{\omega_K}=1$ with a hybrid fluid-elastic
character, and then the higher modes again have an elastic character.
}
\label{fig:BAvsomega}
\end{figure}
%%%%%%%%%%%%%%%%%%%%%%%%%%%%%%%%%%%%%%%%%%%%%%%%%%%%%%%%%%%%%%%%

We can then plot the scaled $l=2$ eigenvalues $\frac{[(\omega_S)^{ln}]^2}{\omega_K^2}$
as a function of $b$; the first 30 of these are shown in Figure \ref{fig:BAvsomega}. 
Concentrating on a given value of $b$, we can see the start of an infinite set of eigenvalues,
which we label with the radial eigenvalue number $n$ in order of increasing frequency. 
The squared frequency $\omega^2$ scales linearly
with $b$ and hence with the shear modulus $\mu$: this
is typical of modes of an elastic solid, which have frequency
$\sim \sqrt{\frac{\mu}{\rho}}$.
The exception to this
is the behaviour of the modes closest to the fluid Kelvin mode $\omega_K$.
Around this value, we see avoided crossings between modes. This is characteristic
of systems where two different types of mode have a similar frequency
\cite{1977Aizenman}: in this case these are the single fluid-like mode
close to the Kelvin mode frequency, and one of the elastic modes of the star.
%
%figure: eigenfns for b=0.01%%%%%%%%%%%%%%%%%%%%%%%%%%%%%%%%%%%%%%
\begin{figure}
\centering     %%% not \center
\subfigure[Plot of the $U$ radial part of the first
ten spheroidal eigenfunctions.
 The hybrid fluid-like mode is marked in solid red; the other eigenfunctions (dashed lines)
have an elastic character.]{\label{fig:u}\includegraphics[width=70mm]{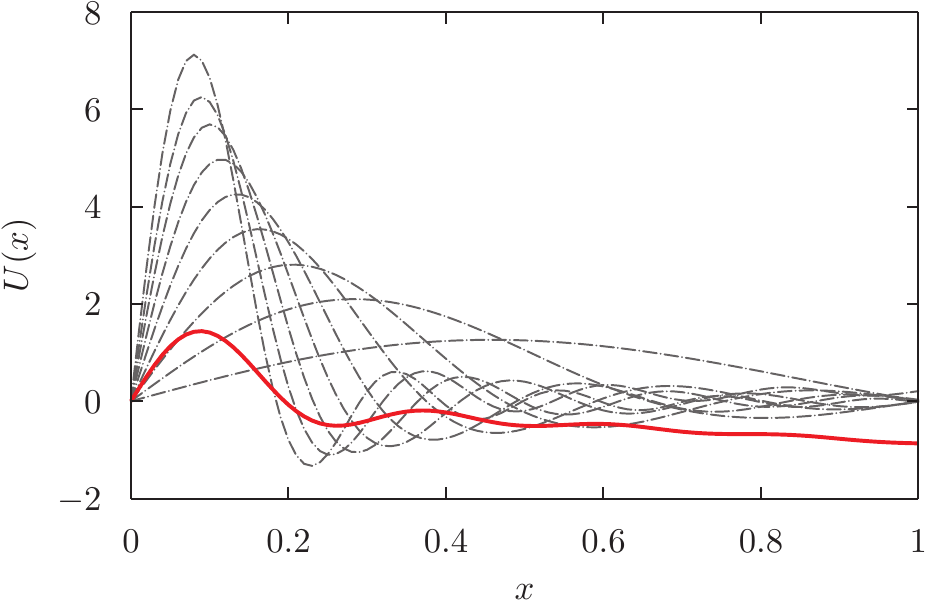}}
\subfigure[Plot of the $V$ radial part of the first
ten spheroidal eigenfunctions, with the hybrid fluid-like mode marked in solid red.]{\label{fig:v}\includegraphics[width=70mm]{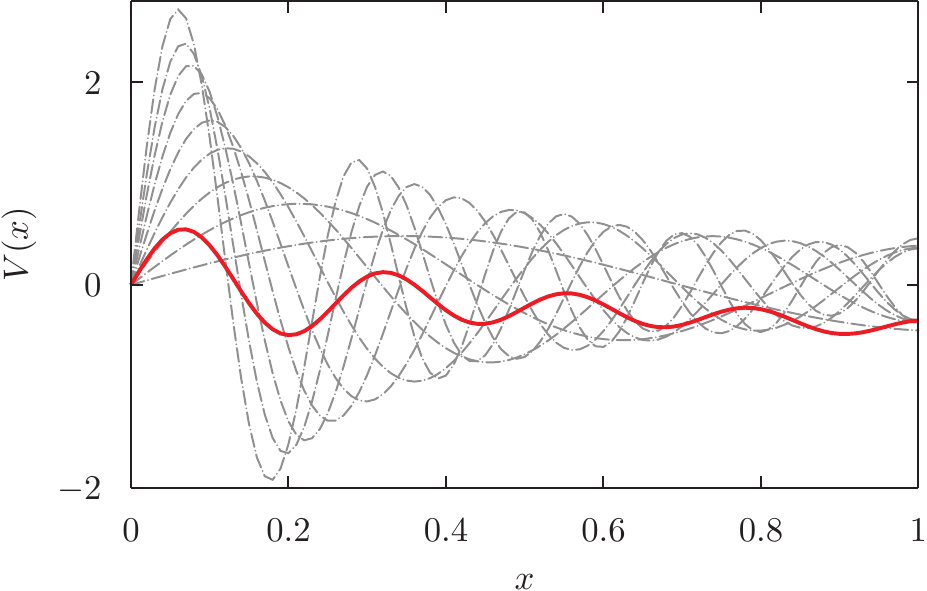}}
\\
\subfigure[Plot of the $W$ radial part of the first ten toroidal eigenfunctions.]{\label{fig:toroidaleigenfns}\includegraphics[width=70mm]{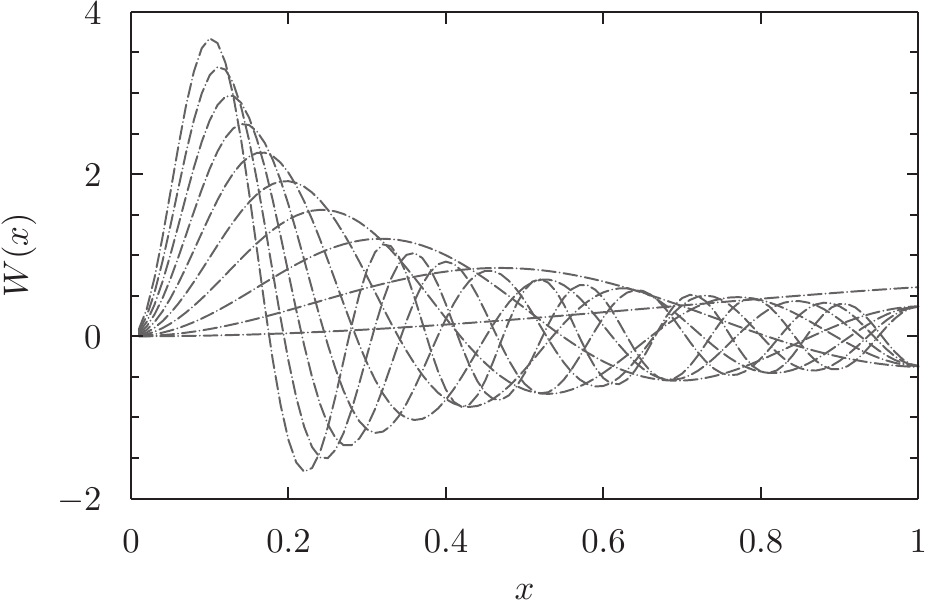}}
\caption[]{Figure showing the $U$, $V$ and $W$ radial parts of the first ten $l=2$ spheroidal and toroidal eigenfunctions
for the case $b = 0.01$, as a function of the fractional radius $x=\frac{r}{R}$.}
\label{fig:eigenfns}
\end{figure}
%%%%%%%%%%%%%%%%%%%%%%%%%%%%%%%%%%%%%%%%%%%%%%%%%%%%%%%%%
%figure: kelvin-like eigenfns%%%%%%%%%%%%%%%%%%%%%%%%%%%%%%%%%%%
\begin{figure}
\centering
\includegraphics[width=70mm]{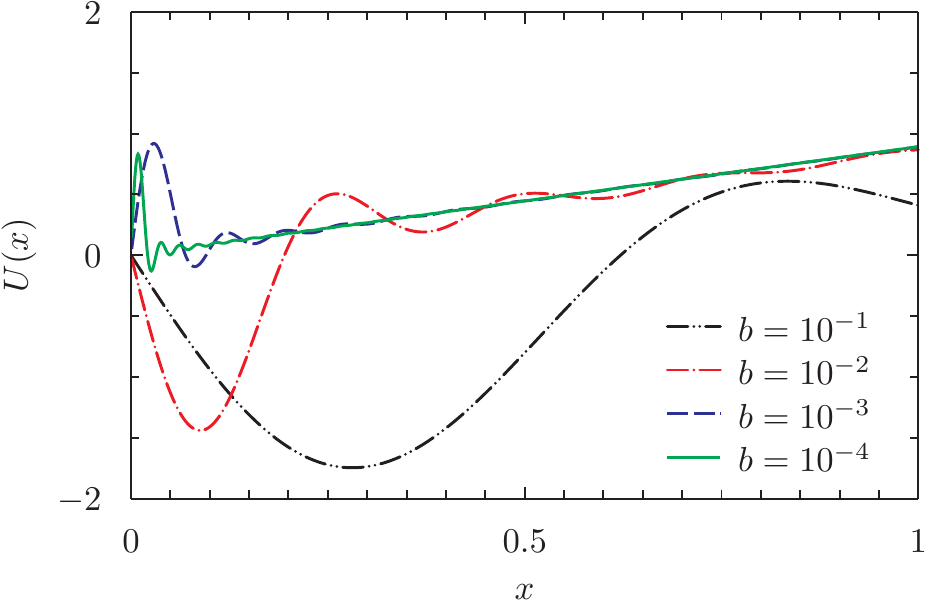}
\caption[Plot showing the hybrid fluid-like mode for a range of values
of $b$.]{Plot showing the $U$ radial part of the $l=2$ hybrid fluid-like mode 
for $b$ varying from $10^{-1}$ to $10^{-4}$. As $b$ is made smaller, the 
form of the eigenfunction becomes closer to the linear eigenfunction of an
incompressible fluid star.}
\label{fig:kelvin}
\end{figure}
%%%%%%%%%%%%%%%%%%%%%%%%%%%%%%%%%%%%%%%%%%%%%%%%%%%%%%%%%%%%%%%%

%figure: continuity%%%%%%%%%%%%%%%%%%%%%%%%%%%%%%%%%%%%%%%%%%%%%
\begin{figure}
\centering
\includegraphics[scale=0.6]{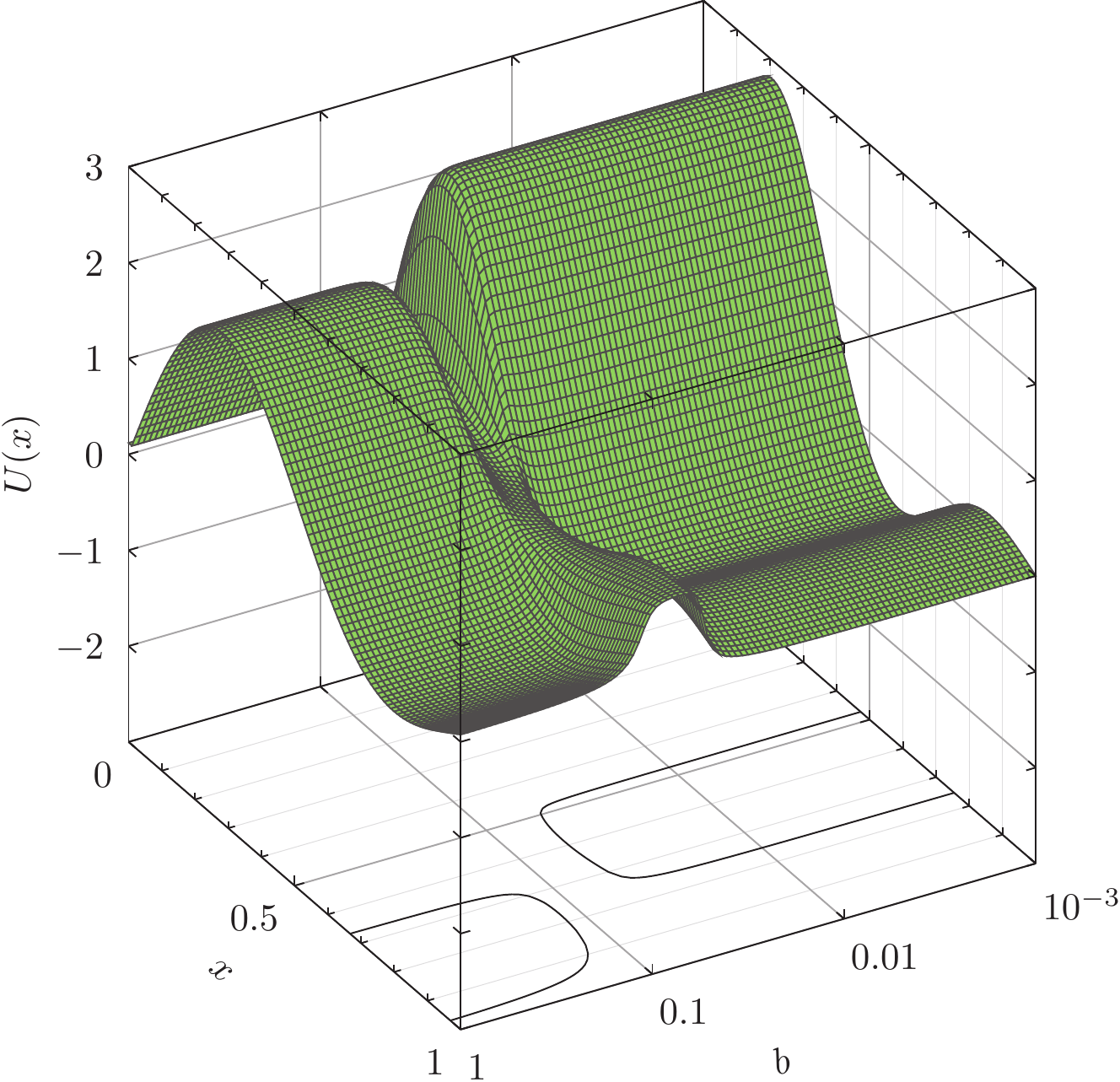}
\caption[Surface plot showing how the $U$ radial part of the $n=3$
eigenfunction varies with $\frac{B}{A}$.]{Surface plot showing how the $U$ radial part of the $n=3$
eigenfunction varies with $b$. The eigenfunction goes
through three distinct phases: in the middle region around $b=0.1$ there
is a transitional area where this mode has a hybrid fluid-elastic 
character.
The contours projected onto the base of the graph show the 
zeroes of the eigenfunction; there are none for the hybrid mode.}
\label{fig:continuity}
\end{figure}
%%%%%%%%%%%%%%%%%%%%%%%%%%%%%%%%%%%%%%%%%%%%%%%%%%%%%%%%%%%%%%%%

Next, we can use these eigenvalues to calculate the corresponding
toroidal and spheroidal eigenfunctions.
Figure \ref{fig:toroidaleigenfns} shows a plot of the first ten
toroidal eigenfunctions for $b=0.01$. 
The lowest, $n=1$ eigenfunction has no nodes; 
each eigenfunction gains one more node with increasing $n$.

The spheroidal eigenfunctions display more complicated behaviour.
As an example, Figures \ref{fig:u} and \ref{fig:v} show the $U$ and $V$ parts of the first
ten eigenfunctions for $b=0.01$. 
The majority of the eigenfunctions, shown as grey dashed lines, are of elastic type.
These form an ordered sequence with the lowest $n=1$ mode having one
stationary point, the $n=2$ mode having two, etc. These modes also have
a very small amplitude at the surface.

The eighth eigenfunction, marked as a solid red line, has a frequency just above
$\omega_K$ and exhibits very different behaviour to the rest of the set.
In particular, it has a much larger amplitude at the surface.
This is consistent with this eigenfunction having a hybrid fluid-elastic
character: the overall shape of the eigenfunction is similar to the
linear $l=2$ Kelvin mode eigenfunction 
for a completely fluid star \eqref{modes:eigenfunction}, while the elastic character 
of the mode is evident in the oscillations superimposed on this.

This behaviour is generic for all values of $b$ in the range
$b = 10^{-6}-1$  that we have studied. As $b$ gets smaller,
the gap between mode frequencies gets smaller, so that the single
hybrid mode is pushed to higher $n$. By $b = 10^{-6}$,
the hybrid mode is at $n=877$.
For lower values of $b$, where the star is closer to a
fluid star, we should expect the hybrid eigenfunction to approach that
for a fluid star
This is indeed what we see in Figure \ref{fig:kelvin}, where the hybrid
mode is plotted for four different values of $b$.
In the $l=2$ case the fluid Kelvin mode eigenfunction \eqref{modes:eigenfunction} has
a linear $U$ component; the hybrid modes approach this form as $b$ becomes
smaller. 

Another way to obtain some insight into the behaviour of the eigenfunctions
is to pick a value of $n$ and track the variation of the eigenfunction
with $b$. As an example, Figure \ref{fig:continuity} is a 
surface plot for $n=3$, showing how the $U$ part eigenfunction changes
between $b=10^{-3}$ and $b=1$. We can see that the 
eigenfunction changes continuously, but goes through three distinct phases.
For the smallest values of $b$ on the right, the three lowest eigenfunctions
are all elastic-type eigenfunctions, and so the $n=3$ eigenfunction is the third
elastic eigenfunction with three stationary points. As $b$ gets 
larger there is a transitional area where the third mode has a hybrid character.
For values of $b$ larger than around $0.1$, the hybrid mode 
is found at $n=1$ or $2$, and so the $n=3$ eigenfunction is again an
elastic-type function, this time the second such function with two
stationary points.
The contours projected onto the base of the graph show the how the
zeroes of the eigenfunction vary with $b$. This makes it clear
that the intermediate, hybrid eigenfunction has a very different character,
with no zeroes.

%%%%%%%%%%%%%%%%%%%%%%%%%%%%%%%%%%%%%%%%%%%%%%%%%
% Modes of a rotating star

\subsection{Rotating star}
\label{subsec:rotating}
To model mode excitation of a glitching star, we will also need to account
for the fact that the star is rotating. 
In this section we will add the effects of rotation in to our
computation of the oscillation modes of our model.

We will assume that the star
is rotating slowly, in the sense that the centrifugal force is small
compared to the gravitational force at the surface of the star; this is a reasonable assumption for most pulsars. This approach was used for
fluid stars by \cite{1949Cowling}, and has also
been developed in the geophysics literature 
(\cite{1961Backus,1961Pekeris,1961MacDonald}).
Here we will mainly use the method of 
\cite{1991Strohmayer2}. However, we will use somewhat different notation 
in our paper, based on that of \cite{1978Friedman},
so it will be helpful to restate some of the main results 
in this notation.

We first write the mode equation for the spherical star \eqref{modes:elastic force modes} 
schematically as

\begin{equation}
\label{xi AC}
-\omega^2 A\boxi + C\boxi = 0.
\end{equation}
In a rotating frame this becomes

\begin{equation}
\label{xi ABC}
-\omega^2A \boxi + i\omega B\boxi + C\boxi = 0,
\end{equation}
where the new operator $B$ is defined by

\begin{equation}
\label{B}
B\boxi \equiv 2\rho\,\boldsymbol{\Omega}\times\dv{\boxi}{t}.
\end{equation}
We next make the slow rotation approximation, defining the small dimensionless quantity

\begin{equation}
\label{modes:veprot}
\vep = \frac{\Omega}{\omega_*} \ll 1,
\end{equation}
where $\omega_* = \sqrt{\frac{GM}{R^3}}$. 
In terms of our small rotational parameter $t$ \eqref{t exact}, we have

\begin{equation}
\label{epsilon t}
\vep = \sqrt{\frac{10}{3}}\sqrt{t}.
\end{equation}
We linearise in this small parameter $\vep$, writing

\begin{align}
\label{xi linearise}
\boxi^\alpha &= \boxi^\alpha_{(0)} + \vep\boxi^\alpha_{(1)}, \\
\label{omega linearise}
\omega^\alpha &= \omega^\alpha_{(0)} + \vep\omega^\alpha_{(1)},
\end{align}
where $\xi^\alpha_{(0)}$ and $\omega^\alpha_{(0)}$ are the eigenfunctions
and corresponding eigenvalues of the nonrotating star, found in the previous
section. 
%To this order in $\vep$, the mode equation \eqref{modes:elastic force modes}
%becomes
%
%\begin{equation}
%\label{xi ABC slowrot}
%A\ddot{\boxi} + \vep B\dot{\boxi} + C\boxi = 0,
%\end{equation}
%where the operators $A$ and $C$ are defined as for the nonrotating star, while
%to first order the operator $B$ \eqref{B} retains only the first, Coriolis term, i.e.
%
%\begin{equation}
%\label{B1}
%\vep B\boxi = 2\rho\Omega \,\left(\hat{z} \times \boxi\right).
%\end{equation} 
%The second, centrifugal force term does not appear at first order in $\vep$,
%meaning that at this order we can continue to treat the rotating star as spherical.
We then look for normal mode solutions of the form $\boxi^\alpha (x,t) = e^{i\omega t}\boxi^\alpha(x)$. 
At zeroth order in $\vep$, we retain the mode equation for the spherical star, 
while at first order we find that

\begin{equation}
\label{first order}
-\left(\omega^{\alpha}_{(0)}\right)^2A_{(0)}\boxi_{(1)}^\alpha 
- 2\omega^{\alpha}_{(0)}\omega_{(1)}^\alpha A_{(0)}\boxi^{\alpha}_{(0)}
+ i\omega^{\alpha}_{(0)}B_{(1)}\boxi^{\alpha}_{(0)}
+ C_{(0)}\boxi_{(1)}^\alpha = 0,
\end{equation}
where

\begin{align}
\label{A0}
(A_{(0)})_{ij} \xi^j &\equiv \rho  \xi^j \\
\label{B0}
(B_{(0)})_{ij} \xi^j &\equiv 2\rho\, \vep_{ijk} \Omega^j \dv{\xi^k}{t} \\
\label{C0}
(C_{(0)})_{ij} \xi^j &\equiv \nabla_j (\Delta \tau)\indices{_i^j} + \rho\nabla_i(\Delta\Phi).
\end{align}
We next write $\boxi_{(1)}$ as a sum over the zeroth order eigenfunctions,

\begin{equation}
\label{xi 1}
\boxi_{(1)}^{\alpha} = \sum_{\beta}\lambda_{(1)}^{\alpha\beta}\boxi_{(0)}^{\beta}.
\end{equation}
Substituting this back into the first order equation \eqref{first order} and making
use of the zeroth order equation, we find that

\begin{equation}
\label{strohmayer ABC}
-\sum_{\beta}
\left(
\left(\omega^{\alpha}_{(0)}\right)^2 
-
\left(\omega^{\beta}_{(0)}\right)^2
\right)
\lambda_{(1)}^{\alpha\beta}A_{(0)}\boxi_{(0)}^{\beta}
- 2\omega^{\alpha}_{(0)}\omega_{(1)}^{\alpha}A_{(0)}\boxi^{\alpha}_{(0)}
+ i\omega^{\alpha}_{(0)}B_{(1)}\boxi^{\alpha}_{(0)}
= 0.
\end{equation}
This equation can then be used to find expressions for the first order rotational
corrections to the eigenfunctions and eigenvalues.
To do this, we define the inner product

\begin{equation}
\label{innerproduct}
\left\langle \boeta, \boxi \right\rangle \equiv \int_V \boeta^*\cdot \boxi\, dV.
\end{equation}
We show in Appendix \ref{appx:orthogonal} that the zeroth order eigenfunctions are
orthogonal with respect to the operator $A_{(0)}$, We will also scale the
eigenfunctions so that they are orthonormal:
\begin{equation}
\label{06}
\left\langle \boxi_{(0)}^{\alpha}, A_{(0)}\boxi_{(0)}^{\beta}\right\rangle = \delta^{\alpha\beta}.
\end{equation}
Taking the inner product of \eqref{strohmayer ABC} and $\xi_{(0)}^\alpha$ and using 
this orthogonality condition, we find an expression for the rotational corrections to 
the eigenvalues,

\begin{equation}
\label{sigma1}
\omega_{(1)}^{\alpha}= \frac{i}{2}\left\langle \xi_{(0)}^{\alpha},B_{(1)}\xi^{\alpha}_{(0)}\right\rangle.
\end{equation}
Similarly, to find the corrections to the eigenfunctions we can take an inner product of
\eqref{strohmayer ABC} and $\xi_{(0)}^\gamma$, $\gamma \neq \alpha$, finding that

\begin{equation}
\label{lambdas}
\lambda_{(1)}^{\alpha\gamma} = 
\frac{i\omega^{\alpha}_{(0)}}{
\left(\omega^{\alpha}_{(0)}\right)^2 
-
\left(\omega^{\gamma}_{(0)}\right)^2}
\left\langle \xi_{(0)}^{\gamma}, B_{(1)}\xi^{\alpha}_{(0)}\right\rangle \,.
\end{equation}
Finally, it can be shown that for $\alpha = \gamma$, the coefficient $\lambda_{(1)}^{\alpha\alpha}$ 
can be chosen to be zero (see, for example, \cite{Rae}).

To calculate explicit formulae for the corrections $\boxi_{(0)}^\alpha$ and $\omega_{(0)}^\alpha$, 
it will be necessary to separate out spheroidal and toroidal corrections. 
The zeroth order spheroidal and toroidal eigenfunctions $\mathbf{S}_{(0)}$ and $\mathbf{T}_{(0)}$
have the forms \eqref{Sln} and \eqref{Tln} respectively.
%For the full, corrected spheroidal eigenfunctions, we write
%
%\begin{equation}
%\label{S full}
%\mathbf{S}^{\alpha} = \mathbf{S}_{(0)}^\alpha + \vep\sum_\gamma\left([SS]_{(1)}^{\alpha\gamma} \mathbf{S}^{\gamma}_{(0)}
%+[ST]_{(1)}^{\alpha\gamma} \mathbf{T}^{\gamma}_{(0)}\right),
%\end{equation}where the $[SS]_{(1)}^{\alpha\gamma}$ and $[ST]_{(1)}^{\alpha\gamma}$ 
%are respectively the coefficients of the spheroidal and toroidal corrections to the
%spheroidal eigenfunctions. Similarly, the full
%toroidal eigenfunctions are
%
%\begin{equation}
%\label{T full}
%\mathbf{T}^{\alpha} = \mathbf{T}_{(0)}^\alpha + \vep\sum_\gamma\left([TS]_{(1)}^{\alpha\gamma} \mathbf{S}^{\gamma}_{(0)}
%+[TT]_{(1)}^{\alpha\gamma} \mathbf{T}^{\gamma}_{(0)}\right).
%\end{equation}
Formulae for the full corrected eigenfunctions are given in \cite{1991Strohmayer2}.
In the $m=0$ case, the eigenvalue corrections $\left(\omega_{(1)} \right)_S^\alpha$ 
and $\left(\omega_{(1)} \right)_T^\alpha$ are zero, and the corrected eigenfunctions
have the form

\begin{equation}
\label{modesrot:Sfull}
\textbf{S}^{nl} = \textbf{S}_{(0)}^{nl} + \vep \sum_{n=1}^{\infty}\left[
 [ST]_{(1)}^{nl,n'l-1}\textbf{T}_{n'l-1} 
 + [ST]_{(1)}^{nl,n'l+1}\textbf{T}_{n'l+1}
\right],
\end{equation}

\begin{equation}
\label{modesrot:Tfull}
\mathbf{T}^{nl} = \mathbf{T}_{(0)}^{nl} + \vep \sum_{n'=1}^{\infty}\left[
[TS]_{(1)}^{nl,n'l-1}\mathbf{S}_{(0)}^{n'l-1} + [TS]_{(1)}^{nl,n'l+1}\mathbf{S}_{(0)}^{n'l+1}
\right],
\end{equation}
where the $[ST]_{(1)}$ are the coefficients of the toroidal corrections to the spheroidal
eigenfunctions and the $[TS]_{(1)}$ are the coefficients of the spheroidal corrections to the 
toroidal eigenfunctions.

To calculate these corrections numerically, we again scale our equations so that
the star has unit radius $R=1$, and
that the $l=2$ fluid Kelvin mode has $\omega_K = 1$.
With this choice, the parameter $\sigma_*$ \eqref{modes:veprot}
becomes $\sigma_* = \frac{\sqrt{5}}{2}$.

The analytic expressions for the corrected eigenfunctions \eqref{modesrot:Sfull}, \eqref{modesrot:Tfull} include
an infinite sum over the radial eigenvalue number $n$.
For numerical work we will need to truncate these eigenfunctions, calculating the sum
only up to some maximum value $N$. We will label these truncated eigenfunctions as
$\boldsymbol{S}^\alpha_N$ and $\boldsymbol{T}^\alpha_N$.

The value of $N$ we choose to cut off at will depend on $b$.
It will turn out that in the special case where the angular velocity of the star after the
glitch is zero, the majority of the energy released in the glitch goes into the
hybrid fluid-elastic mode. This effect becomes more pronounced as $b$ becomes smaller.
As rotation is only a small correction, this will largely hold true even in the rotating case. 
This means that the cutoff $N$ should be chosen to be somewhat larger than the $n$ number
of the hybrid mode.

\begin{figure}
\centering     %%% not \center
\includegraphics[width=80mm]{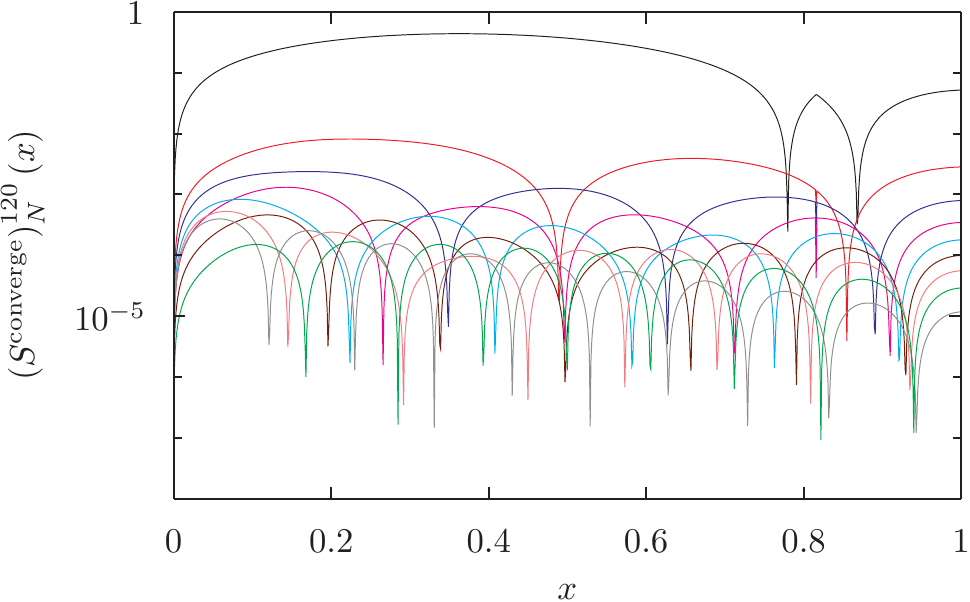}
%\\
%\subfigure[Corrections to spheroidal eigenfunctions for $l=2,\,n=2$]{\label{fig:spherN2}\includegraphics[width=95mm]{figures/spherN2.pdf}}
%\\
%\subfigure[Corrections to spheroidal eigenfunctions for $l=2,\,n=3$]{\label{fig:spherN3}\includegraphics[width=95mm]{figures/spherN3.pdf}}
\caption[Figure showing the convergence of the rotational corrections
to the spheroidal $l=2$ eigenfunctions]{Figure showing the convergence of the rotational corrections
to the spheroidal $l=2$, $n=1$ eigenfunctions for $b=10^{-1}$,
as the corrections are truncated at progressively higher values of the 
radial eigenvalue $N$.
In particular, the $\phi$ component of the functions $\left(S^{\text{converge}}\right)^{n20}_N$
\eqref{Sconverge} is plotted for $N = 1$ (plotted in black) up to $N=9$ (plotted in green).
}
\label{fig:spher}
\end{figure}
We will investigate the cases $b = 10^{-1}, 10^{-2}, 10^{-3}$. In these cases the hybrid
mode is at $n=2$, $7$ and $27$ respectively. We will choose our cutoffs for each
of these cases to be at $N=10$, $15$ and $40$.
To check that we have truncated at a high enough value $N$, we have checked that the partial
sums $\boldsymbol{S}^\alpha_N$ and $\boldsymbol{T}^\alpha_N$ converge.
As an example, we show this here for the spheroidal $l=2$, $n=1$ eigenfunction, with
$b=10^{-1}$, by plotting the function

\begin{equation}
\label{Sconverge}
\left(\boldsymbol{S}^{\text{converge}}\right)^{120}_n(x) \equiv \left|\frac{\boldsymbol{S}^{120}_n(x) -
\boldsymbol{S}^{120}_{10}(x)}{\boldsymbol{S}^{120}_{10}(x)}\right|
\end{equation}
for $k = 1, \dots, 10$. For the spheroidal corrections, only the $\phi$ component of 
$\left(\boldsymbol{S}^{\text{converge}}\right)^{120}_n(x)$ is nonzero. This component
is plotted in Figure \ref{fig:spher}, with $\left(\boldsymbol{S}^{\text{converge}}\right)^{120}_1(x)$
in black and $\left(\boldsymbol{S}^{\text{converge}}\right)^{120}_{10}(x)$ in green,
showing convergence as more modes are added to the sum.

%%%%%%%%%%%%%%%%%%%%%%%%%%%%%%%%%%%%%%%%%%%%%%%%%%%%%%%%%%%%%%%%%%
% PROJECTION
%%%%%%%%%%%%%%%%%%%%%%%%%%%%%%%%%%%%%%%%%%%%%%%%%%%%%%%%%%%%%%%%%%

\section{Projecting the initial data}
\label{sec:projection}
We are now in a position to project our initial data 
for the displacement and velocity fields against
the eigenfunctions of Star D \eqref{modesrot:Sfull}, \eqref{modesrot:Tfull} after the glitch.  
This initial data has the form 

\begin{align}
\label{xiDC ref}
\boxi^{\text{DC}} &= 
\left(\frac{5z}{R^2} \left(3r^3-8R^2r\right)\right)P_2\,\boldsymbol{\hat{r}} 
+ \left(\frac{5z}{R^2} \left(\frac{5}{2} r^4-4R^2r^2\right)\right)\babla P_2,
\\
\label{xidotDC ref}
\dot{\boxi}^{\text{DC}} &= -\sqrt{\frac{3}{8}}\,\sqrt{t}\,z r \left(\boldsymbol{\hat{r}}\times\babla P_1\right),
\end{align}
where we have rewritten the velocity initial data \eqref{arb:vel} to show that it has
a pure toroidal form \eqref{Tln}. We can in fact think of the velocity data as a zero-frequency
$l=1$ toroidal eigenfunction. %!!was there a ref to this?
There are two free parameters in the initial data: $z$ \eqref{energies:delta OmegaBD}, 
related to the size of the glitch, and $t$ \eqref{t exact},
related to the angular velocity of Star B. 
As $z$ just produces an overall scaling of the
initial data, we shall set $z=1$ in the following numerical work, and only investigate the change
in amplitudes produced as we vary $t$ and $b$. As in Section \ref{sec:modes}, we will 
also scale the initial data so that the radius of the star $R=1$, and the $l=2$ fluid Kelvin mode
$\omega_K=1$.

Before considering the full problem, we can get some useful
insight from the special case where the star spins down to zero angular velocity
($\Omega_B = 0$) before the `glitch', modelled as a sudden loss of
strain from the star. Although this is not a realistic model -- we have
lost the key observational feature of the glitch, the change in rotation rate
of the star -- there are good reasons to discuss this first. For a start, 
it is easier to interpret the results for this simpler case, and certain 
features of these will carry over to the more realistic case where the
star is still rotating before the glitch. 
We can also check that the results for the rotating problem converge to the nonrotating one
as rotation becomes small.

Another reason for considering the nonrotating problem first
is that the rotation of the star introduces significant complications
to the projection scheme. One of these
is that the eigenfunctions of the rotating star are no longer orthogonal. 
Another is the existence of the zero eigenvalue $l=1$ toroidal mode 
\eqref{xidotDC ref} that our velocity initial data is built from. As this
mode is zero frequency, it will not have the $e^{i\omega t}$
time-dependence of the other eigenfunctions and we will have to consider it
separately. We will discuss these problems in Section \ref{subsec:timedep}.

%%%%%%%%%%%%%%%%%%%%%%%%%%%%%%%%%%%%%%%%%%%%%%%%%
% GLITCH AT ZERO SPIN

\subsection{Special case: glitch at zero spin}
\label{subsec:glitchzero}

For the special case where $\Omega_B = 0$, the velocity
field part of the initial data \eqref{xidotDC ref} is zero. 
For the displacement field, although there is no change in angular velocity of
the star at the glitch, we still have a finite value of the parameter $z$ 
\eqref{energies:delta OmegaBD},

\begin{equation}
\label{z omB zero}
z\vert_{\Omega_B = 0} = \frac{5}{2\pi G\rho}\, b \Omega_A^2,
\end{equation}
and the displacement field initial data takes the same form \eqref{xiDC ref} as for the general,
rotating case. Our only free parameter in this $\Omega_B = 0$ case is $z$, which we 
have set to unity.  

\begin{figure}
\centering     %%% not \center
\subfigure[$b=10^{-1}$]{\label{fig:b1}\includegraphics[width=45mm]{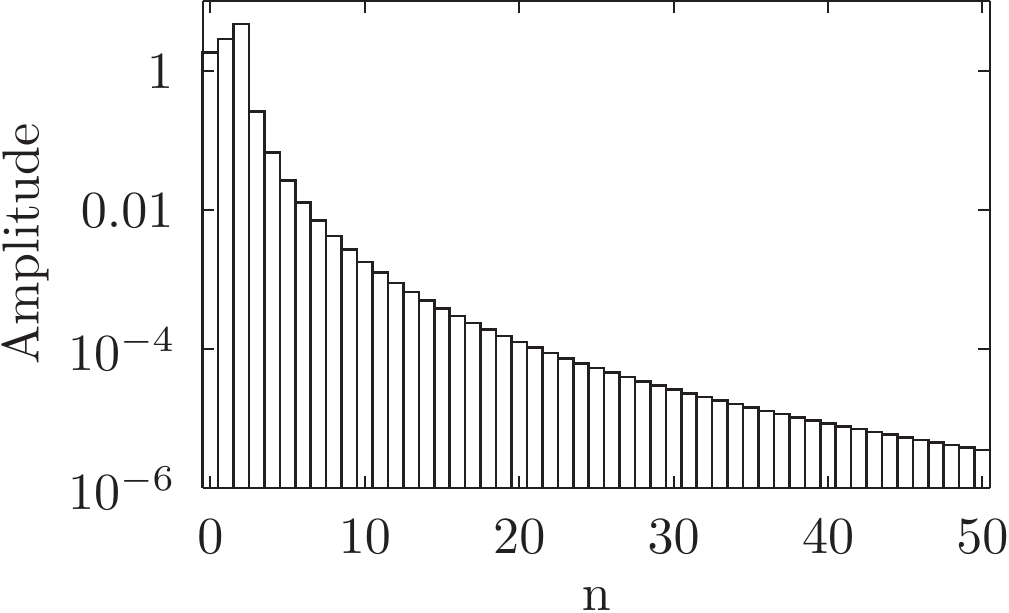}}
\subfigure[$b=10^{-2}$]{\label{fig:b2}\includegraphics[width=45mm]{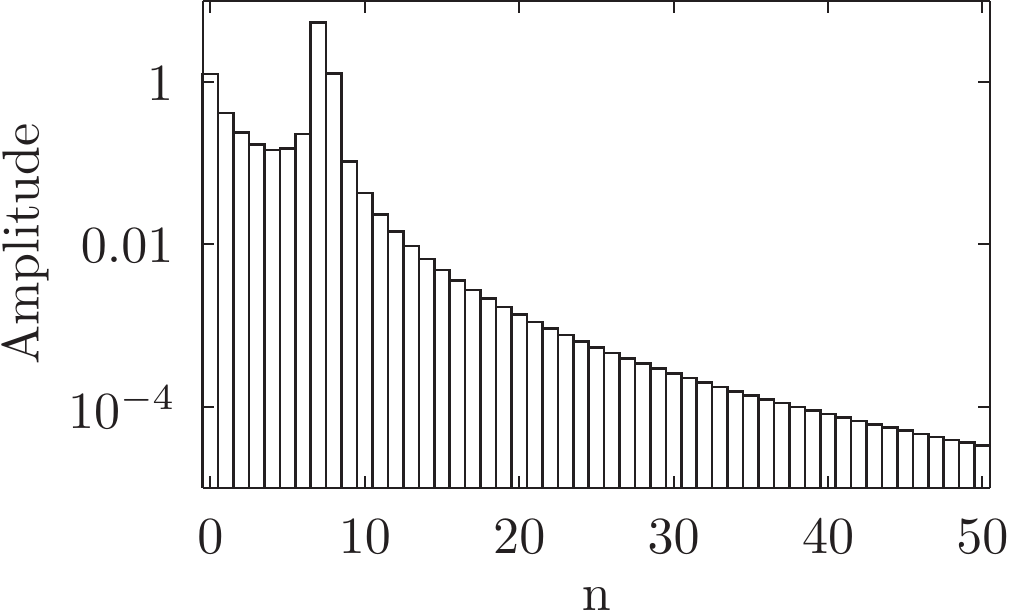}}
\subfigure[$b=10^{-3}$]{\label{fig:b3}\includegraphics[width=45mm]{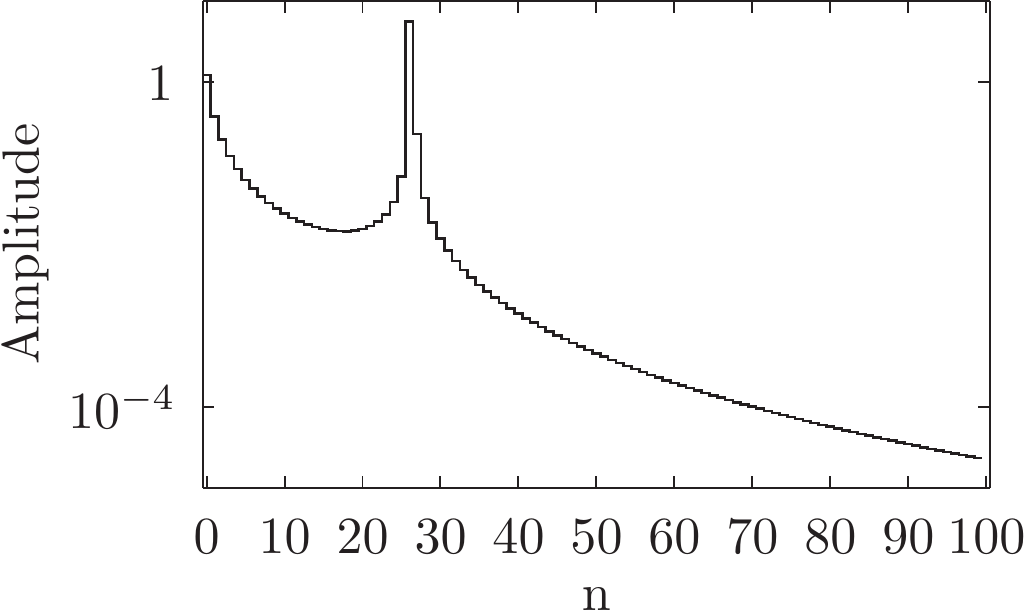}}
\\
\subfigure[$b=10^{-4}$]{\label{fig:b4}\includegraphics[width=45mm]{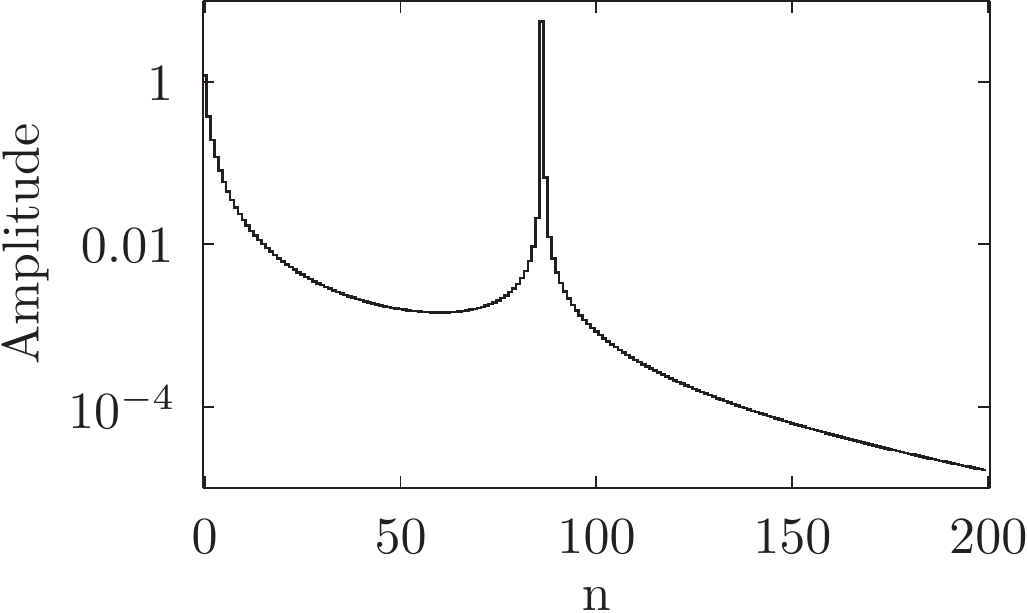}}
\subfigure[$b=10^{-5}$]{\label{fig:b5}\includegraphics[width=45mm]{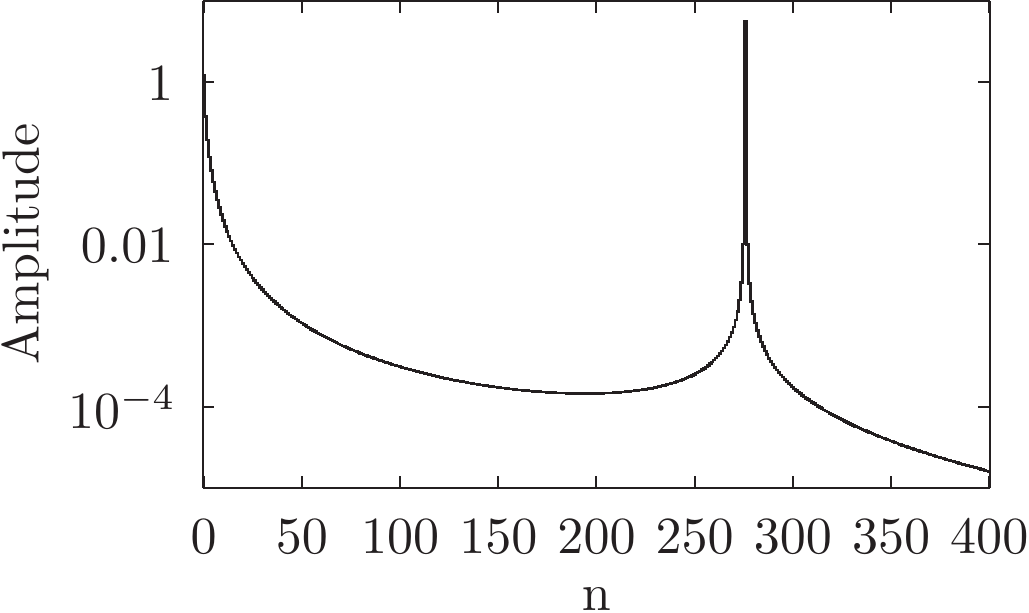}}
\subfigure[$b=10^{-6}$]{\label{fig:b6}\includegraphics[width=45mm]{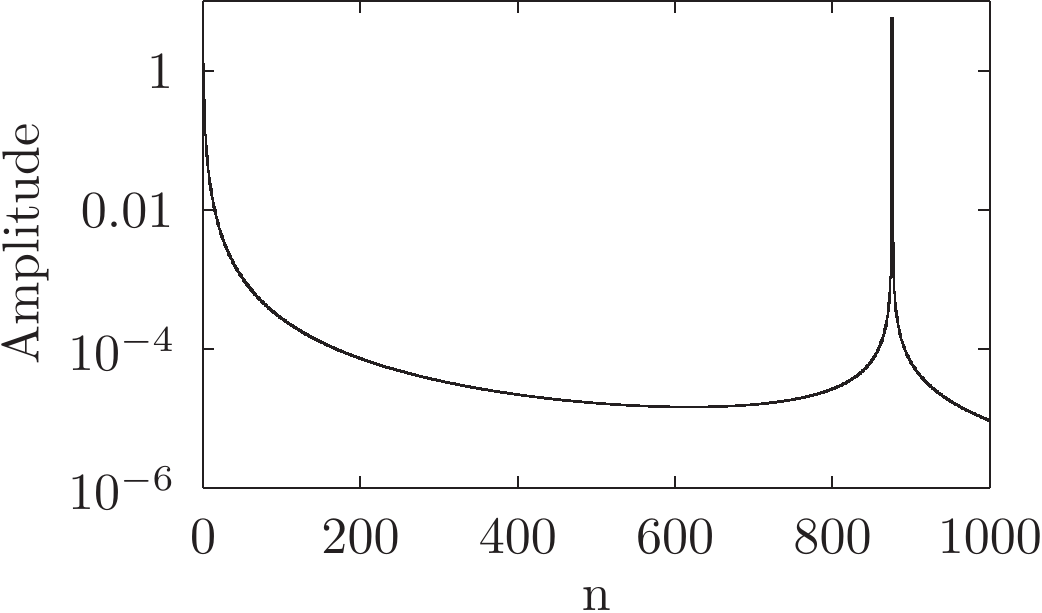}}
\caption[Figure showing the results of the projection for different values
of $b$.]{Figure showing the results of the projection for different values
of $b$. For clarity, individual vertical bars are not displayed for plots (c) to (f).
The $x$-axis shows the radial number of the mode, while the $y$-axis shows the amplitude of that mode. 
The hybrid fluid-elastic mode has the largest amplitude; this becomes more pronounced as $b$ is made smaller.\color{black}
}
\label{fig:wholelot}
\end{figure}

The eigenfunctions we are projecting against are those of the nonrotating star
\eqref{Uln}, \eqref{Vln}, with time-dependence

\begin{equation}
\label{eigenfns nonrot}
\boxi_{(0)}^\alpha(x,t) 
\equiv \boxi^\alpha_{(0)}(x)e^{i\omega^\alpha t}.
\end{equation}
The initial data is then a sum over these modes, with each
eigenfunction $\boxi^\alpha$ excited by 
some amplitude $b^\alpha$. The initial data is real, so to ensure
that the sum over the eigenfunctions is also real we write it as

\begin{align}
\label{xi ID nonrot}
\boxi^{\text{ID}}(x)  &= \frac{1}{2}\sum_\alpha \left[
b^\alpha \boxi^\alpha_{(0)} (x) + b^{*\alpha} \boxi^{*\alpha}_{(0)}(x)
\right],
\\
\label{dot xi ID nonrot}
\dot{\boxi}^{\text{ID}}(x) &= \frac{1}{2}\sum_\alpha \left[
i\omega^\alpha b^\alpha \boxi^\alpha_{(0)} (x) 
-i\omega^\alpha b^{*\alpha} \boxi^{*\alpha}_{(0)} (x)
\right].
\end{align}
The complex conjugate is not really necessary at the moment, where we are
projecting against the real eigenfunctions of a nonrotating star. 
However, it will be
required later when we extend to the rotating case and the eigenfunctions
have an imaginary part, so for consistency we keep it in here.
We will scale our eigenfunctions so that they are orthonormal with respect to 
the inner product \eqref{innerproduct} defined in Section \ref{sec:modes}:

\begin{equation}
\label{zero:orthonormal}
\left\langle \boxi^\beta_{\text{(0)}}, \boxi^\alpha_{\text{(0)}} \right\rangle = \delta^{\alpha\beta}.
\end{equation}
By taking the inner products of $\boxi^{\text{ID}}$ and $\dot{\boxi}^{\text{ID}}$
with an eigenfunction $\boxi^\alpha$, we find that the amplitudes $b^\alpha$ satisfy

\begin{equation}
\label{bbeta}
b^\alpha = -\frac{i}{\omega^\alpha}\left(
\left\langle \boxi^\alpha_{\text{(0)}},\dot{\boxi}^{\text{ID}}\right\rangle
+i\omega^\alpha 
\left\langle \boxi^\alpha_{\text{(0)}},\boxi^{\text{ID}}\right\rangle
\right).
\end{equation}
%Physically, we are more interested in the energy in each mode 
%than its amplitude. As the modes we are interested in are all $m=0$
%modes,
%there is a point in each oscillation where the star is spherical
%and all the energy is kinetic. This means that the energy in each mode is
%
%\begin{equation}
%\label{zero:Ekinetic}
%E^\alpha = \frac{1}{2}(\omega^\alpha)^2|b^\alpha|^2,
%\end{equation}
%where we have used the fact that the eigenfunctions are orthonormal.
%
Figure \ref{fig:wholelot} shows the results of the projection for 
different values of $b$, ranging from the high value of
$b=0.1$ \ref{fig:b1} down to the physical range of 
$b=10^{-5}$ \ref{fig:b5} and $b=10^{-6}$
\ref{fig:b6}. 
The $x$-axis shows the radial number of the mode, while the $y$-axis
shows the amplitude of that mode.
In all of these plots,
the hybrid fluid-elastic mode discussed 
in Section \ref{sec:modes} has the greatest amplitude; this is the mode with the frequency
closest to the fundamental Kelvin mode of a purely fluid star.
For the higher values of $b$, the lowest order modes are also excited to significant amplitudes: these are the shear modes
with the lowest number of radial nodes. However, for physical values
of $b$ ($b \sim 10^{-5}$), only the fluid-elastic mode is appreciably excited. \color{black}

%figure: projection (zero rot case) %%%%%%%%%%%%%%%%%%%%%%%%%%%
%%%%%%%%%%%%%%%%%%%%%%%%%%%%%%%%%%%%%%%%%%%%%%%%%%%%%%%%%%%%%%%
Finally, as a check we can also show that we are reproducing the initial data correctly
by reconstructing the sum over the eigenfunctions \eqref{eigenfns nonrot}
with our calculated amplitudes $b^\alpha$. 
The results of this are shown in Appendix \ref{appx:reconstruct}.

%%%%%%%%%%%%%%%%%%%%%%%%%%%%%%%%%%%%%%%%%%%%%%%%%
% ZERO EIG

\subsection{The general case: glitch at arbitrary spin}
\label{subsec:timedep}

We will now move on to the projection of the initial data in the general case, 
where Star B is rotating before the glitch. Here we will have to deal with the zero eigenvalue $l=1$ toroidal mode 
\eqref{xidotDC ref} that our velocity initial data is built from. As this
mode is zero frequency, it will not have the $e^{i\omega t}$
time-dependence of the other modes and we will have to consider it
separately.
For the purposes of this section, we will label the zero-frequency eigenfunction $\boxi_{(0)}^1$,
with eigenvalue $\omega_{(0)}^1$. 
To find the time-dependence of $\boxi_{(0)}^1(x,t)$, we go back to the governing equation for the zeroth order
eigenfunctions,

\begin{equation}
\label{zeroorder}
A_{(0)}\ppv{\boxi^\alpha_{(0)}}{t} +  C_{(0)}\boxi^\alpha_{(0)} = 0.
\end{equation}
In the case of our zero frequency mode,
the eigenfunction describes rigid rotation, so that its velocity field is constant in 
time, 

\begin{equation}
\label{ddv bxi}
\ppv{\boxi^1_{(0)}}{t} = 0.
\end{equation}
The solutions of this are  $\boxi^1_{(0)}(x,t) = \boxi^1_{(0)}(x)T(t)$ with

\begin{equation}
\label{arb:T}
T(t) = C^1t + D^1,
\end{equation}
$C^1$ and $D^1$ real. The governing equation for the zeroth order modes \eqref{zeroorder} becomes just

\begin{equation}
\label{arb:C0xi0}
C_{(0)} \boxi^1_{(0)} = 0.
\end{equation}
We can also look at the first-order corrections 
$\boxi^1_{(1)}=\sum_\beta \lambda_{(1)}^{1\beta}\boxi_{(0)}^\beta$ to this mode. These
satisfy the equation

\begin{equation}
\label{arb:C0xi1}
C_{(0)} \boxi^1_{(1)} = 0,
\end{equation}
which has the same form as the equation for the zeroth order modes.
This means that we can absorb them into the zeroth order mode $\boxi^0_{(1)}$, 
so that the corrections $\lambda_{(1)}^{1\beta} =0$. 
%
%For the other modes with $\alpha \geq 2$, we take the case where
%$\lambda$ is negative, $\lambda^\alpha = -(\omega^\alpha)^2$, and retain the 
%time-dependence
%
%\begin{equation}
%\label{}
%T^\alpha(t) = \text{Re}\left[
%c^\alpha e^{i\omega^\alpha t}
%\right],
%\end{equation}
%$c^\alpha$ a complex constant. To show explicitly that the solution
%is real, we will instead write
%
%\begin{equation}
%\label{}
%T(t) = \frac{1}{2}\left(b^\alpha e^{i\omega^\alpha t} + (b^\alpha)^*e^{-i\omega^\alpha t}\right)
%\end{equation}
%for a new complex constant $b^\alpha$. 
%For the zero frequency eigenvalue $\omega^1$, we take
%$\lambda^1 = 0$, so that the time-dependence is
%
%\begin{equation}
%\label{arb:T}
%T(t) = C^1t + D^1,
%\end{equation}
%$C^1$ and $D^1$ real.
We will now describe a scheme to allow us to project the initial data against the eigenfunctions
$\boxi$ of the rotating star, which are not orthogonal, by using the orthogonality of the 
zeroth order eigenfunctions $\boxi_{(0)}$. We start by writing out the time-dependence explicitly
for the displacement field $\boxi(x,t)$:

\begin{equation}
\label{xi x t}
\boxi(x,t) = \left[ Ct + D\right]\boxi_{(0)}^1(x)
 + \frac{1}{2}\sum_{\alpha=2}^\infty \left[
b^\alpha \boxi^\alpha (x)e^{i\omega^\alpha t} + b^{*\alpha}\boxi^{*\alpha} (x)e^{-i\omega^\alpha t}
\right].
\end{equation}
Here we have separated out the zero eigenvalue mode -- note that it
we can just use its zeroth order form because it has no first order corrections.
To make sure that the data for the other modes is real we have added on the complex
conjugate. The complex coefficients $b^\alpha$ contain the amplitude
and phase of each mode. 
It will be convenient to write the zero frequency eigenfunction
$\boxi_{(0)}^1$ in a similar way to the other eigenfunctions by defining

\begin{align}
\label{C}
C^1 &= \frac{i}{2}\left(b^1 -  \left(b^1\right)^* \right) ,\\
\label{D}
D^1 &= \frac{1}{2}\left(b^1 + \left(b^1\right)^*\right), 
\end{align}
where $b^1$ is a complex constant, so that \eqref{xi x t} becomes

\begin{equation}
\label{08}
\boxi(x,t) = \left[ \frac{i}{2}\left(b^1 -  \left(b^1\right)^* \right)t
+\frac{1}{2}\left( b^1 + \left(b^1\right)^*\right)\right]\boxi_{(0)}^1(x)
 + \frac{1}{2}\sum_{\alpha=2}^\infty \left[
b^\alpha \boxi^\alpha (x)e^{i\omega^\alpha t} + b^{*\alpha}\boxi^{*\alpha} (x)e^{-i\omega^\alpha t}
\right].
\end{equation}
Differentiating this, the velocity field is

\begin{equation}
\label{09}
\dot{\boxi}(x,t) = \frac{i}{2}\left( b^1 - \left(b^1\right)^*\right)\boxi_{(0)}^1(x) + 
\frac{1}{2}\sum_{\alpha=2}^\infty \left[
i\omega_{(0)}^\alpha b^\alpha \boxi^\alpha (x,t) -i\omega_{(0)}^\alpha b^{*\alpha} \boxi^{*\alpha} (x,t)
\right],
\end{equation}
where we have used the fact that there
are no rotational corrections to the zeroth order eigenvalues $\omega_{(0)}$
in the $m=0$ case we are interested in.
We can combine $\boxi_{(0)}^1$ with the other eigenfunctions 
by making the definition

\begin{equation}
\label{010}
\tilde{\omega}_{(0)}^\alpha = 
\begin{cases}
1 & \alpha = 1, \\
\omega_{(0)}^\alpha & \text{otherwise}.
\end{cases}
\end{equation}
Specialising to $t=0$, we can then use this along with our definitions of $C$ \eqref{C} and $D$ \eqref{D}
to write the initial data in the neat form

\begin{align}
\label{xi ID}
\boxi^{\text{ID}}(x) \equiv \boxi(x,0) &= \frac{1}{2}\sum_{\alpha = 1}^\infty \left[
b^\alpha \boxi^\alpha (x) + \text{c.c.}
\right], \\
\label{dot xi ID}
\dot{\boxi}^{\text{ID}}(x)\equiv \dot{\boxi}(x,0) &= \frac{1}{2}\sum_{\alpha = 1}^\infty \left[
i\tilde{\omega}_{(0)}^\alpha b^\alpha \boxi^\alpha (x) + \text{c.c.}
\right].
\end{align}
Next we expand the eigenfunctions in the sum to first order in rotation \eqref{xi linearise}. 
Introducing the notation

\begin{equation}
\label{Lambda intro}
\boxi^\alpha 
= \sum_{\beta=1}^\infty\left(\delta^{\alpha\beta} + \lambda_{(1)}^{\alpha\beta}\right) \boxi^\beta_{\text{(0)}}
\equiv \sum_{\beta=1}^\infty \Lambda^{\alpha\beta} \boxi^\beta_{\text{(0)}},
\end{equation}
we then have

\begin{align}
\label{011}
\boxi^{\text{ID}} (x) &= 
\sum_{\alpha=1}^{\infty}\left[
\frac{b^{\alpha}}{2} \left(
\sum_{\beta=1}^{\infty}\Lambda^{\alpha\beta}\boxi_{(0)}^\beta
\right) +
\frac{(b^{\alpha})^*}{2}
\left(
\sum_{\beta=1}^{\infty}(\Lambda^*)^{\alpha\beta}
\boxi_{(0)}^\beta
\right)
\right], \\
\label{012}
\dot{\boxi}^{\text{ID}} (x) &= \sum_{\alpha=1}^{\infty} \left[
i\tilde{\omega}^\alpha \frac{b^{\alpha}}{2} \left(
\sum_{\beta=1}^{\infty}\Lambda^{\alpha\beta}\boxi_{(0)}^\beta
\right)
-i\tilde{\omega}^\alpha \frac{(b^{\alpha})^*}{2} 
\left(
\sum_{\beta=1}^{\infty}(\Lambda^*)^{\alpha\beta}
\boxi_{(0)}^\beta
\right)
\right],
\end{align}
where the corrections to $\boxi_{(0)}^1$ are $\lambda^{1\beta} = 0$.
We can take an inner product with another zeroth order
eigenfunction $\boxi_{(0)}^\gamma$ to obtain

\begin{align}
\label{arb:1}
\left\langle \boxi_{(0)}^\gamma,\boxi^{\text{ID}} (x) \right\rangle &= \sum_{\alpha=1}^{\infty}\left[
\frac{b^{\alpha}}{2} \left(
\Lambda^{\alpha\gamma}
\right) +
\frac{(b^{\alpha})^*}{2}
\left(
(\Lambda^*)^{\alpha\gamma}
\right)
\right],
\\
\label{arb:2}
\left\langle \boxi_{(0)}^\gamma ,\dot{\boxi}^{\text{ID}} (x)\right\rangle
&= \sum_{\alpha=1}^{\infty} \left[
i\tilde{\omega}^\alpha \frac{b^{\alpha}}{2} \left(
\Lambda^{\alpha\gamma}
\right)
-i\tilde{\omega}^\alpha \frac{(b^{\alpha})^*}{2} 
\left(
(\Lambda^*)^{\alpha\gamma}
\right)
\right].
\end{align}
In practice, when carrying out a projection numerically we will only 
consider a finite sum, cutting off at some sufficiently large $\beta = N$.
We can then view \eqref{arb:1} and \eqref{arb:2} as vectors
of $N$ equations, so that we can rewrite them as

\begin{align}
\label{013}
\mathbf{x} &= \frac{1}{2}\left(\Lambda^T \mathbf{b} + \left(\Lambda^T\right)^*\mathbf{b}^*\right), \\
\label{014}
\mathbf{\dot{x}} &= \frac{1}{2}\left(\Omega^T \mathbf{b} + \left(\Omega^T\right)^*\mathbf{b}^*\right),
\end{align}
where the vectors $\mathbf{x}$ and $\mathbf{\dot{x}}$ are defined by

\begin{align}
\label{015}
x^\beta &\equiv \left\langle \boxi^\beta,\boxi^{\text{ID}}(x)\right\rangle \\
\label{016}
\dot{x}^\beta &\equiv \left\langle \boxi^\beta,\dot{\boxi}^{\text{ID}}(x)\right\rangle 
\end{align}
for $\beta = 1,\dots , N$, and the matrix $\Omega$ is built from $\Lambda$ \eqref{Lambda intro} as

\begin{align}
\label{Omega}
\Omega \equiv D\Lambda, \\ 
\label{Diag}
D\equiv\text{diag}\,[i\tilde{\omega}^1, i\tilde{\omega}^2,\dots, i\tilde{\omega}^N].
\end{align}
Solving for $\mathbf{b}$, we have

\begin{equation}
\label{arb:b}
\mathbf{b} = 2\text{A}^{-1}\left\lbrace
\left[(\Lambda^*)^{\text{T}}\right]^{-1}\mathbf{x}
-\left[(\Omega^*)^{\text{T}}\right]^{-1}\mathbf{\dot{x}}
\right\rbrace
\end{equation}
where $\text{A}$ is the matrix

\begin{equation}
\label{A}
\text{A}=\left[(\Lambda^*)^{\text{T}}\right]^{-1}\Lambda^{\text{T}}
-\left[(\Omega^*)^{\text{T}}\right]^{-1}\Omega^{\text{T}}.
\end{equation}
These coefficients $\mathbf{b}$ can then be used to reconstruct the initial data, as

\begin{align}
\label{reconstruct xiCD}
\boxi^{\text{reconstruct}}(N) &= \frac{1}{2}\sum_{\alpha=1}^N\left[
b^\alpha\Lambda^{\alpha\beta} \boxi_{(0)}^\beta + (b^*)^\alpha\left(\Lambda^*\right)^{\alpha\beta}\boxi_{(0)}^\beta
\right],
\\
\label{reconstruct xidotCD}
\dot{\boxi}^{\text{reconstruct}}(N) &= \frac{1}{2}\sum_{\alpha=1}^N\left[
i\omega^\alpha b^\alpha\Lambda^{\alpha\beta} \boxi_{(0)}^\beta - i\omega^\alpha(b^*)^\alpha\left(\Lambda^*\right)^{\alpha\beta}\boxi_{(0)}^\beta
\right].
\end{align}
To make use of this projection scheme, we first need to construct the 
matrix $\Lambda$. To do this, we will need to decide which modes
$\alpha = (n,l,m)$ we are keeping in the projection, and  choose an
ordering for the mode indices $\alpha$.
Looking at the initial data we are projecting, we can see that
the displacement field \eqref{xiDC ref} is spheroidal $l=2$ and $m=0$. 
From the form of the full, corrected spheroidal eigenfunctions \eqref{modesrot:Sfull},
this can be expected to 
couple to $l=1$ and $l=3$ toroidal $m=0$ modes. The velocity field 
\eqref{xidotDC ref} is
built from the toroidal $l=1$, $m=0$ zero frequency mode, which has no
first order corrections. However from the form of the corrected toroidal
eigenfunctions \eqref{modesrot:Tfull}, the $l=2$ spheroidal eigenfunctions
can be expected to couple to this mode.
Motivated by this, we will choose to consider only $m=0$ and also to cut off 
at $l=3$, including only $l=1,2,3$ (the $l=0$ radial modes are all zero 
for an incompressible star). Counting both spheroidal and toroidal eigenfunctions, this gives us
$6N$ eigenfunctions in total. This is illustrated in Figure \ref{rotationmodes}.

%figure: rotation coupling schematic%%%%%%%%%%%%%%%%%%%%%%%%%%%
\begin{figure}
\centering
\includegraphics[scale=0.65]{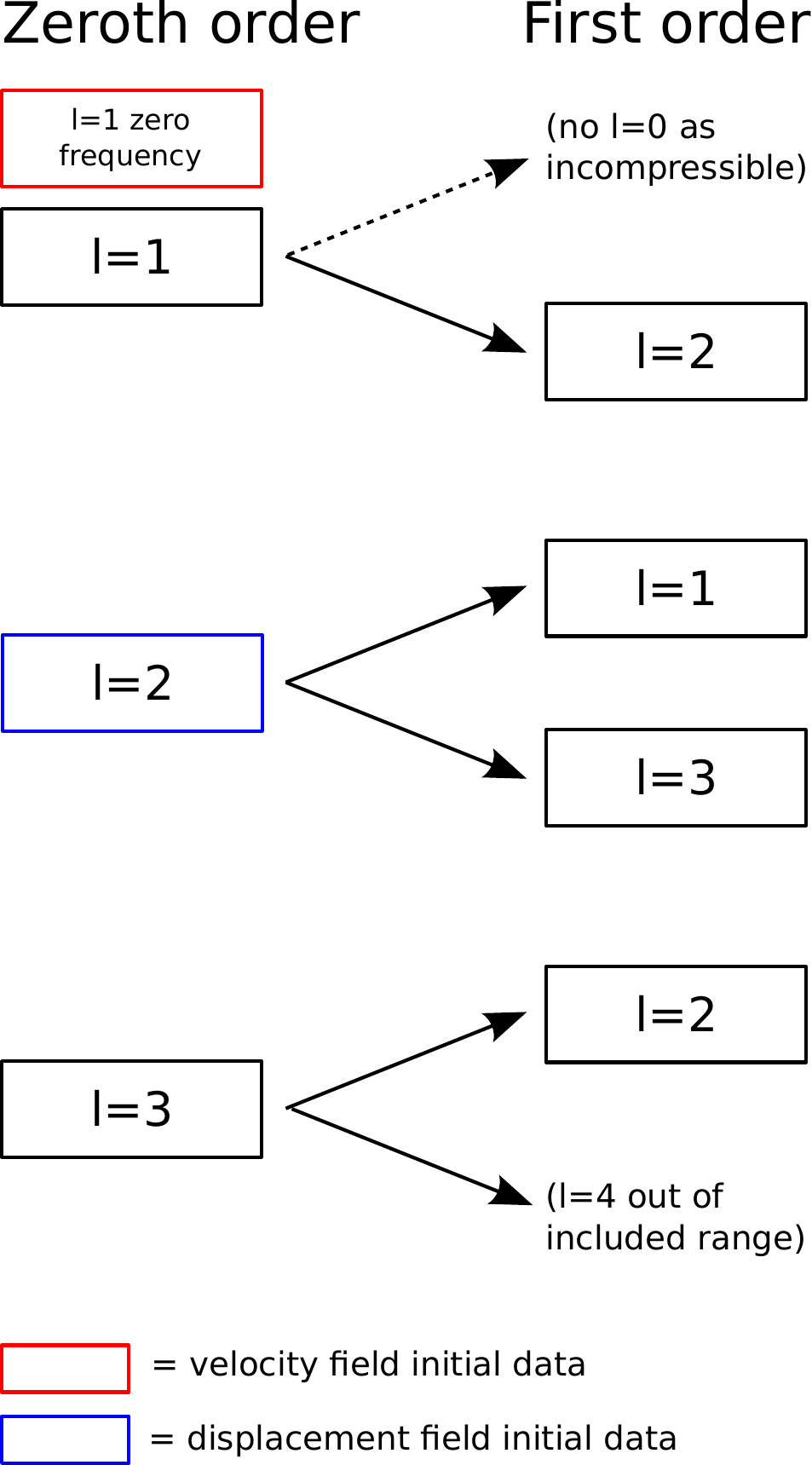}
\caption[Diagram showing how the zeroth order modes couple to 
other modes at first order in rotation.]{Schematic showing how the zeroth order modes couple to 
other modes at first order in rotation. The displacement field has an
$l=2$ form and couples to $l=1$ and $l=3$; this is shown in blue.
The velocity field, shown in read, has the form of a zero frequency 
$l=1$ mode; it has no rotational corrections. The other modes in 
the range we are considering are also shown. The $l=1$ modes only couple to 
$l=2$, while for the $l=3$ modes, only the $l=2$ coupling falls into 
the considered range.}
\label{rotationmodes}
\end{figure}
%%%%%%%%%%%%%%%%%%%%%%%%%%%%%%%%%%%%%%%%%%%%%%%%%%%%%%%%%%%%%%%

We can then write each of the zeroth order eigenfunctions we plan
to use as one entry of a vector of length $6N$, which schematically looks like

\begin{equation}
\label{vector6N}
\left(\begin{array}{c} \null[S1]_N \\
                       \null[T1]_N \\
                       \null[S2]_N \\ 
                       \null[T2]_N \\                                                        
                       \null[S3]_N \\                            
                       \null[T3]_N \\                           
\end{array}\right).
\end{equation}
Here $[S1]_N$ is a vector of the first $N$ spheroidal $l=1$ eigenfunctions, 
$[T1]_N$ is a vector of the first $N$ toroidal $l=1$ eigenfunctions, etc. 
The corrected eigenfunctions
are then obtained by acting on this vector with the $6N \times 6N$ matrix $\Lambda$
\eqref{Lambda intro}, which has the form
\begin{equation}
\label{Lambda 6N}
\left(\begin{array}{cccccc}
I_N & 0 & 0 & \vep[ST12]_N & 0 & 0 \\ 
0 & I_N & \vep[TS12]_N & 0 & 0 & 0 \\
0 & \vep[ST21]_N & I_N & 0 & 0 &  \vep[ST23]_N \\
 \vep{[TS21]_N} & 0 & 0 & I_N &  \vep[TS23]_N & 0 \\
0 & 0 & 0 & \vep[ST32]_N & I_N & 0 \\
0 & 0 & \vep[TS32]_N & 0 & 0 & I_N 
\end{array} \right).
\end{equation}
Each entry represents an $N\times N$ matrix. Most off-diagonal entries
are zero apart from a few blocks.
For $l=2$ these are the $l=1$ and $3$ toroidal
corrections to the spheroidal eigenfunctions, $[ST21]_N$ and $[ST23]_N$,
and the $l=1$ and $3$ spheroidal corrections to the toroidal 
eigenfunctions, $[TS21]_N$ and $[TS23]_N$. 
For $l=1$ there are only $l=2$ corrections; this is also true
for $l=3$ as we are not including $l=4$.
%The structure of the matrix
%is shown in more detail in Figure \ref{lambda} for the case $b=0.1$ !!change?, where the absolute
%value of the matrix coefficients is plotted (a darker colour indicates
%a larger value of the coefficient).

%\begin{figure}
%\centering
%\includegraphics[scale=0.5]{figures/lambda.pdf}
%\caption[Plot of the matrix $\Lambda$]{Plot of the matrix $\Lambda$, which we are using to 
%act on a vector of our zeroth order eigenfunctions to obtain the rotationally
%corrected eigenfunctions. Each coloured box represents a component
%of the matrix; a darker orange indicates a larger absolute value
%of the component.}
%\label{lambda}
%\end{figure}

%%%%%%%%%%%%%%%%%%%%%%%%%%%%%%%%%%%%%%%%%%%%%%%%%
% RESULTS

\subsection{Results of the projection}
\label{subsec:results}

%figure: amplitude barcharts%%%%%%%%%%%%%%%%%%%%%%%%%%%%%%%%%%%
\begin{figure}
\centering     %%% not \center
\subfigure[$l=1$ toroidal, $\frac{\sqrt{t}}{b} = 10^{\frac{1}{2}}$]{\label{fig:BA11}\includegraphics[width=43mm]{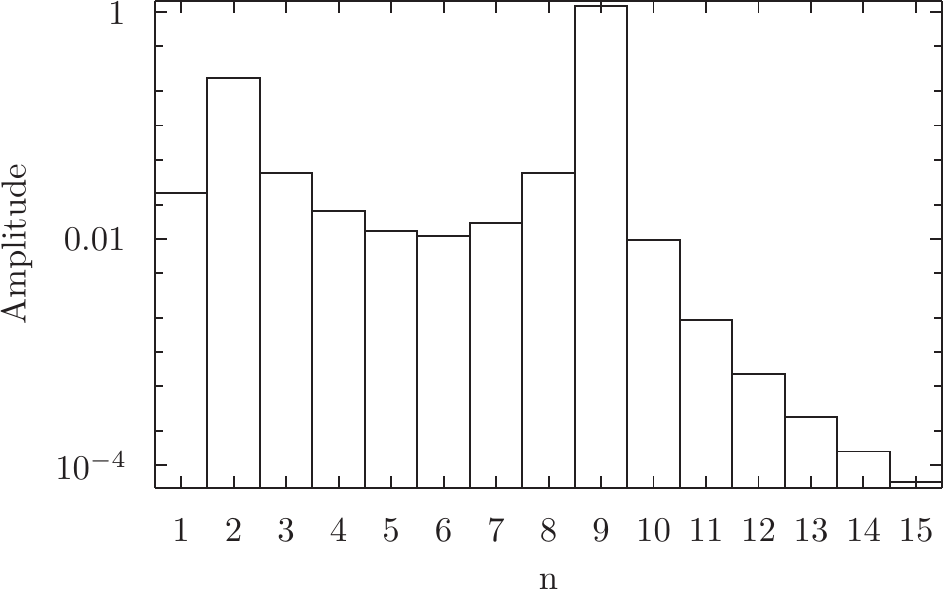}}
\subfigure[$l=2$ spheroidal, $\frac{\sqrt{t}}{b} = 10^{\frac{1}{2}}$]{\label{fig:BA12}\includegraphics[width=43mm]{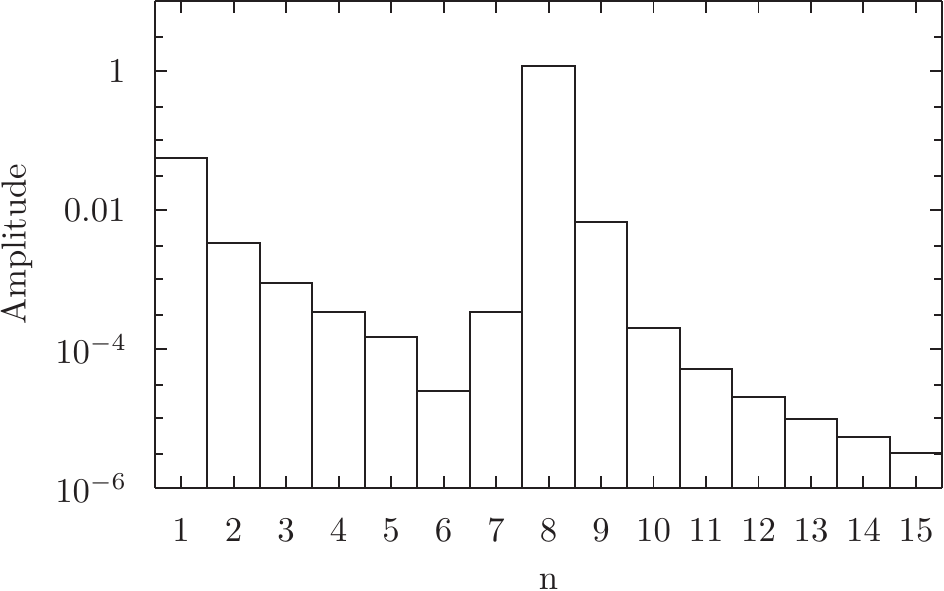}}
\subfigure[$l=3$ toroidal, $\frac{\sqrt{t}}{b} = 10^{\frac{1}{2}}$]{\label{fig:BA13}\includegraphics[width=43mm]{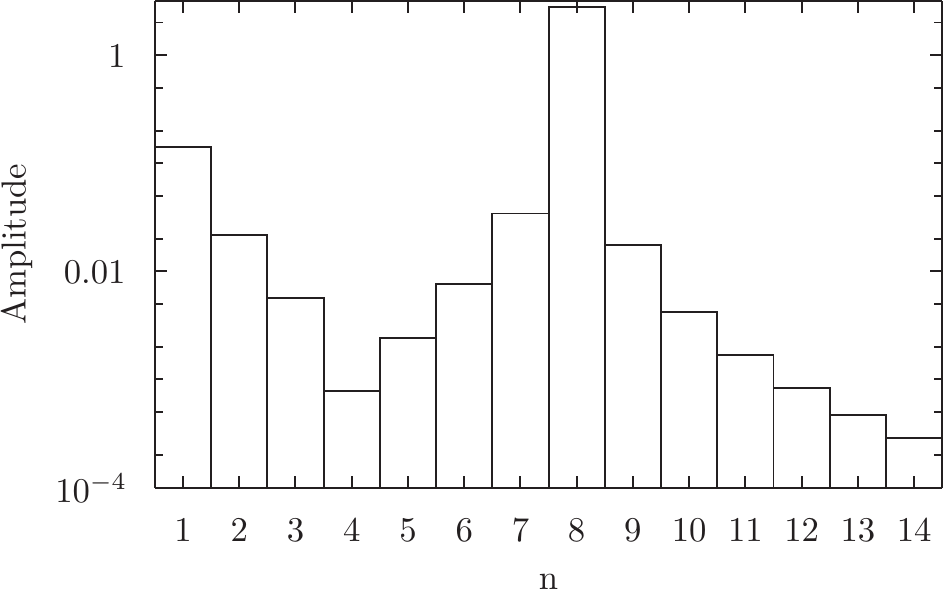}}
\\
\subfigure[$l=1$ toroidal, $\frac{\sqrt{t}}{b} = 1$]{\label{fig:BA21}\includegraphics[width=43mm]{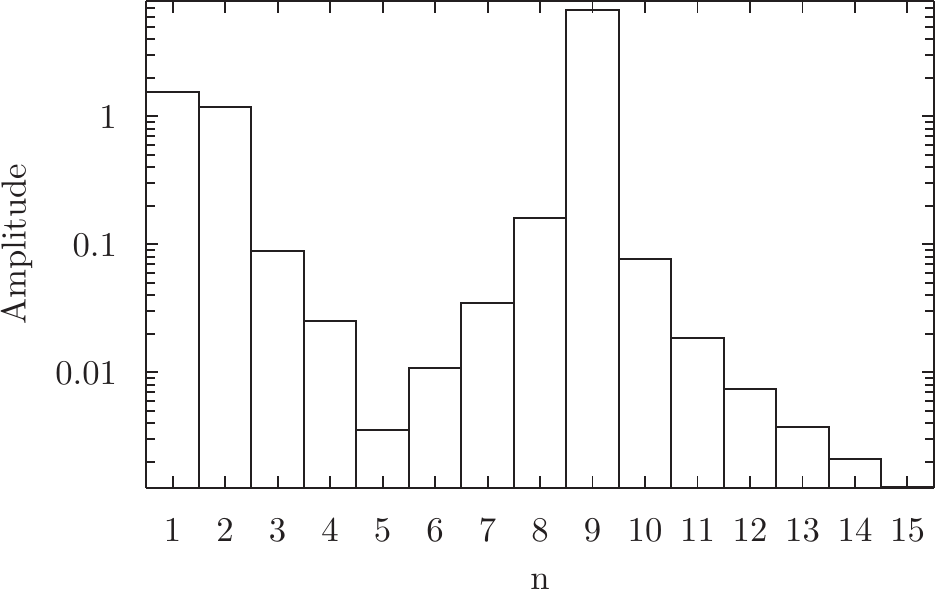}}
\subfigure[$l=2$ spheroidal, $\frac{\sqrt{t}}{b} = 1$]{\label{fig:BA22}\includegraphics[width=43mm]{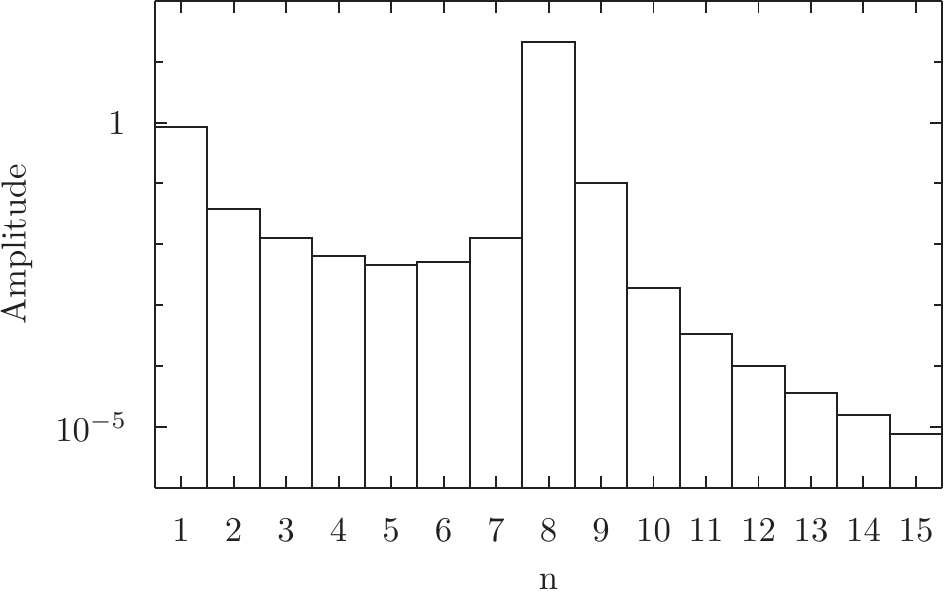}}
\subfigure[$l=3$ toroidal, $\frac{\sqrt{t}}{b} = 1$]{\label{fig:BA23}\includegraphics[width=43mm]{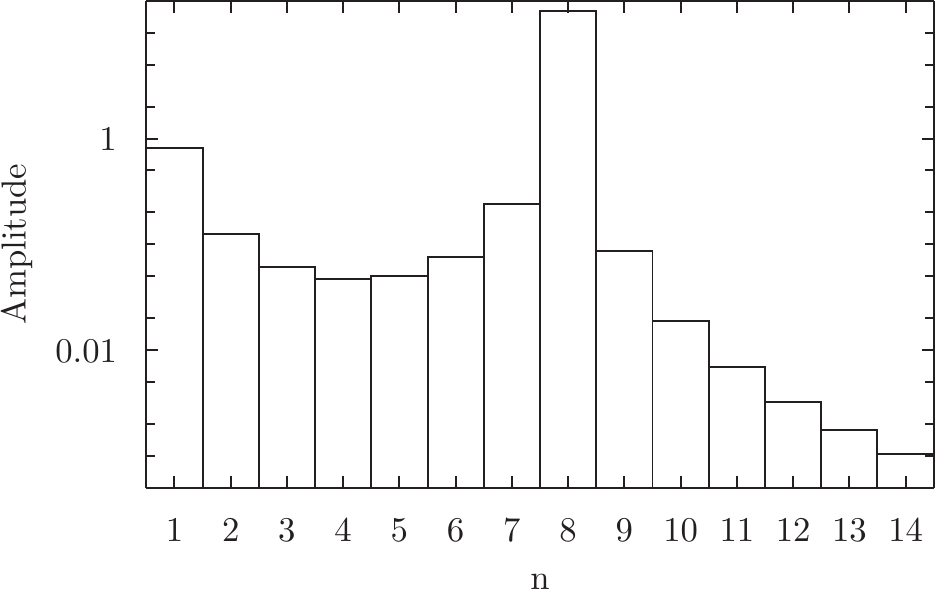}}
\\
\subfigure[$l=1$ toroidal, $\frac{\sqrt{t}}{b} = 10^{-\frac{1}{2}}$]{\label{fig:BA31}\includegraphics[width=43mm]{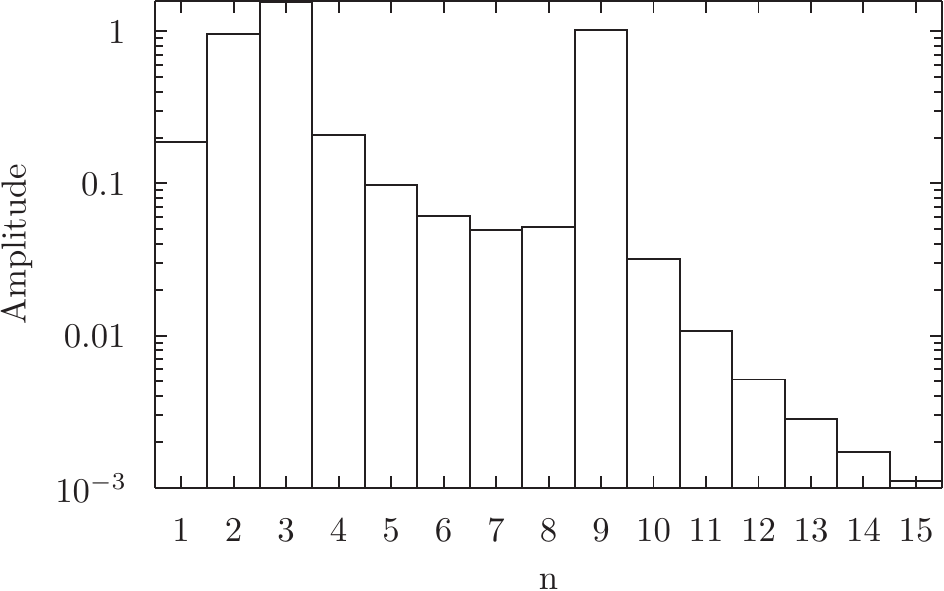}}
\subfigure[$l=2$ spheroidal, $\frac{\sqrt{t}}{b} = 10^{-\frac{1}{2}}$]{\label{fig:BA32}\includegraphics[width=43mm]{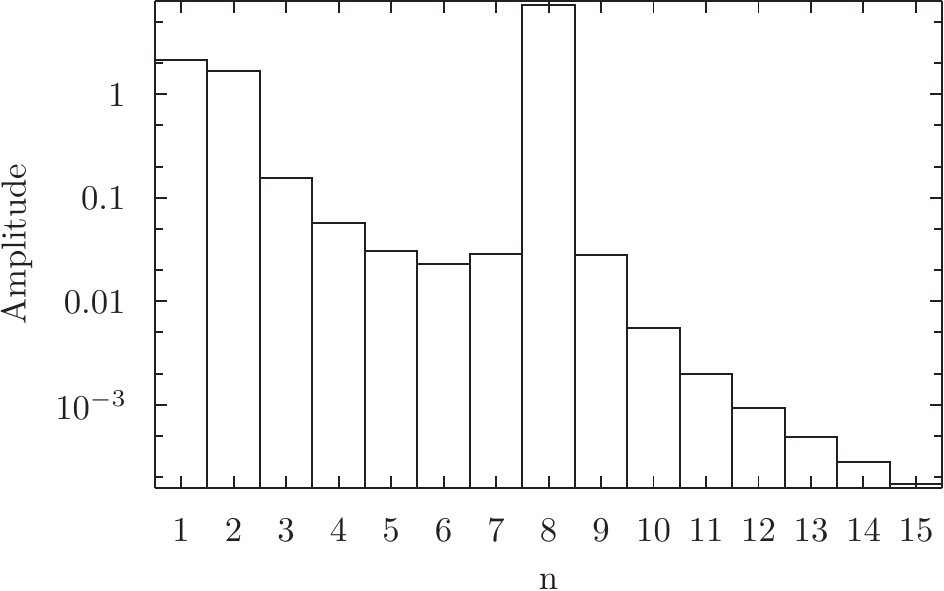}}
\subfigure[$l=3$ toroidal, $\frac{\sqrt{t}}{b} = 10^{-\frac{1}{2}}$]{\label{fig:BA33}\includegraphics[width=43mm]{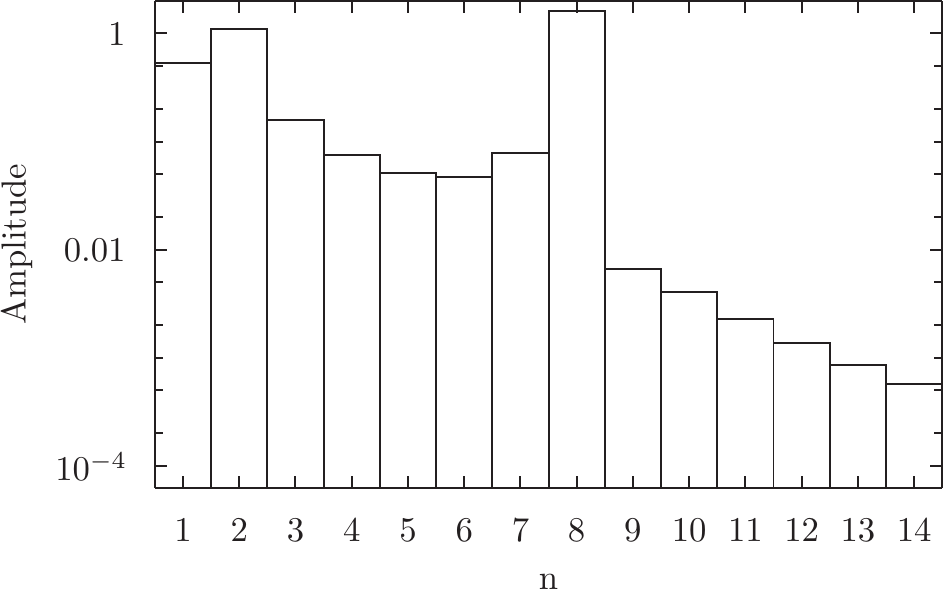}}
\\
\subfigure[$l=1$ toroidal, $\frac{\sqrt{t}}{b} = 10^{-1}$]{\label{fig:BA41}\includegraphics[width=43mm]{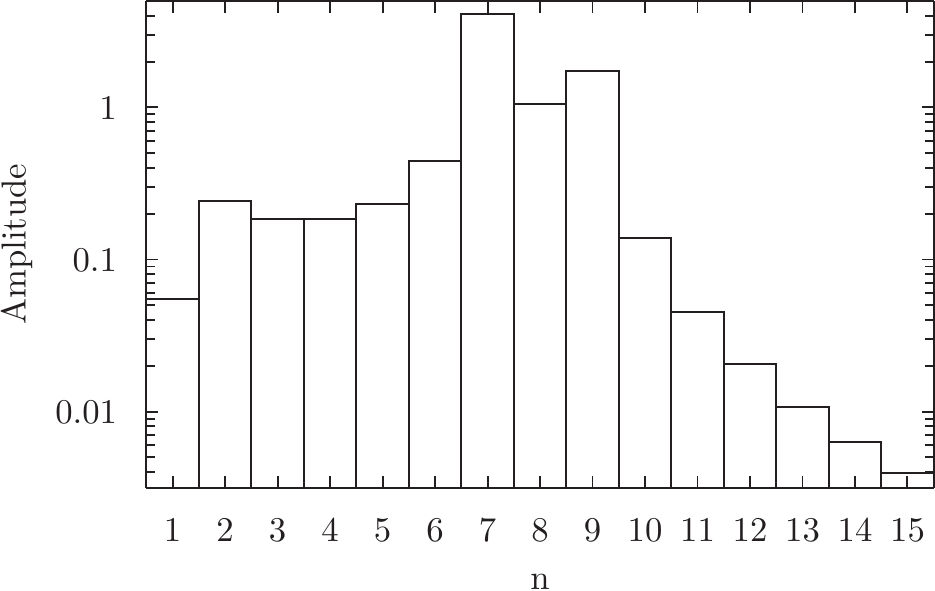}}
\subfigure[$l=2$ spheroidal, $\frac{\sqrt{t}}{b} = 10^{-1}$]{\label{fig:BA42}\includegraphics[width=43mm]{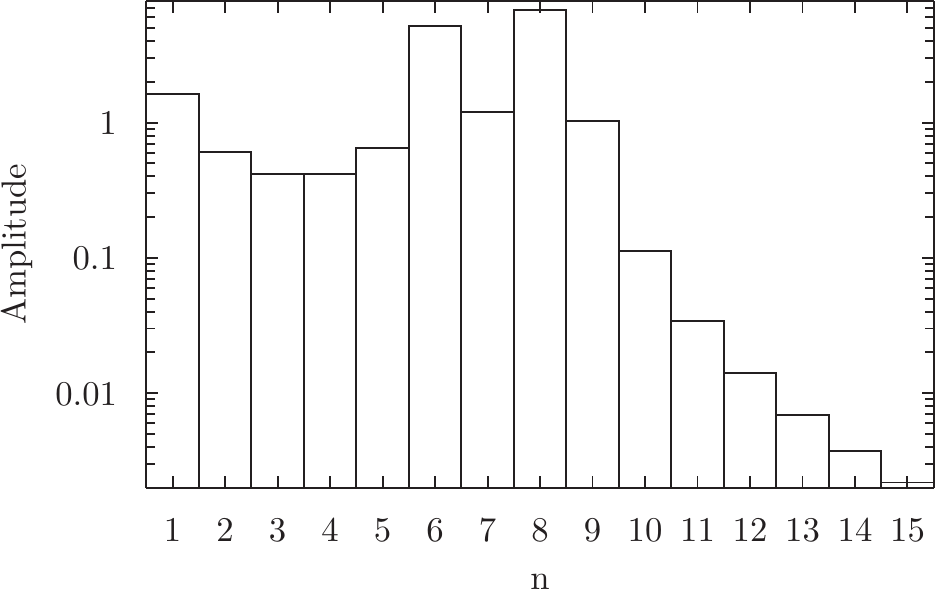}}
\subfigure[$l=3$ toroidal, $\frac{\sqrt{t}}{b} = 10^{-1}$]{\label{fig:BA43}\includegraphics[width=43mm]{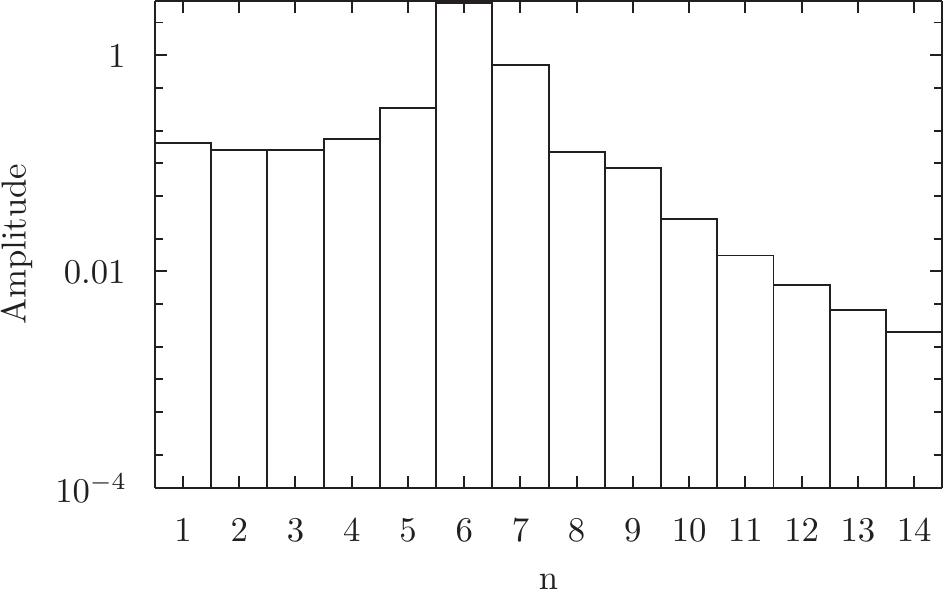}}
\\
\subfigure[$l=1$ toroidal, $\frac{\sqrt{t}}{b} = 10^{-\frac{3}{2}}$]{\label{fig:BA51}\includegraphics[width=43mm]{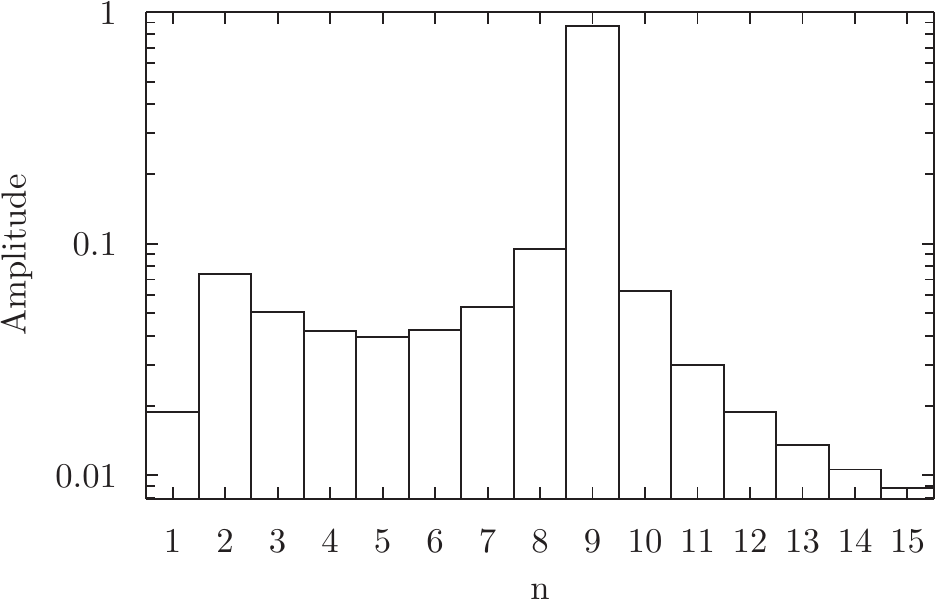}}
\subfigure[$l=2$ spheroidal, $\frac{\sqrt{t}}{b} = 10^{-\frac{3}{2}}$]{\label{fig:BA52}\includegraphics[width=43mm]{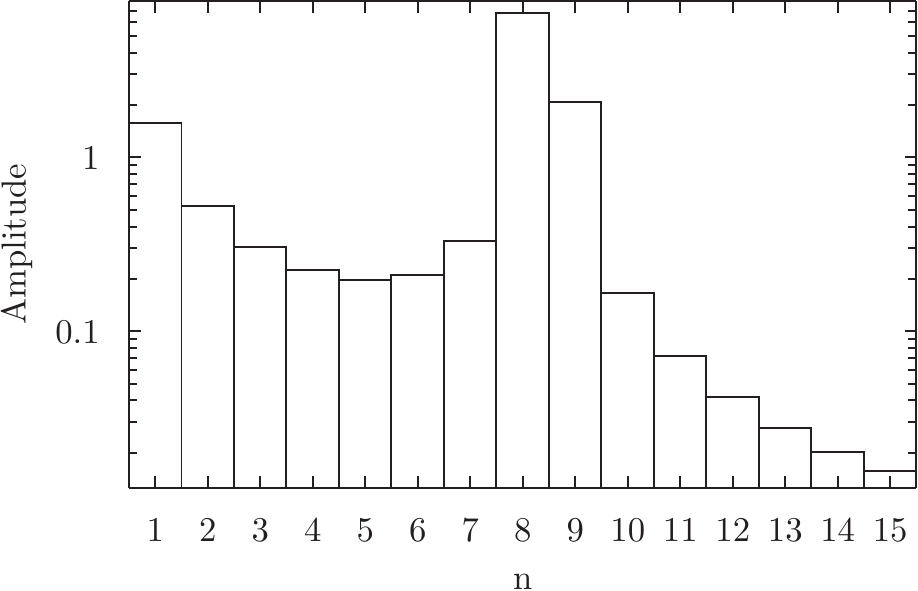}}
\subfigure[$l=3$ toroidal, $\frac{\sqrt{t}}{b} = 10^{-\frac{3}{2}}$]{\label{fig:BA53}\includegraphics[width=43mm]{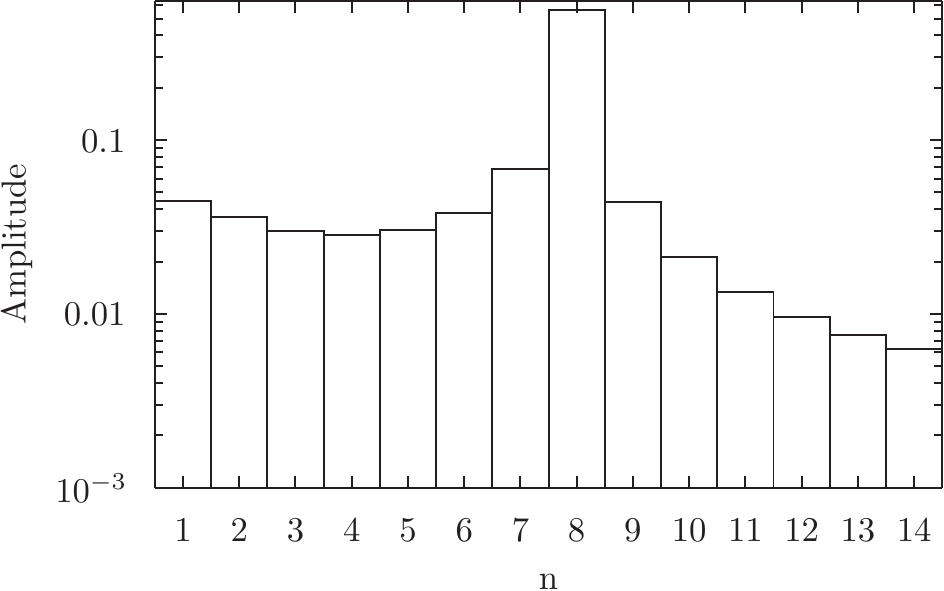}}
\\
\subfigure[$l=1$ toroidal, $\frac{\sqrt{t}}{b} = 10^{-2}$]{\label{fig:BA61}\includegraphics[width=43mm]{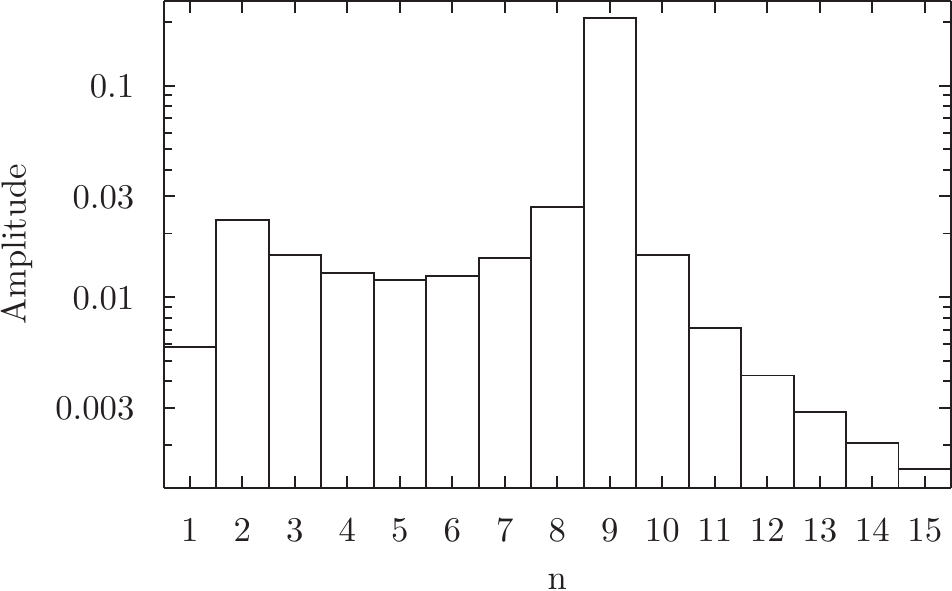}}
\subfigure[$l=2$ spheroidal, $\frac{\sqrt{t}}{b} = 10^{-2}$]{\label{fig:BA62}\includegraphics[width=43mm]{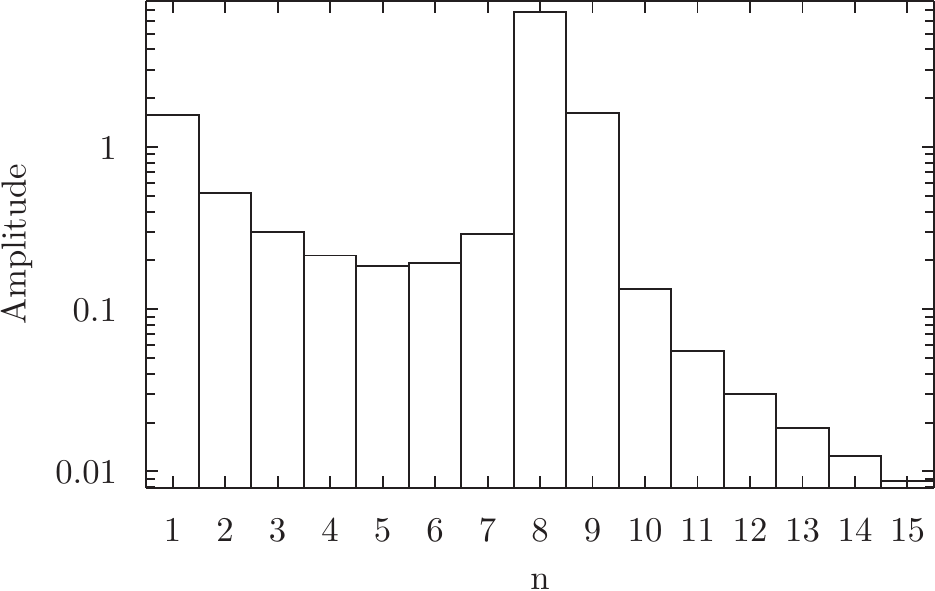}}
\subfigure[$l=3$ toroidal, $\frac{\sqrt{t}}{b} = 10^{-2}$]{\label{fig:BA63}\includegraphics[width=43mm]{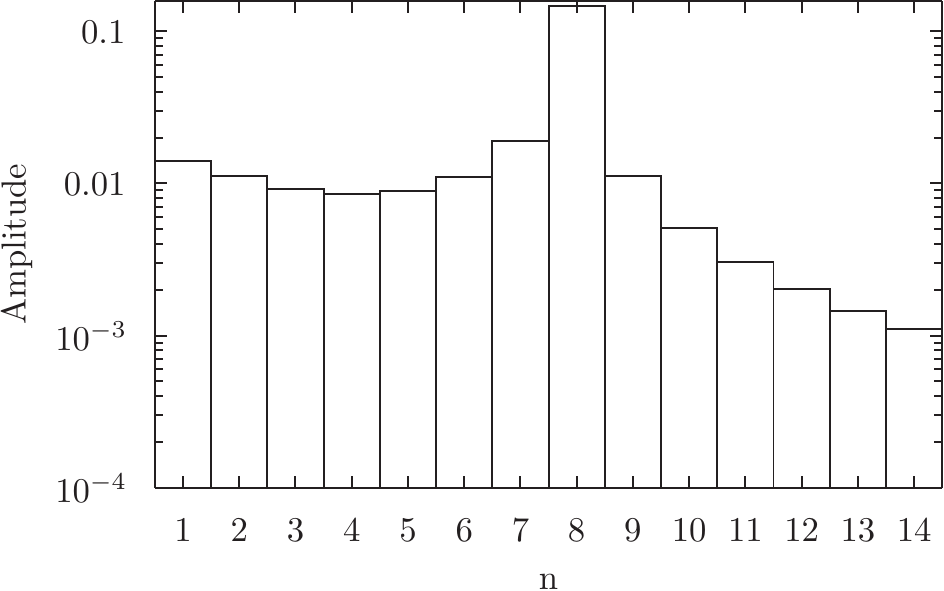}}
\caption[]{Figures showing the amplitude of each excited mode for the case $b=10^{-2}$ and 
$\log_{10} t = -1, \dots, -6$, where we have
cut off at radial eigenvalue $N=15$.}
\label{fig:barcharts t}
\end{figure}
%%%%%%%%%%%%%%%%%%%%%%%%%%%%%%%%%%%%%%%%%%%%%%%%%%%%%%%%%%%%%%%
%figure: continuous plots%%%%%%%%%%%%%%%%%%%%%%%%%%%%%%%%%%%%%%
\begin{figure}
\centering     %%% not \center
\subfigure[$l=1$ toroidal, $n=1$]{\label{fig:l1n1}\includegraphics[width=40mm]{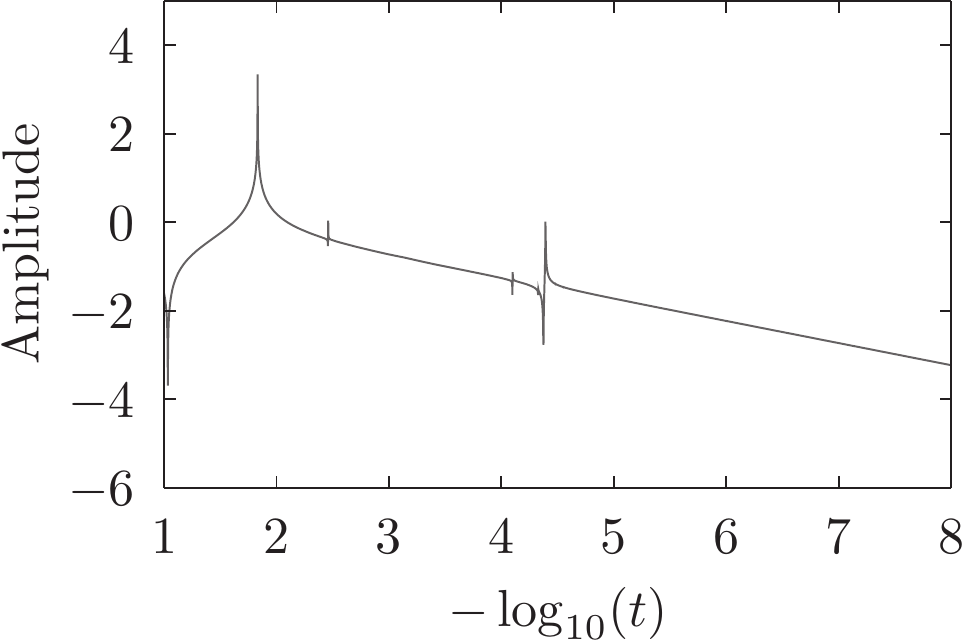}}
\subfigure[$l=2$ spheroidal, $n=1$]{\label{fig:l2n1}\includegraphics[width=40mm]{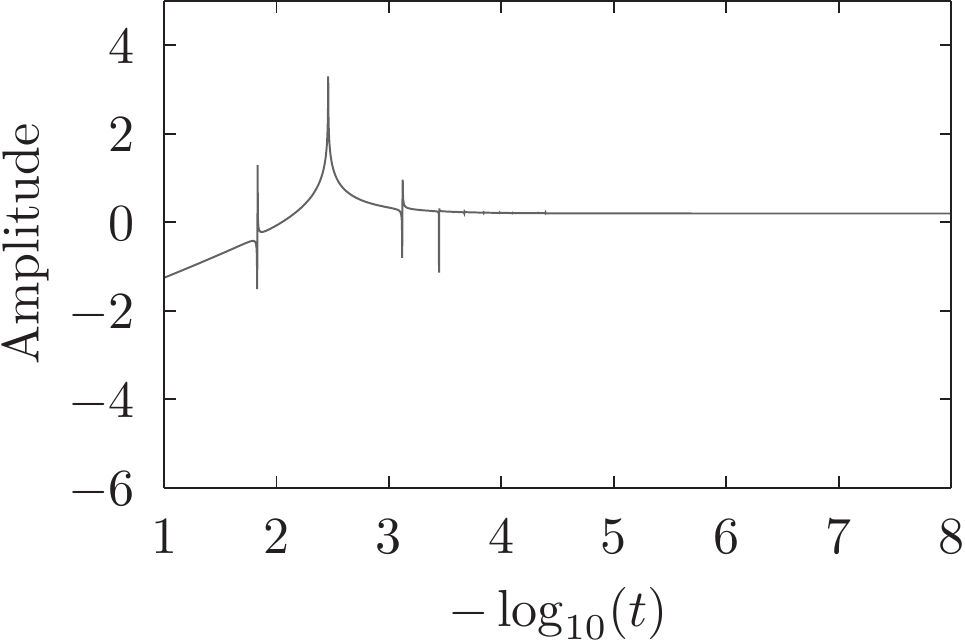}}
\subfigure[$l=3$ toroidal, $n=1$]{\label{fig:l3n1}\includegraphics[width=40mm]{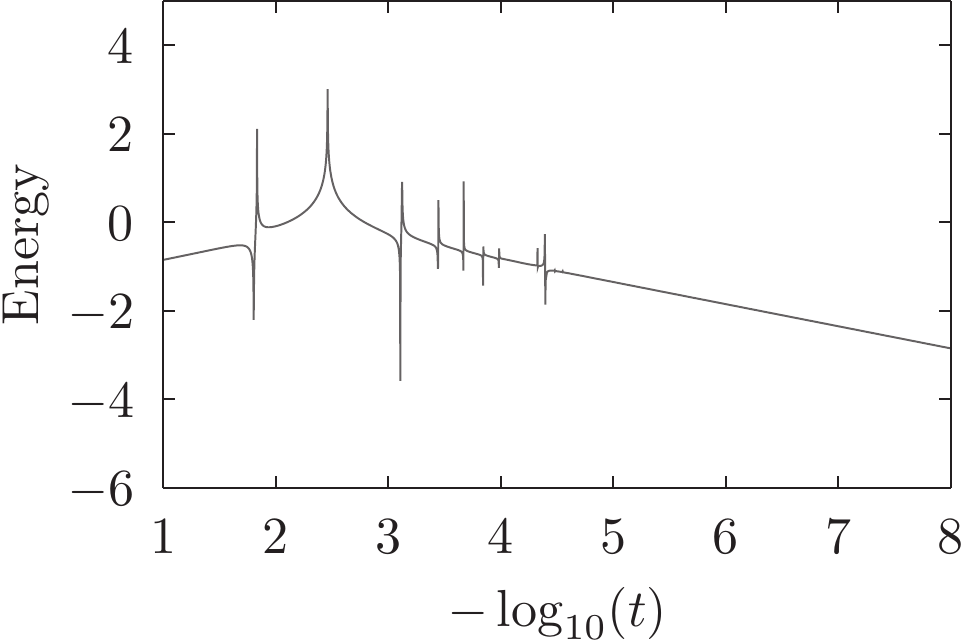}}
\\
\subfigure[$l=1$ toroidal, $n=2$]{\label{fig:l1n2}\includegraphics[width=40mm]{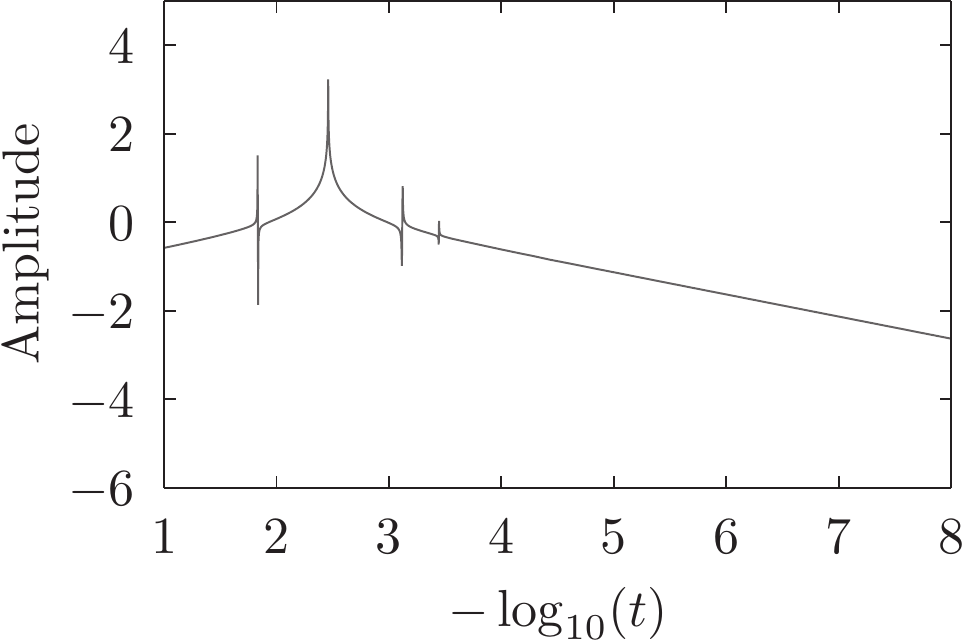}}
\subfigure[$l=2$ spheroidal, $n=2$]{\label{fig:l2n2}\includegraphics[width=40mm]{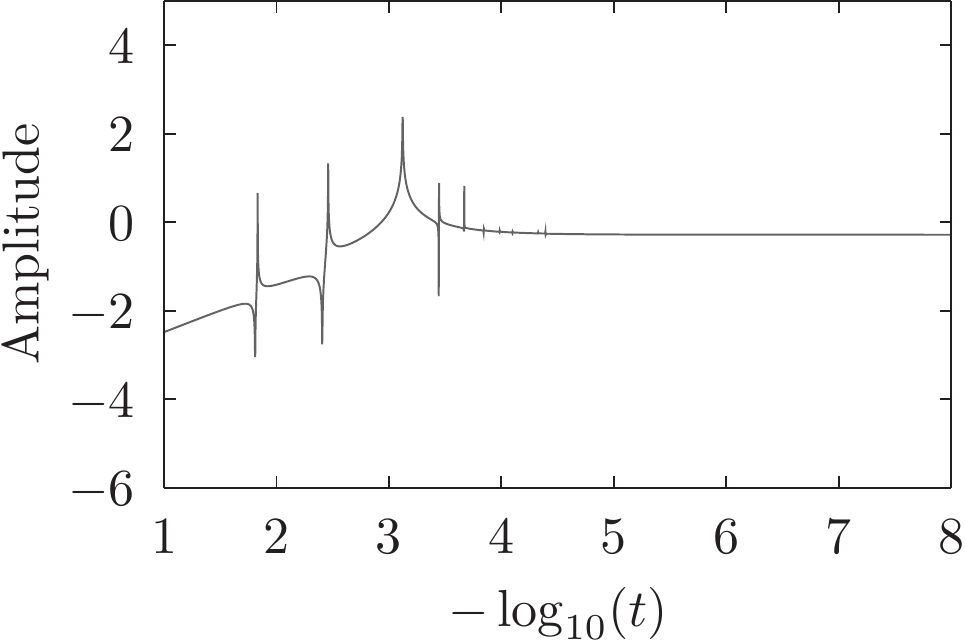}}
\subfigure[$l=3$ toroidal, $n=2$]{\label{fig:l3n2}\includegraphics[width=40mm]{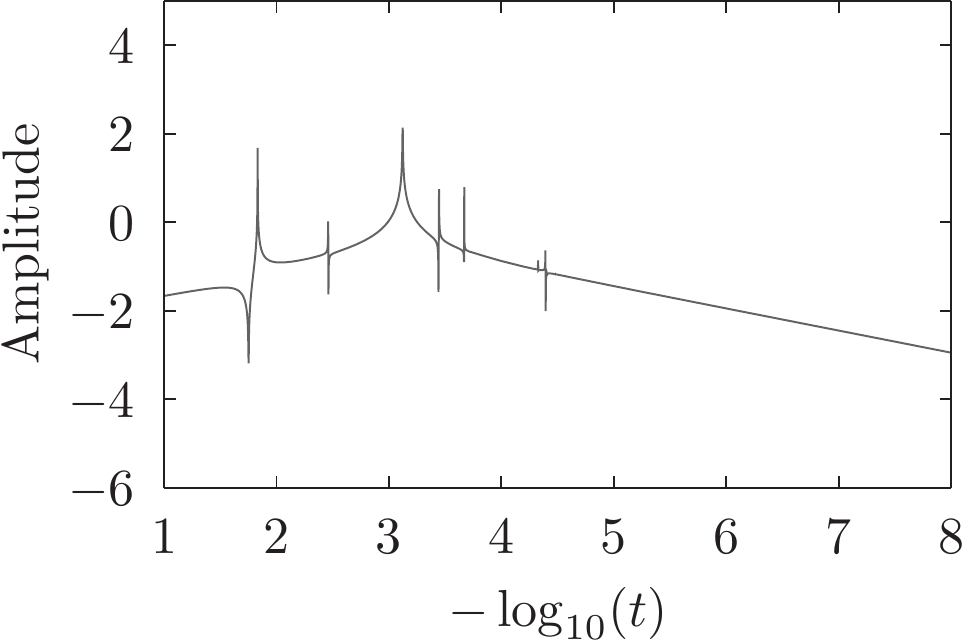}}
\\
\subfigure[$l=1$ toroidal, $n=3$]{\label{fig:l1n3}\includegraphics[width=40mm]{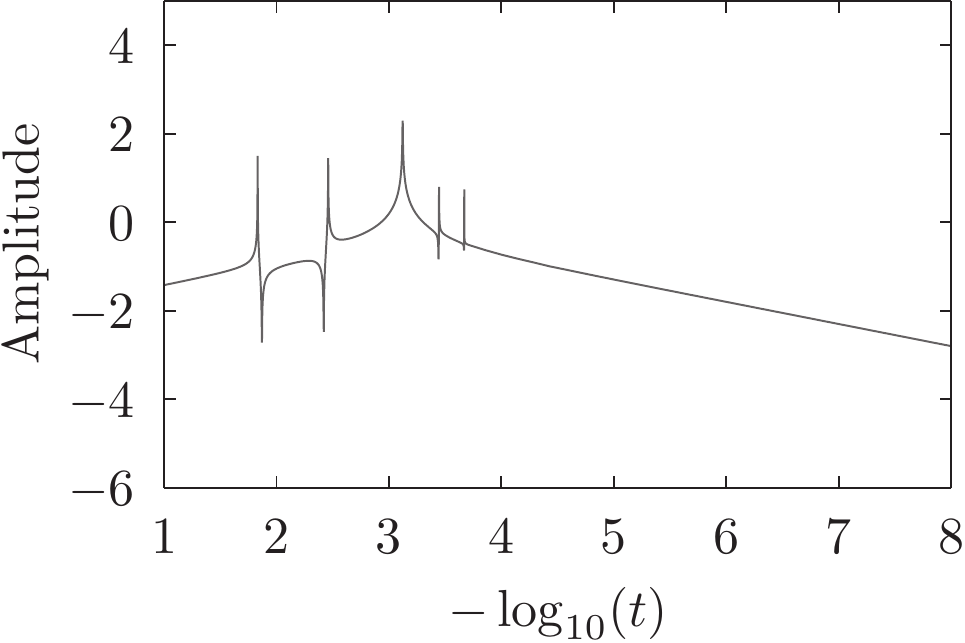}}
\subfigure[$l=2$ spheroidal, $n=3$]{\label{fig:l2n3}\includegraphics[width=40mm]{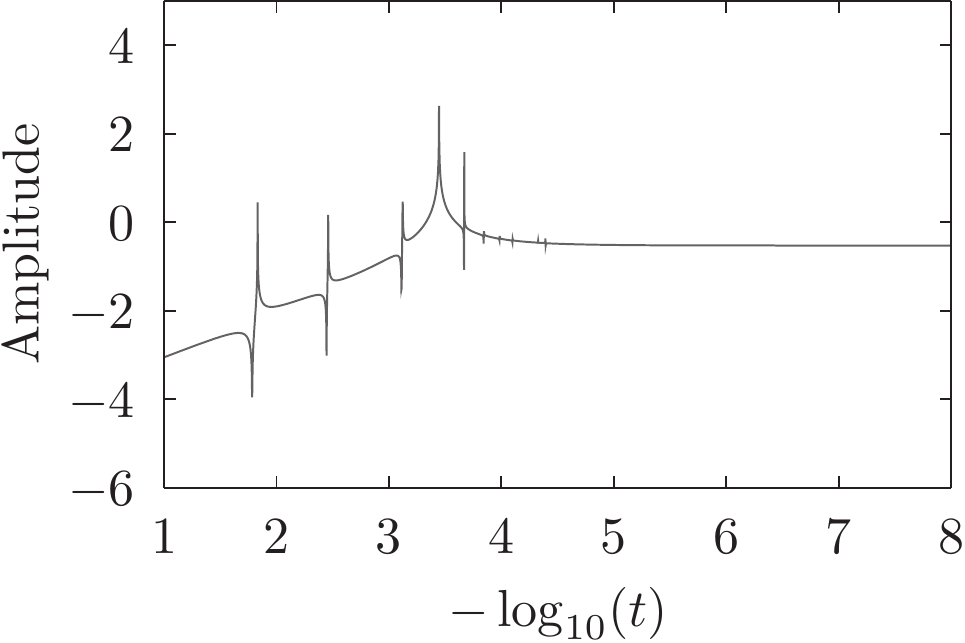}}
\subfigure[$l=3$ toroidal, $n=3$]{\label{fig:l3n3}\includegraphics[width=40mm]{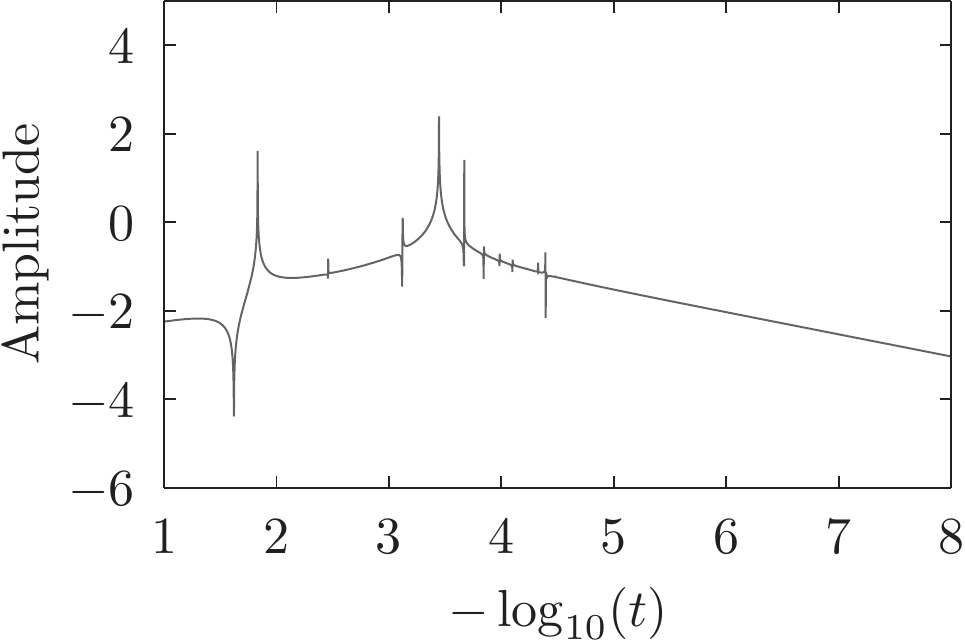}}
\\
\subfigure[$l=1$ toroidal, $n=7$]{\label{fig:l1n4}\includegraphics[width=40mm]{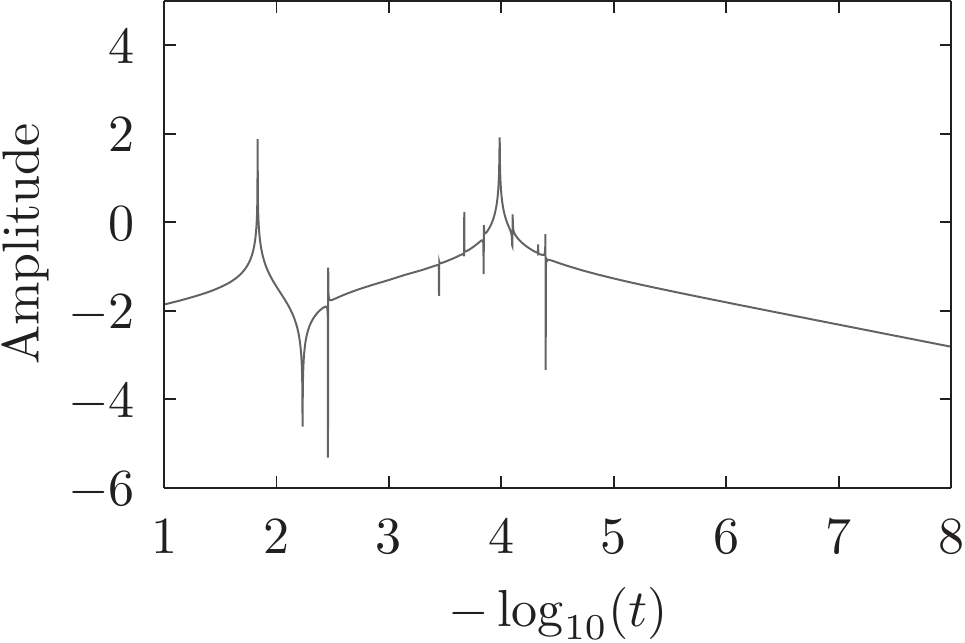}}
\subfigure[$l=2$ spheroidal, $n=7$]{\label{fig:l2n4}\includegraphics[width=40mm]{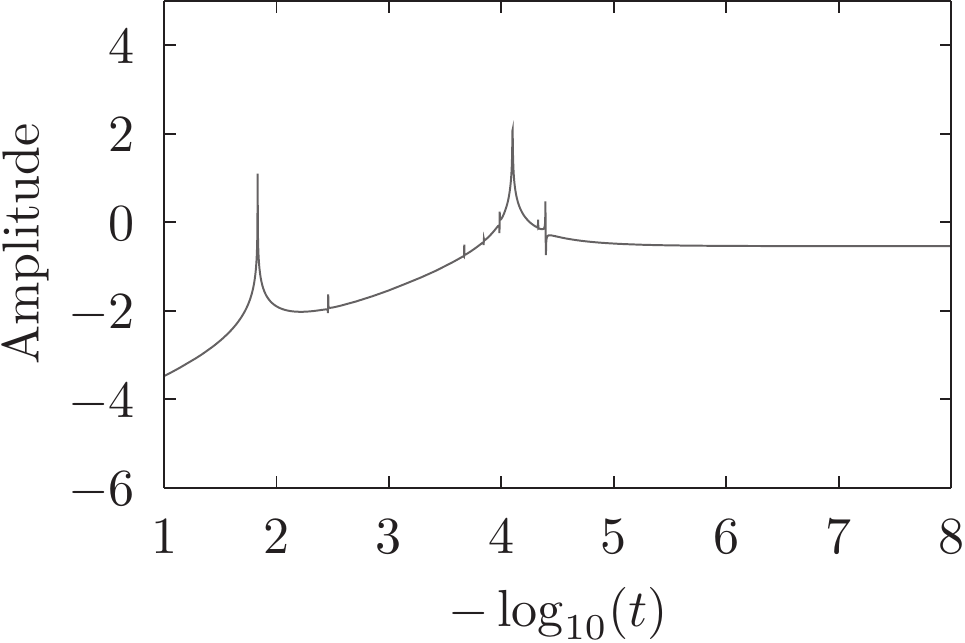}}
\subfigure[$l=3$ toroidal, $n=7$]{\label{fig:l3n4}\includegraphics[width=40mm]{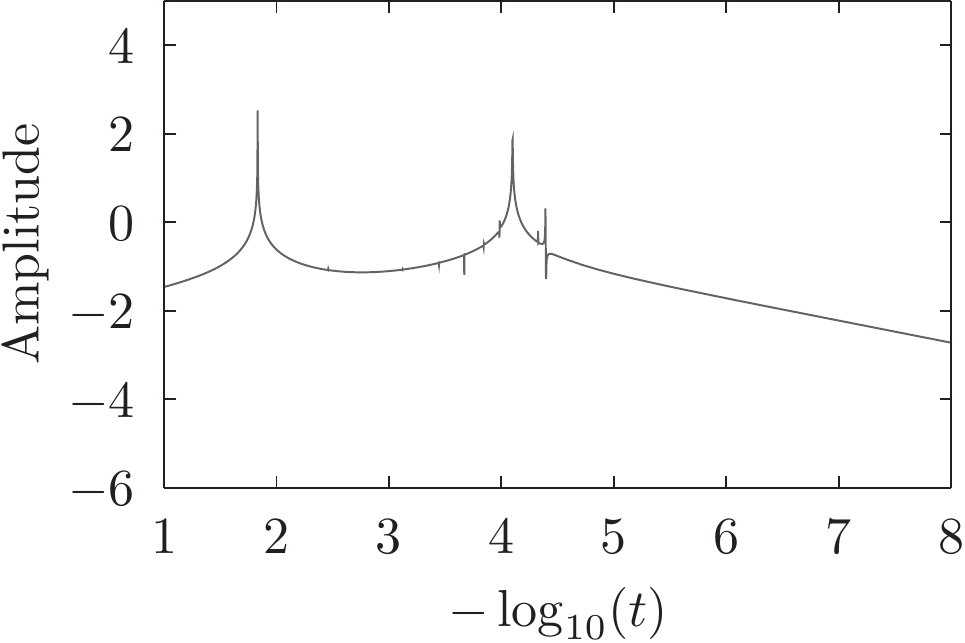}}
\\
\subfigure[$l=1$ toroidal, $n=8$]{\label{fig:l1n5}\includegraphics[width=40mm]{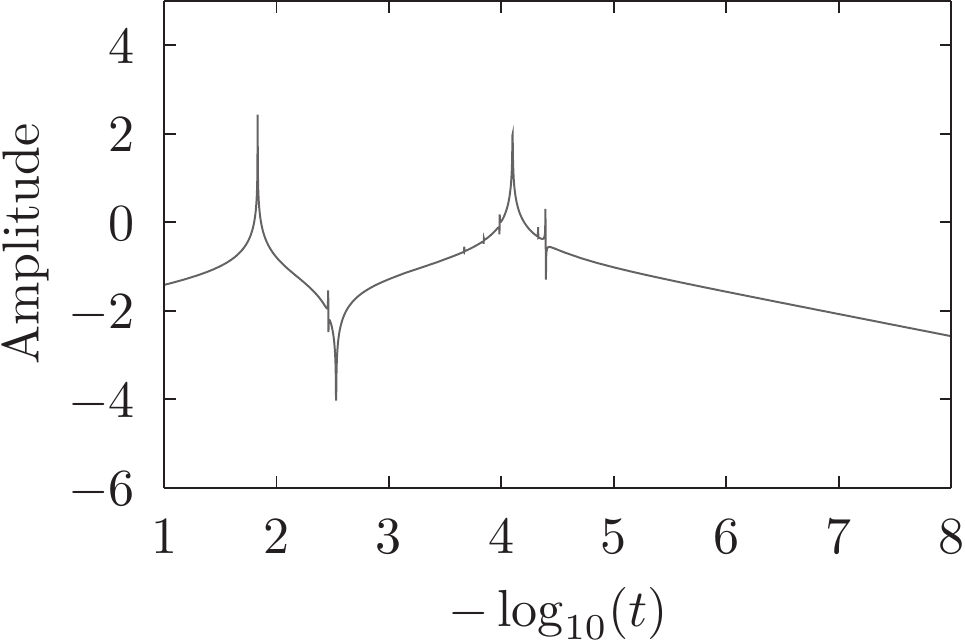}}
\subfigure[$l=2$ spheroidal, $n=8$]{\label{fig:l2n5}\includegraphics[width=40mm]{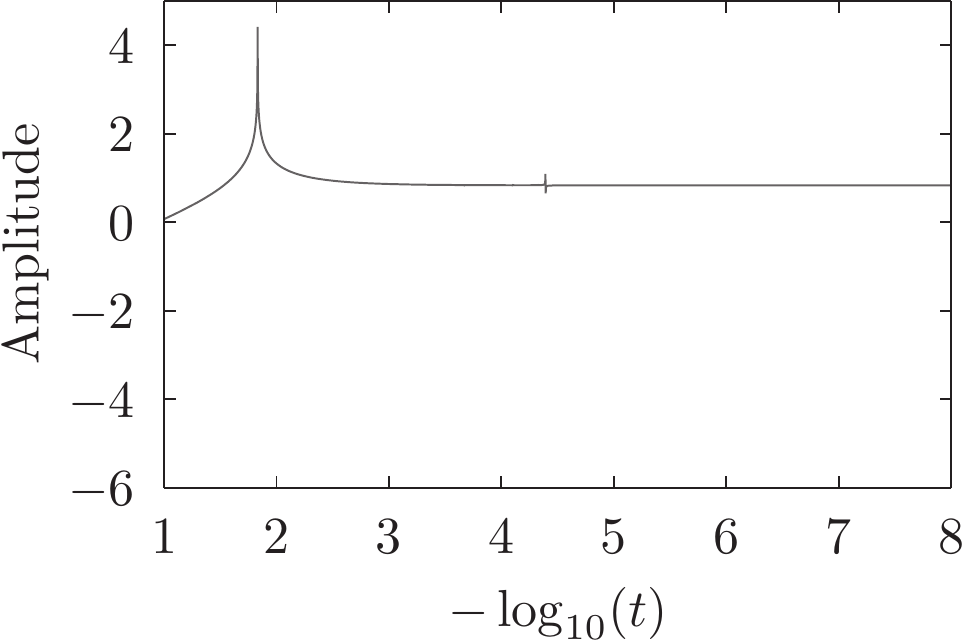}}
\subfigure[$l=3$ toroidal, $n=8$]{\label{fig:l3n5}\includegraphics[width=40mm]{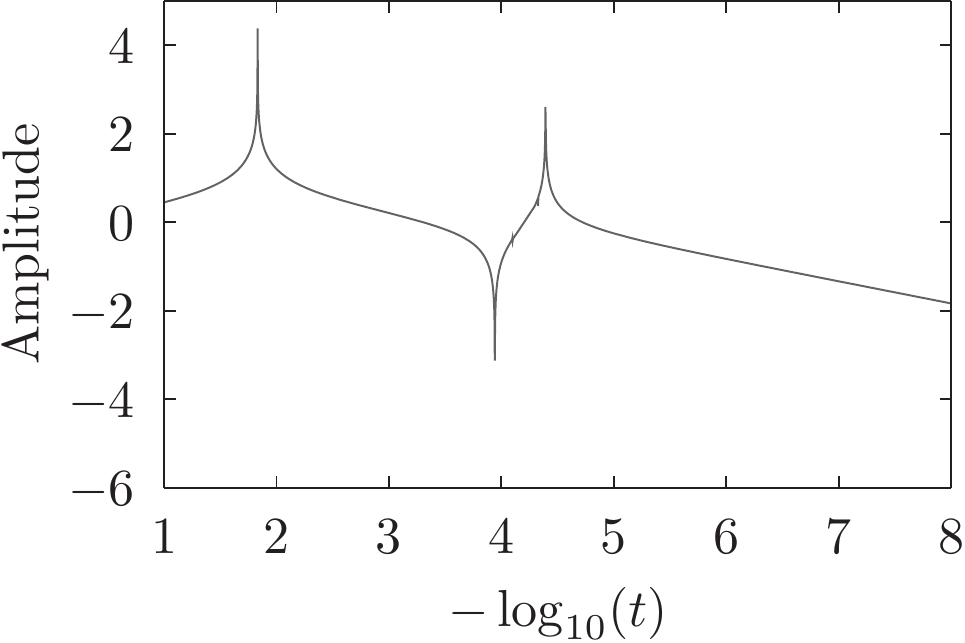}}
\\
\subfigure[$l=1$ toroidal, $n=9$]{\label{fig:l1n6}\includegraphics[width=40mm]{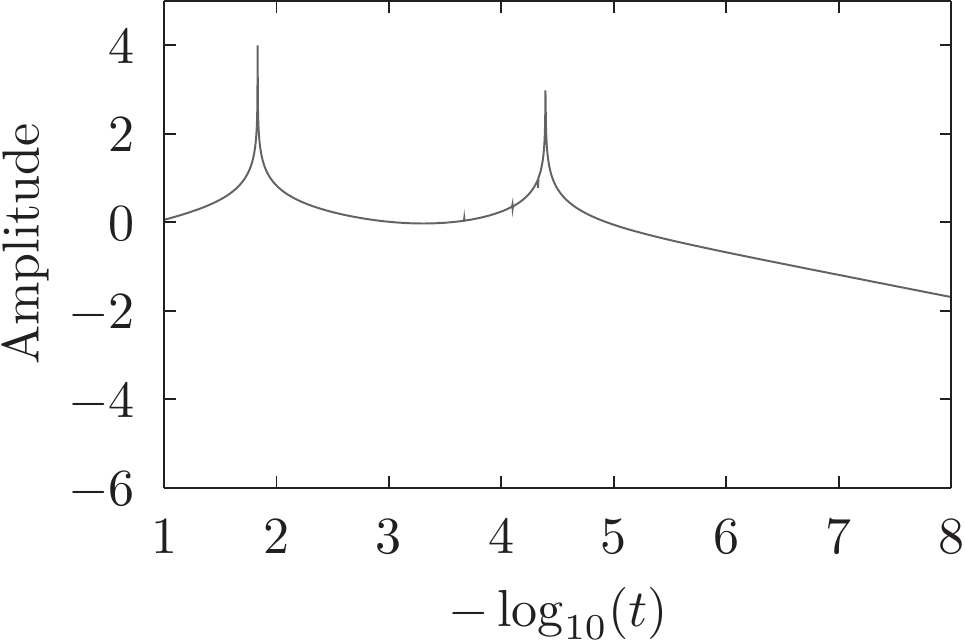}}
\subfigure[$l=2$ spheroidal, $n=9$]{\label{fig:l2n6}\includegraphics[width=40mm]{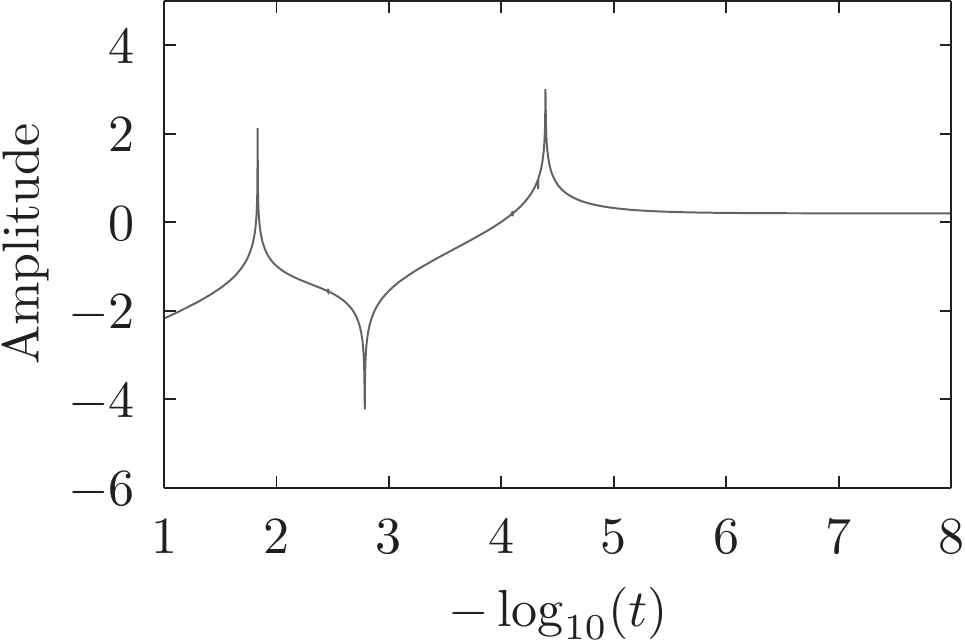}}
\subfigure[$l=3$ toroidal, $n=9$]{\label{fig:l3n6}\includegraphics[width=40mm]{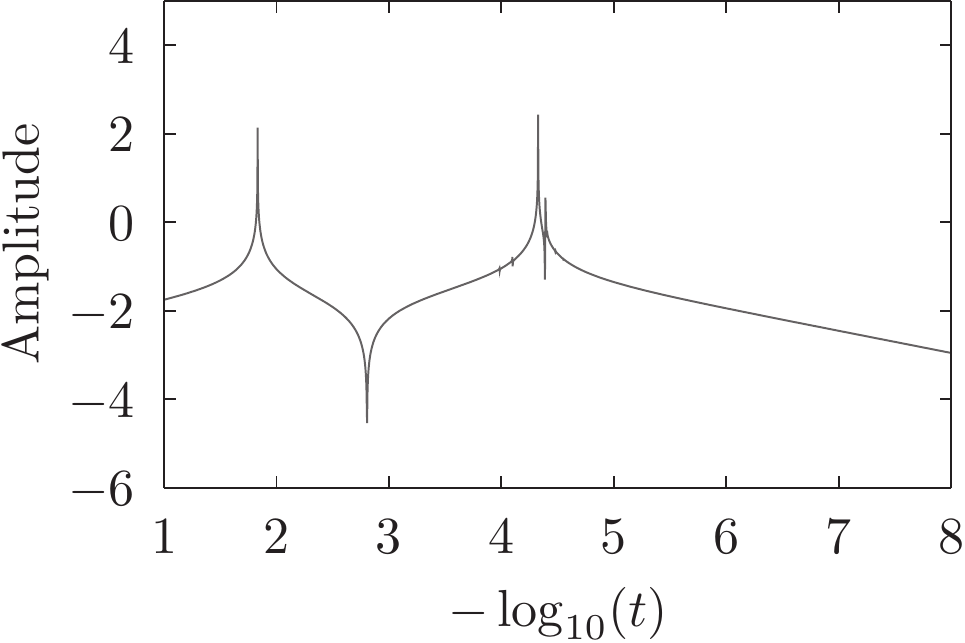}}
\caption[]{Figures showing how the amplitude in each mode varies with $t$ for $b = 10^{-2}$.
The lowest order modes are shown ($n=1,2,3$), along with those around the $n=8$ hybrid mode
($n=7,8,9$). Amplitude is plotted on the $y$-axis,
while the $x$-axis shows $-\log_{10}t$. }
\label{fig:continuous t}
\end{figure}
%%%%%%%%%%%%%%%%%%%%%%%%%%%%%%%%%%%%%%%%%%%%%%%%%%%%%%%%%%%%%%%
We carried out the projection using this numerical scheme for the cases $b=10^{-1}$, $10^{-2}$, $10^{-3}$.
As with the calculation of the rotational corrections of the eigenfunctions, we cut off at $N=10, 15$ and $40$
respectively, with these values chosen so that the hybrid fluid-elastic mode is within the range for each case.
The increasingly high $n$ value of this mode for small $b$,
and the fact that the calculation of the $6N \times 6N$ $\Lambda$ matrix is computationally intensive, meant
that we did not investigate any smaller values of $b$.

For given $b$, we also need to consider what parameter range of $t$ our scheme is valid for.
We have used the slow rotation approximation to calculate the modes.
Considering the form of the eigenvalue 
equation for the rotating star \eqref{xi ABC}, we can expect this to be a good approximation when
$\omega^2 A$ is large compared to $\vep\omega B$, 
i.e when $\frac{\Omega_B}{\omega}\ll 1$. We have $\Omega_B \sim \sqrt{t}$,
but we also have to take into consideration which mode $\omega$ we are looking at. 
The elastic modes scale with $b$, so that for the lowest modes with small $n$ value we need
$\frac{\sqrt{t}}{b}\ll 1$. The hybrid fluid-like mode has a scaling $\sqrt{G\rho}$ --
we have scaled our numerical results so that this is of order unity, so for this 
mode we only need $\sqrt{t}\ll 1$. 

The results for $b =10^{-2}$ are shown in detail as an illustrative example.
In this case, we only have $\frac{\sqrt{t}}{b}<10^{-1}$ for $t< 10^{-4}$), so cannot expect our results
to be reliable for the lowest $n$ modes when $t$ is larger than this.    
Around the $n=8$ hybrid fluid-elastic mode, we can instead expect the 
results to be reliable when $\sqrt{t}$ is small, i.e. from around $t=10^{-2}$
onwards. 

Figure \ref{fig:barcharts t} shows the calculated amplitudes of the the modes for 
values of $\log_{10}t = -1, -2, \dots, -6$. These amplitudes are nonzero for the spheroidal $l=2$ 
and toroidal $l=1$ and $l=3$ cases. For the spheroidal $l=2$ and toroidal $l=1$ cases, the first 15 modes are shown,
while for the toroidal $l=3$ modes the cutoff is at $n =14$, because our numerical 
scheme does not reproduce the highest $l=3$ mode correctly. To test this, we tried cutting 
off at lower values of $N$, and found that the $l=3$, $n=N$ mode is then not correct; to obtain
the $n=15$ mode we would have to cut off at $N=16$. 

As a first consistency check, we can see that as $t$ is made smaller, the results 
reproduce that of the non-rotating case: below $t=10^{-5}$, the 
the $l=2$ spheroidal amplitudes become very similar to those when $t=0$, 
shown in Figure \ref{fig:BA22},
while the $l=1$ and $l=3$ amplitudes
appear to roughly scale with $\sqrt{t}$, as would be expected from the size of the
rotational corrections \eqref{epsilon t} to the first order eigenfunctions.

For $t \geq 10^{-4}$, the modes excited vary considerably from the non-rotating 
case, and there is no clear pattern visible in the figure. 
We can obtain greater insight by instead focussing on one
mode at a time, and tracking the amplitude as a continuous function of $t$; this is shown
in Figure \ref{fig:continuous t} for the lowest $n$ modes ($n= 1,2,3$) and also
for some values of $n$ around the hybrid mode ($n=7,8,9$). The amplitude is plotted on the $y$-axis,
while the $x$-axis shows $-\log_{10}t$.  

For the $l=2$ modes, we can again see that as $t$ is made smaller, the amplitude converges to 
that of the non-rotating case, while the $l=1$ and $l=3$ modes drop off with $\sqrt{t}$
as expected. For each mode, this happens once $\frac{\sqrt{t}}{b} \ll 1$.
However, for higher values of $t$, there are a number of `spikes' in 
the amplitude in each mode, small parameter ranges of $t$ over which the 
amplitude changes rapidly.
We currently have no clear explanation for the existence of these spikes, 
but  have checked these are stable under the choice of cutoff $N$: as $N$ is 
made smaller, some of the detailed features are lost, but the larger spikes remain. 
We also see somewhat different behaviour for the hybrid $n=8$ mode, with fewer
spikes in the amplitude.
%%%%%%%%%%%%%%%%%%%%%%%%%%%%%%%%%%%%%%%%%%%%%%%%%%%%%%%%%%%%
% fig: plots of continuous t for b=10^-1 and 10^-3
\begin{figure}
\centering     %%% not \center
\subfigure[$l=2$ spheroidal, $n=1$]{\label{fig:amppts121}\includegraphics[width=40mm]{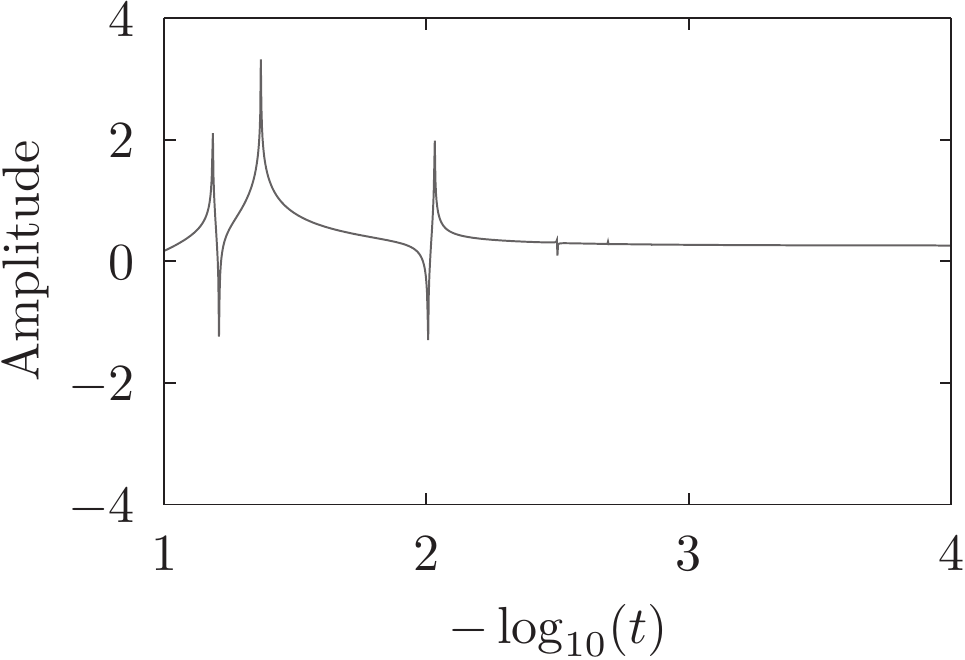}}
\subfigure[$l=2$ spheroidal, $n=2$]{\label{fig:amppts122}\includegraphics[width=40mm]{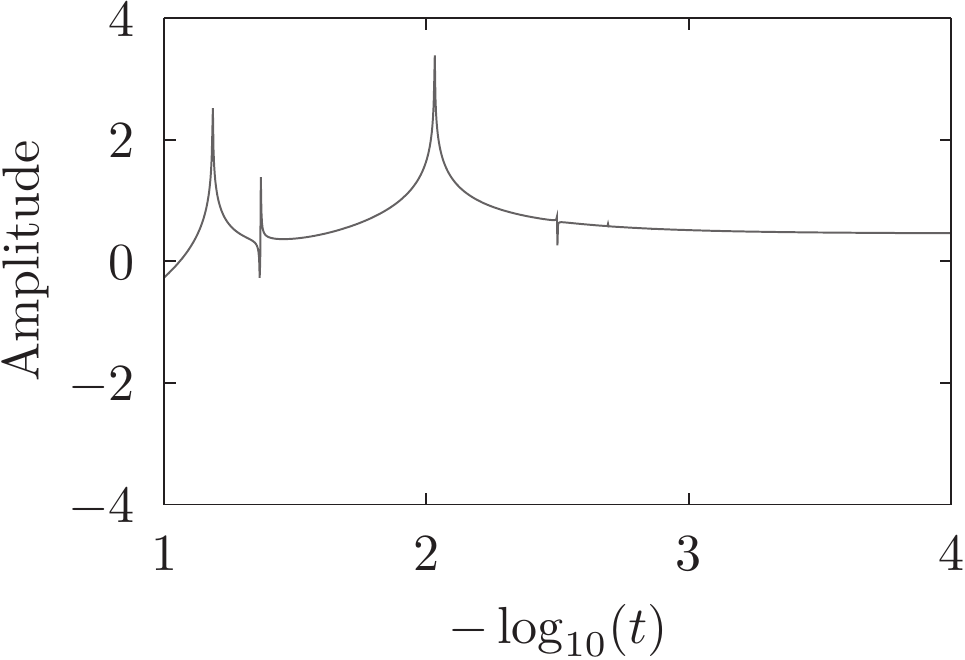}}
\subfigure[$l=2$ spheroidal, $n=3$]{\label{fig:amppts123}\includegraphics[width=40mm]{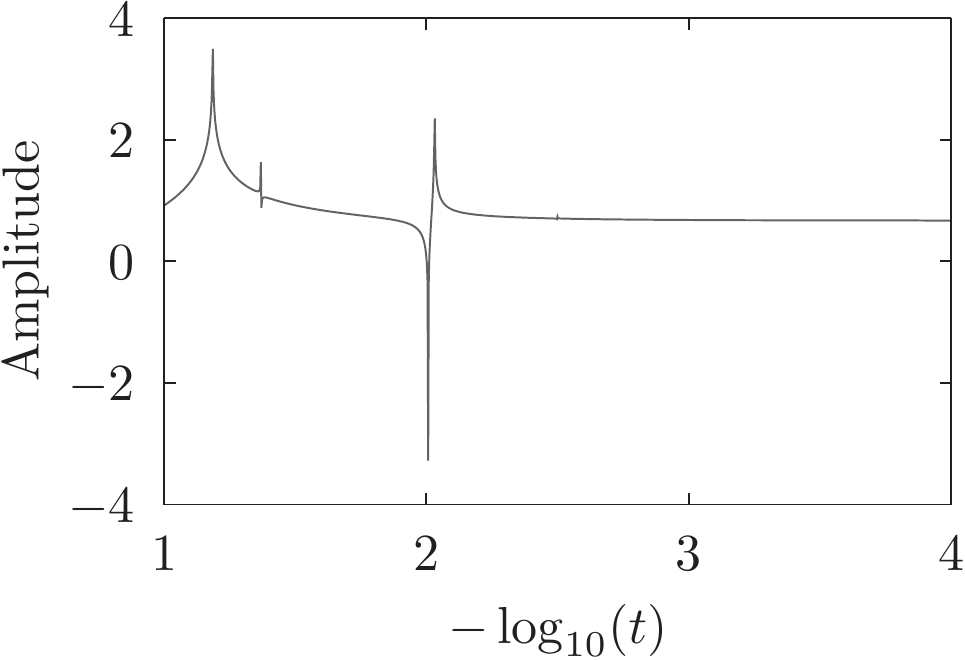}}
\\
\subfigure[$l=2$ spheroidal, $n=1$]{\label{fig:amppts321}\includegraphics[width=40mm]{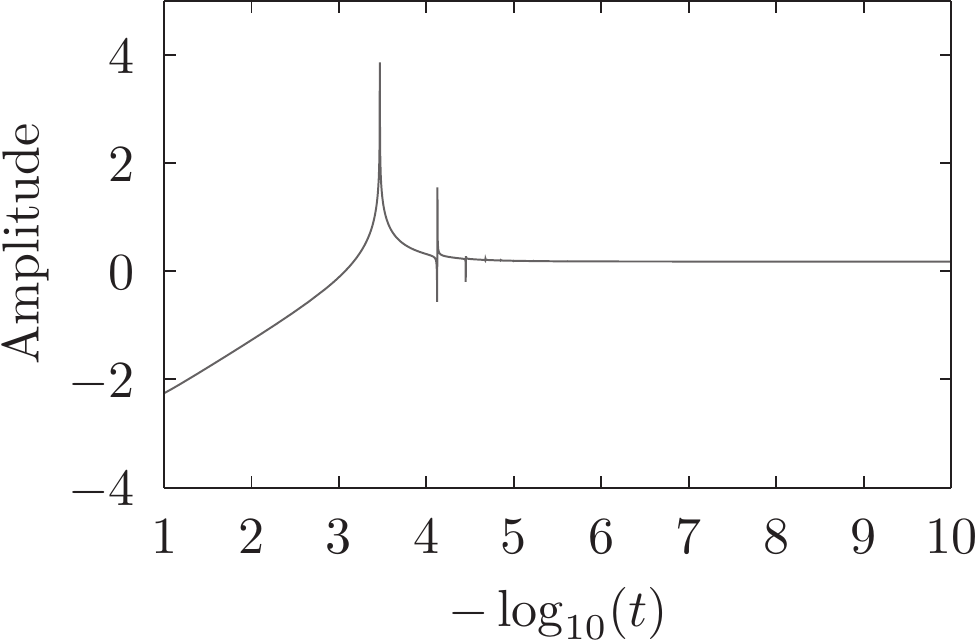}}
\subfigure[$l=2$ spheroidal, $n=2$]{\label{fig:amppts322}\includegraphics[width=40mm]{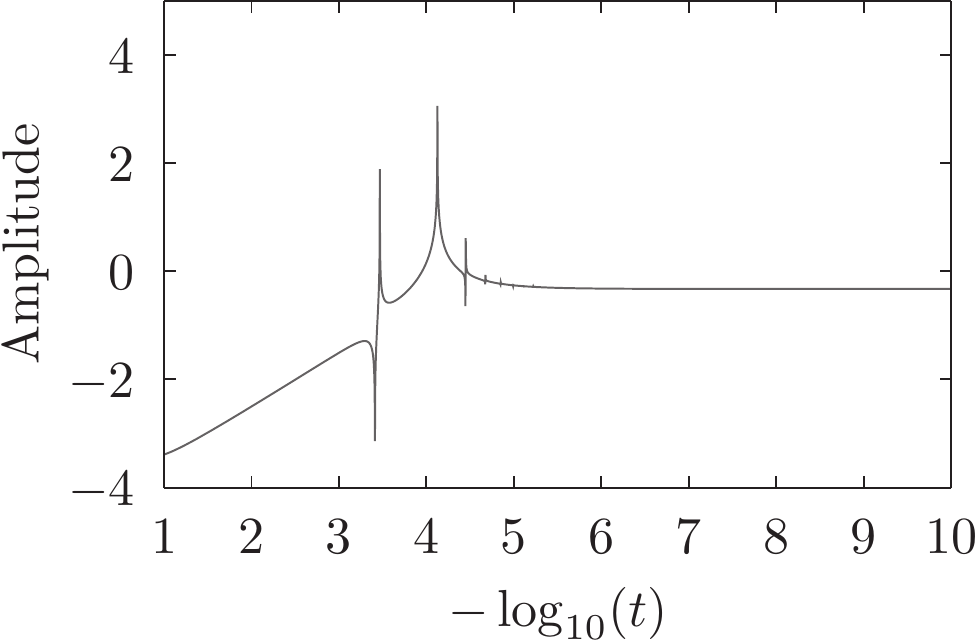}}
\subfigure[$l=2$ spheroidal, $n=27$]{\label{fig:amppts3227}\includegraphics[width=40mm]{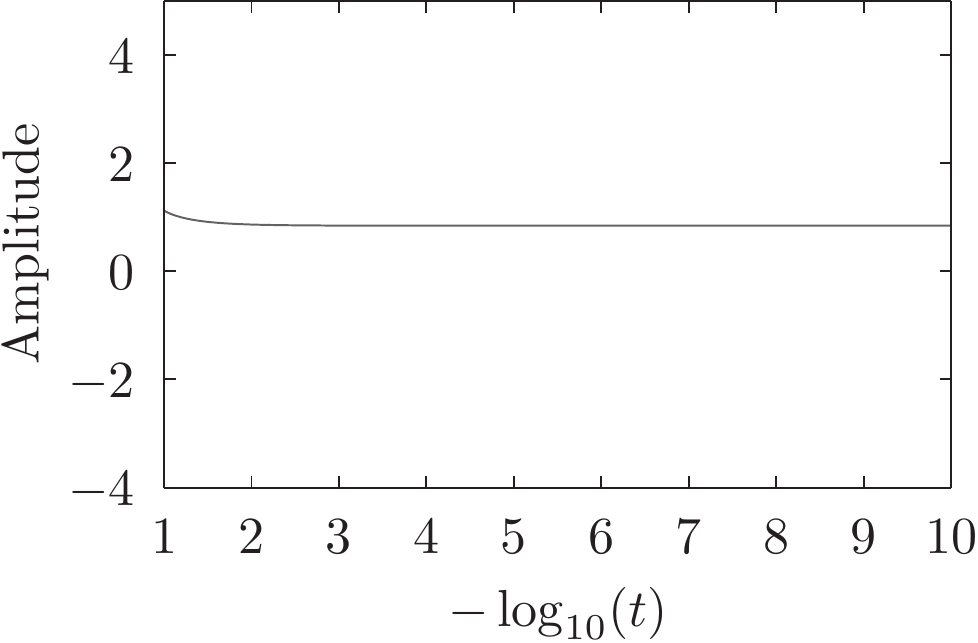}}
\caption[]{Figures showing how the amplitudes of the $l=2$ spheroidal modes vary with $t$ for $b=10^{-1}$ and
$10^{-3}$. The lowest order $n=1,2$ modes are shown in each case, along with the 
hybrid mode
($n=3$ for $b=10^{-1}$, and $n=27$ for $b=10^{-3}$). The amplitude is plotted on the $y$-axis,\color{black}
while the $x$-axis shows $-\log_{10}t$. }
\label{fig:continuous t b13}
\end{figure}
%%%%%%%%%%%%%%%%%%%%%%%%%%%%%%%%%%%%%%%%%%%%%%%%%%%

A similar picture holds for $b=10^{-1}$ and $b=10^{-3}$. Figure \ref{fig:continuous t b13} 
shows a few representative
plots of the amplitude of the $l=2$ spheroidal modes for these cases.
For  $b=10^{-1}$, we have plotted the amplitudes of the $n=1,2$ modes and the $n=3$
hybrid mode. Similarly, for $b=10^{-3}$ we plot $n=1$ and $2$, and the hybrid 
mode $n=27$. In both cases we again see convergence to the non-rotating case once $\frac{\sqrt{t}}{b} \ll 1$.
For $b=10^{-3}$, as with $b=10^{-2}$, the amplitude of the hybrid mode shows different, smoother behaviour
as $t$ is varied; this is not so apparent for $b=10^{-1}$, where the hybrid mode is
less distinct from the other modes of the star.

A final important check is that we can use the calculated amplitudes to reconstruct the initial data. 
This is shown in Appendix \ref{appx:reconstruct}.

%%%%%%%%%%%%%%%%%%%%%%%%%%%%%%%%%%%%%%%%%%%%%%%%%%%%%%%%%%%%%%%%%%
% CONCLUSIONS
%%%%%%%%%%%%%%%%%%%%%%%%%%%%%%%%%%%%%%%%%%%%%%%%%%%%%%%%%%%%%%%%%%

\section{Conclusions}
\label{sec:conclusions}

We have made order-of-magnitude estimates of the maximum energy released in a starquake, finding 
maximum oscillation amplitudes of $\frac{\delta r}{R}\sim 10^{-6}$ and characteristic strains 
of $\sim 4\times 10^{-24}$ Hz$^{-\frac{1}{2}}$
in the upper limit where all energy lost in the glitch is put into oscillations. 
For the case of a completely solid quark star, we find larger maximum amplitudes of
$\frac{\delta r}{R}\sim 10^{-2}$ can be sustained, with corresponding characteristic strains 
of $\sim 5\times 10^{-22}$ Hz$^{-\frac{1}{2}}$. In each case, these estimates
are for a large change in the angular velocity of the star between glitches. 

We have also developed a toy model for starquakes, in which all strain is 
suddenly lost from the star at the glitch.
We first considered a simplified version in which the star is not rotating 
before the glitch, and found that the excited modes are $l=2$ spheroidal oscillations.
Out of these, the oscillation mode preferentially excited is a hybrid fluid-elastic mode
similar to the Kelvin mode of an incompressible fluid star. The addition of
rotation introduces excitation of $l=1$ and $l=3$ toroidal modes of the star.
The hybrid mode is still strongly excited, but other modes of the star 
with an elastic character are also excited to a high level over small parameter
ranges of the rotation rate; we currently have no clear explanation for this
behaviour, but have tested it is stable under changing the number of radial 
modes we include in the set we project against. 

To move from our current toy model to a more realistic model of a starquake, there
are a number of extensions we need to consider. 
First, we would need to include the fluid
core of the star in our model. This would complicate the spectrum
of oscillation modes by adding extra ones connected to the
fluid-elastic interface. We could also consider using a more realistic
equation of state rather than our incompressible model, introducing
new $\textit{p}$- and $\textit{g}$-modes to the mode spectrum.

It would also be necessary to improve the model
for the starquake itself. At the moment we have an acausal mechanism
in which all strain is lost from the star instantaneously. In a realistic
scenario, the quake would propagate across the star over time, possibly in 
the form of surface cracks. The new timescale introduced here could 
strongly affect which modes of the star are excited. However, this 
timescale is not well constrained observationally.

The oscillations produced by a starquake would be expected to shake
the magnetosphere at the surface of the star, which could lead to 
radio emission connected with the starquake,
possibly at a level that could be resolved with the new generation 
of radio telescopes.
It would be interesting to make some estimates of this with a toy model 
of how the pulsar could shake magnetic field lines in the 
magnetosphere.

%%%%%%%%%%%%%%%%%%%%%%%%%%%%%%%%%%%%%%%%%%%%%%%%%%%%%%%%%%%%%%%%%%
% ACKNOWLEDGEMENTS
%%%%%%%%%%%%%%%%%%%%%%%%%%%%%%%%%%%%%%%%%%%%%%%%%%%%%%%%%%%%%%%%%%

\section{Acknowledgements}

LK acknowledges support via an STFC PhD studentship, while DIJ acknowledges support from the STFC via grant number ST/H002359/1.  The authors also acknowledge partial support  from `NewCompStar', COST Action MP1304.

%%%%%%%%%%%%%%%%%%%%%%%%%%%%%%%%%%%%%%%%%%%%%%%%%%%%%%%%%%%%%%%
%%%%%%%%%%%%%%%%%%%%%%%%%%%%%%%%%%%%%%%%%%%%%%%%%%%%%%%%%%%%%%%
% APPENDICES
%%%%%%%%%%%%%%%%%%%%%%%%%%%%%%%%%%%%%%%%%%%%%%%%%%%%%%%%%%%%%%%
%%%%%%%%%%%%%%%%%%%%%%%%%%%%%%%%%%%%%%%%%%%%%%%%%%%%%%%%%%%%%%%

\appendix

%%%%%%%%%%%%%%%%%%%%%%%%%%%%%%%%%%%%%%%%%%%%%%%%%%%%%%%%%%%%%%%%%%
% APPENDIX - EQUILIBRIA
%%%%%%%%%%%%%%%%%%%%%%%%%%%%%%%%%%%%%%%%%%%%%%%%%%%%%%%%%%%%%%%%%%

%\section{Equilibria of a rotating elastic star}
%\label{appx:equilibria}

%%%%%%%%%%%%%%%%%%%%%%%%%%%%%%%%%%%%%%%%%%%%%%%%%%%%%%%%%%%%%%%%%%
% APPENDIX - OSCILLATION MODES 
%%%%%%%%%%%%%%%%%%%%%%%%%%%%%%%%%%%%%%%%%%%%%%%%%%%%%%%%%%%%%%%%%%

\section{Reconstruction of the initial data}
\label{appx:reconstruct}
In this Appendix, we demonstrate that the amplitudes we calculate for the excited modes can 
indeed be used to reconstruct the initial data, both in the non-rotating special case and for the
full problem for the rotating star.

\begin{figure}
\centering     %%% not \center
\subfigure[$U$ part of initial data]{\label{fig:uinit}\includegraphics[width=70mm]{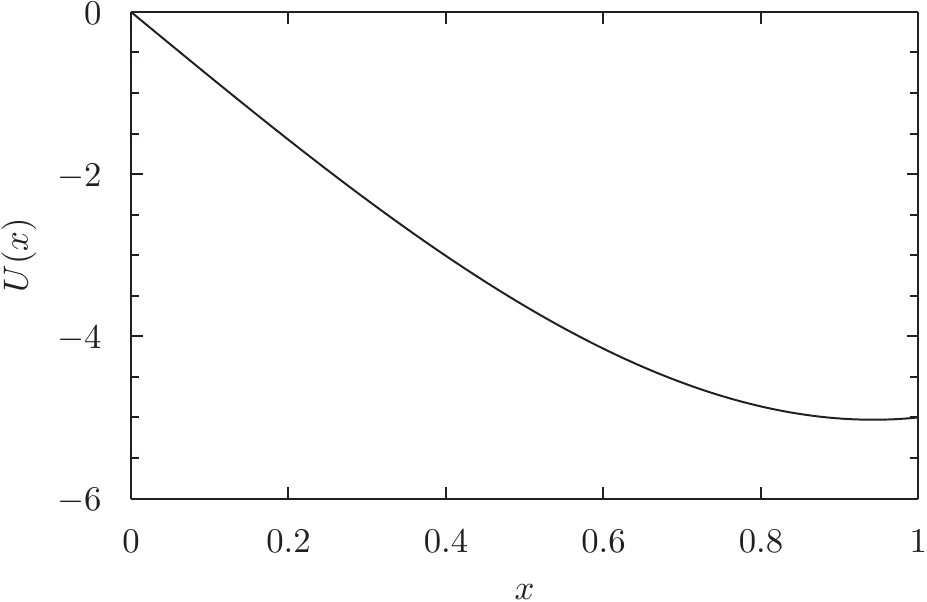}}
\subfigure[$V$ part of initial data]{\label{fig:vinit}\includegraphics[width=70mm]{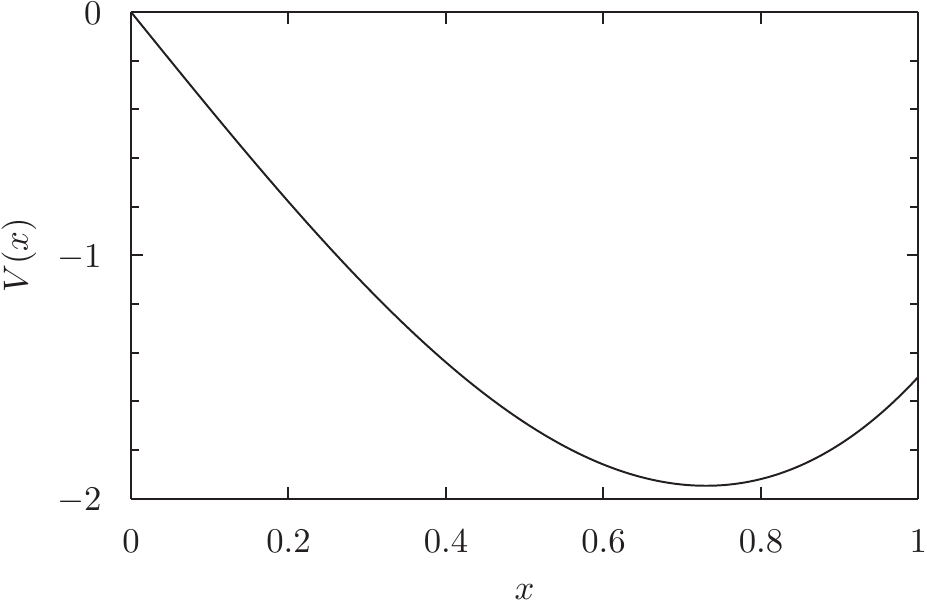}}
\\
\subfigure[Reconstructing the $U$ initial data]{\label{fig:ureconstruct}\includegraphics[width=70mm]{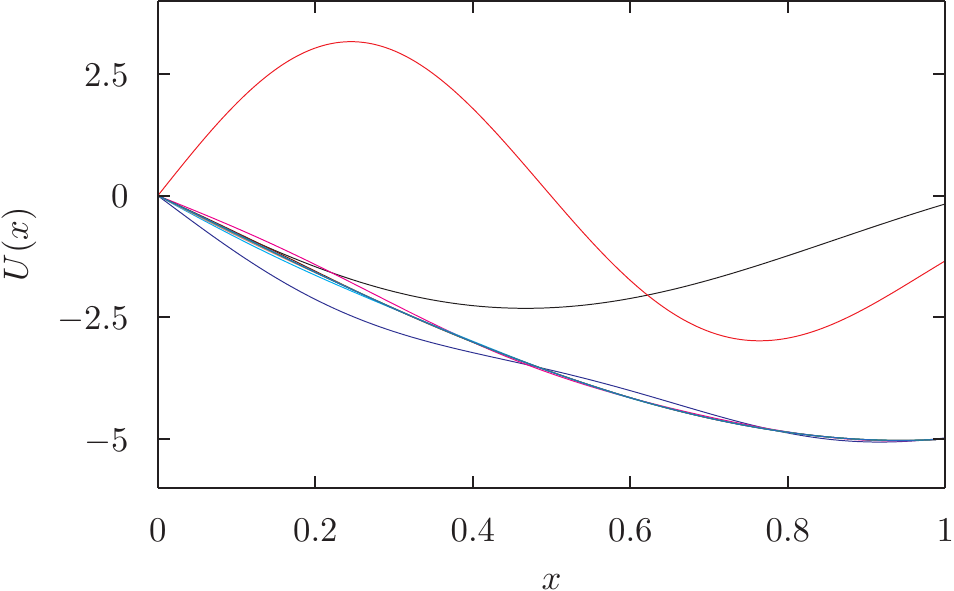}}
\subfigure[Reconstructing the $V$ initial data]{\label{fig:vreconstruct}\includegraphics[width=70mm]{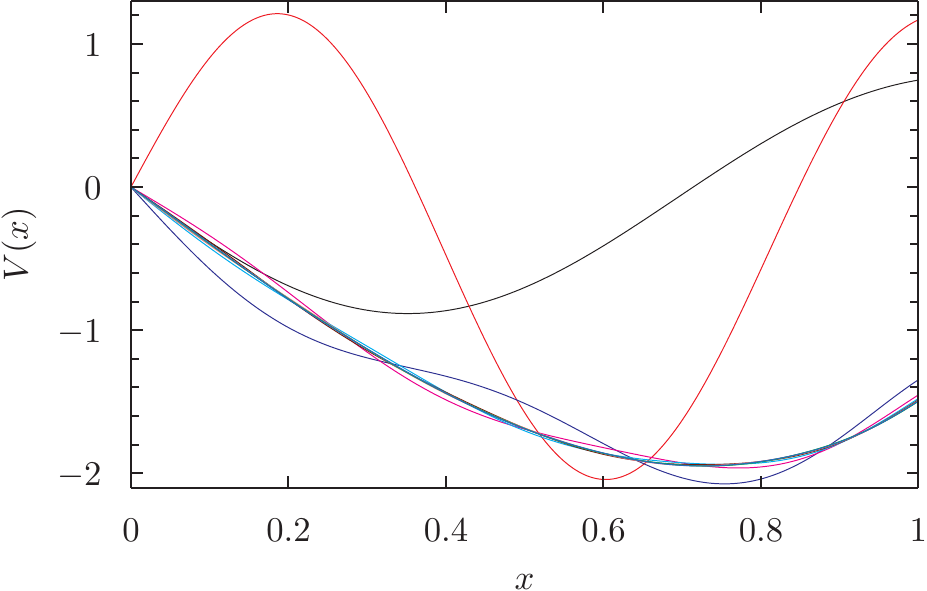}}
\\
\subfigure[Convergence for $U$]{\label{fig:uconverge}\includegraphics[width=70mm]{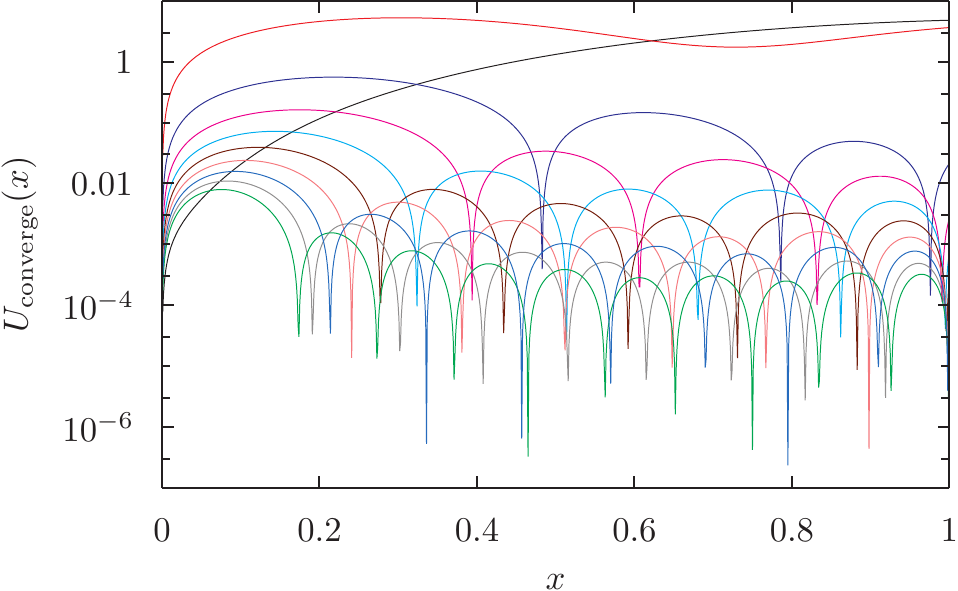}}
\subfigure[Convergence for $V$]{\label{fig:vconverge}\includegraphics[width=70mm]{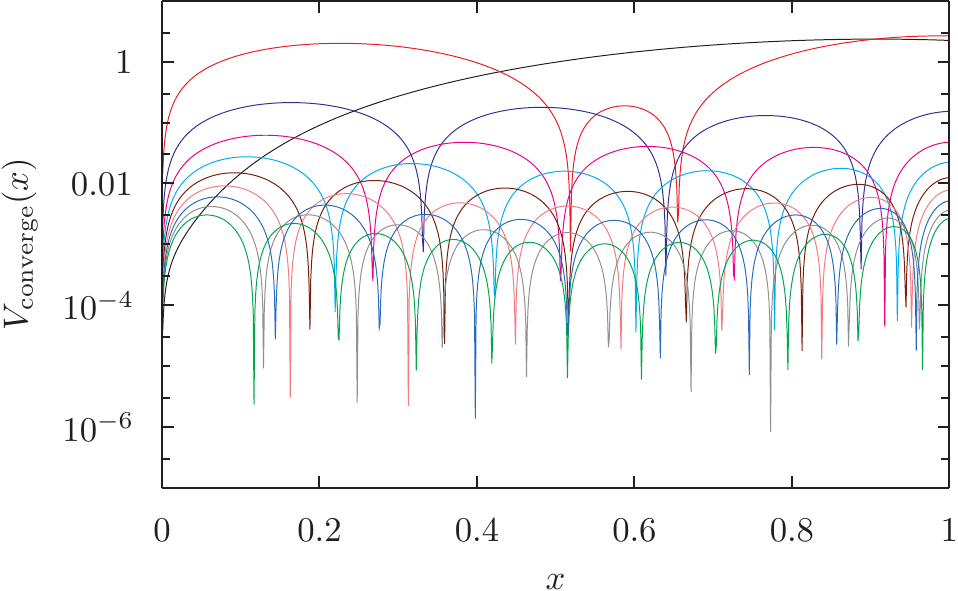}}
\caption[Figure showing the reproduction of the initial data as a
sum of eigenfunctions]{Figure showing the reproduction of the initial data as a
sum of eigenfunctions for the case $b=0.1$. 
The $U$ and $V$ parts of the initial data are shown 
in the top row. In the middle row, the partial sums $U^{\text{partial}}(N,x)$
and $V^{\text{partial}}(N,x)$ are shown for $N=1$ (black line)
up to $N= 10$ (green line); the
largest contribution is from the $N=3$ eigenfunction (dark blue), which is the 
fluid-elastic hybrid mode. The bottom row plots 
 the
absolute value of the difference between the partial sums and the initial 
data, $U^{\text{converge}}(N,x)$ and $V^{\text{converge}}(N,x)$.
}
\label{fig:reproduce}
\end{figure}

We show this first for the non-rotating case.
We should expect the reconstruction to converge to
the initial data as we add more modes in to the sum.
Figure \ref{fig:reproduce} shows the results of this for the illustrative case
$b=0.1$. The $U$ \ref{fig:uinit} and $V$ \ref{fig:vinit} parts of the initial data are plotted
in the top row. 
The middle row shows the results of summing progressively more eigenfunctions
with our calculated amplitudes: defining the partial sum

%%%%%%%%%%%%%%%%%%%%%%%%%%%%%%%%%%%%%%%%%%%%%%%%%%%%%%%%%%%%%%%
%figure: reconstruct initial data (zero rot) %%%%%%%%%%%%%%%%%%

\begin{equation}
\label{xi partial}
\boxi^{\text{partial}}(N,x)  = \frac{1}{2}\sum_{\alpha =1}^{N} \left[
b^\alpha \boxi^\alpha_{(0)} (x) + b^{*\alpha} \boxi^{*\alpha}_{(0)}(x)
\right],
\end{equation}
we have plotted $U^{\text{partial}}(N,x)$ \ref{fig:ureconstruct}
and $V^{\text{partial}}(N,x)$ \ref{fig:vreconstruct} for $N=1,\dots,10$.
The largest contribution is from the $N=3$ mode: this is the hybrid
fluid-elastic mode for $b=0.1$.

To test how the partial sums converge with increased $N$, in the last row
we have plotted the $U$ \ref{fig:uconverge} and $V$ \ref{fig:vconverge}
parts of the absolute value of the 
difference between $\boxi^{\text{partial}}(N,x)$ and the initial data,

\begin{equation}
\label{07}
\boxi^{\text{converge}}(N,x)= \left|\boxi^{\text{partial}}(N,x) - \boxi^{\text{ID}}(x) \right|
\end{equation}
We can see that the contributions become progressively smaller as
$N$ increases.

Next we consider the full problem with rotation, where we also have to reconstruct the 
velocity initial data as well as that for the displacement. 
The velocity field initial data $\dot{\boxi}^{\text{DC}}$ \eqref{xidotDC ref} 
is made up from only one zeroth order eigenfunction,
the zero frequency $n=1$ toroidal mode, and this mode has no corrections at first 
order in the rotation. This means that the velocity initial data can be reconstructed
from one eigenfunction only, and this can be done with close to numerical precision. 

\begin{figure}
\centering
\includegraphics[scale=1]{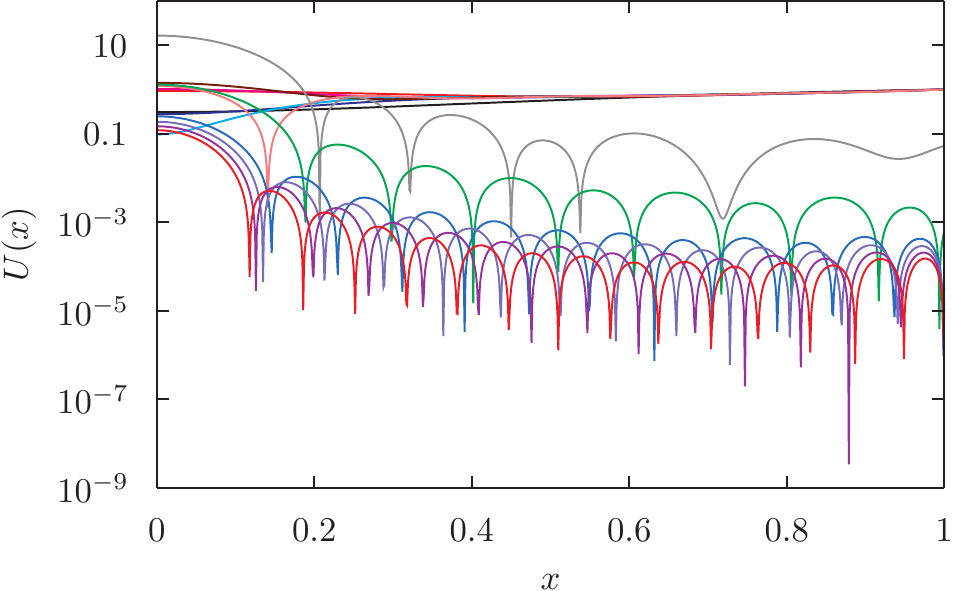}
\caption{Figure showing the reconstruction of the initial data in the cases $b=10^{-1}, 10^{-2}$
and $10^{-3}$, with $t=10^{-2}$. The $U$ part of the fractional difference between the initial data and its reconstruction, 
$\vert U^{\text{reconstruct}}(n) - U^{\text{CD}} \vert /U^{\text{CD}}$,
is plotted for values of $n$ between $1$ and the cutoff $N$. The initial data is not reproduced correctly until 
the hybrid mode is added, and the reconstruction then converges to the initial data as $N$ is made larger.}
\label{fig:reconstruct u}
\end{figure}

As an example of the reconstruction for the
displacement field initial data, Figure \ref{fig:reconstruct u} plots the $U$
part of the fractional difference between the initial data and its reconstruction 
\eqref{reconstruct xiCD}, $\vert U^{\text{reconstruct}}(n) - U^{\text{CD}} \vert /U^{\text{CD}}$,
for $b=10^{-2}$ and $t = 10^{-2}$, with values of $n$ between $1$ and the cutoff $N$.
We see that as with the special case of the glitch at zero spin,
the initial data is not reproduced correctly until the hybrid mode is added, 
and the reconstruction then converges to the initial data as $N$ is made larger.

%%%%%%%%%%%%%%%%%%%%%%%%%%%%%%%%%%%%%%%%%%%%%%%%%%%%%%%%%%%%%%%
%figure: reconstruct initial data%%%%%%%%%%%%%%%%%%%%%%%%%%%%%%
\color{black}

\section{Orthogonality of eigenfunctions for a spherical elastic star}
\label{appx:orthogonal}
Here we show that the eigenfunctions for our background model Star S,
a homogeneous, incompressible elastic star, are orthogonal.
The corresponding result has been shown for the cases of nonradial
oscillations of stationary, perfect fluid stars 
(\cite{1964Chandrasekhar,1967Lynden-Bell,1978Friedman})
and also for the incompressible fluid star (\cite{1964ChandraLebovitz}).
We will follow a similar method here, and show that our operator
$C_{(0)}$ \eqref{C0} is symmetric,

\begin{equation}
\label{00}
\left\langle \boldsymbol{\eta}, C_{(0)}\boxi \right\rangle = \left\langle \xi, C_{(0)}\boldsymbol{\eta} \right\rangle.
\end{equation}
given the boundary conditions \eqref{traction mode}.
Writing the inner product out in full,

\begin{equation}
\label{modes:mode integral}
\left\langle \boldsymbol{\eta}, C_{(0)}\boxi \right\rangle = \int_V\nabla_j  \tau\indices{_i^j}\eta^i dV
+ \int_V\rho(\nabla_i\delta\Phi)\eta^i dV.
\end{equation}
It is then sufficient to show that both terms of the right hand side
are symmetric in $\xi^i$ and $\eta^i$, given our boundary conditions.
We will only show this for the first term, as the second term is identical to 
that for the incompressible fluid star.
Integrating by parts,

\begin{equation}
\label{modes:by parts}
\int_V\nabla_j \tau\indices{_i^j}\eta^i dV = -\int_V (\nabla_j\eta^i)\tau\indices{_i^j} dV
+ \int_{\partial V} \eta^i \tau\indices{_i^j} dS_j.
\end{equation}
For the surface term, we can use the boundary conditions \eqref{traction mode}
to substitute in $\tau_{ij}$, so that

\begin{equation}
\label{modes:4}
\int_{\partial V} \eta^i \tau\indices{_i^j} dS_j 
= \int_{\partial V} \eta^r \xi^r \dv{p}{r} \, dS_r,
\end{equation}
where we have used the spherical symmetry of the background star. This
term is symmetric in $\xi$ and $\eta$.
For the volume term, we expand out $\tau\indices{_i^j}$ so that

\begin{equation}
\label{modes:5}
-\int_V (\nabla_j\eta^i)\tau\indices{_i^j} dV = -\int_V (\nabla_j\eta^i)\left(-\delta p \delta\indices{_i^j}
 + \mu\nabla_i\xi^j + \mu\nabla^j\xi_i\right) dV
\end{equation}
The first term of this is zero by incompressibility, while the third 
term is already symmetric.
To deal with the second term,
we will integrate half by parts with respect to $\nabla_i$, and 
half with respect to $\nabla_j$, so that

\begin{equation}
\begin{split}
\label{modes:6}
-\int_V \mu(\nabla_j\eta^i)\left(\nabla_i\xi^j\right) dV = 
\frac{1}{2}\left[\int_V \eta^i\nabla_j(\mu \nabla_i\xi^j)dV
-\int_{\partial V} \mu \eta^i\nabla_i\xi^j dS_j\right.
\\
\left.+\int_V \xi^j\nabla_i(\mu \nabla_j\eta^i)dV
-\int_{\partial V} \mu \xi^j\nabla_j\eta^i dS_i\right]
\end{split}
\end{equation}
Using the product rule on the volume terms, this becomes

\begin{equation}
\begin{split}
\label{modes:7}
-\int_V \mu(\nabla_j\eta^i)\left(\nabla_i\xi^j\right) dV = 
\frac{1}{2}\left[\int_V (\nabla_j\mu)\eta^i(\nabla_i\xi^j)dV 
+\int_V \mu\eta^i\nabla_j \nabla_i\xi^j dV\right.
\\
+\int_V (\nabla_i \mu)\xi^j( \nabla_j\eta^i)dV
+\int_V \mu\xi^j\nabla_i \nabla_j\eta^i dV
\\
\left.-\int_{\partial V} \mu \eta^i\nabla_i\xi^j dS_j
-\int_{\partial V} \mu \xi^j\nabla_j\eta^i dS_i\right].
\end{split}
\end{equation}
The shear modulus $\mu$ is constant over the surface of the star,
and so its derivative $\nabla_i\mu$ is non zero only at the 
surface,

\begin{equation}
\label{modes:nabla mu}
\nabla_i\mu = -\mu\delta(r-R)\hat{r}_i.
\end{equation}
The terms containing $\nabla_i\mu$ then cancel with the two surface terms to leave

\begin{equation}
\label{modes:8}
-\int_V \mu(\nabla_j\eta^i)\left(\nabla_i\xi^j\right) dV = 
\int_V \mu\eta^i\nabla_j \nabla_i\xi^j dV + \int_V \mu\xi^j\nabla_i \nabla_j\eta^i.
\end{equation}
These two terms are zero, as can be seen by commuting
$\nabla_i$ and $\nabla_j$ and using the incompressibility condition,
so that finally we have

\begin{equation}
\label{modes:9}
\int_V\nabla_j \tau\indices{_i^j}\eta^i dV =
\int_{\partial V} \eta^r \xi^r \dv{p}{r} \, dS_r
-\int_V \mu(\nabla_j\eta^i)(\nabla^j\xi_i).
\end{equation}
We have now shown that $C_{(0)}$ is a symmetric operator, and so the
eigenvalues of a self-gravitating elastic incompressible
star are orthogonal with respect to $A_{(0)}\equiv \rho$:

\begin{equation}
\label{01}
\langle \eta, A_{(0)}\xi \rangle = 0, \quad \omega^2_\eta \neq  \omega^2_\xi.
\end{equation}

%First, though, we will show that the eigenfunctions
%of the star in our glitch model are orthogonal. This is the case
%of a completely solid, elastic incompressible star,
%discussed in Section \ref{elastic modes}. 
%In this case the mode equation \eqref{xi AC}
%can again be written in the form
%
%\begin{equation}
%\label{}
%A\indices{_i_j}\omega^2 \xi^j + C\indices{_i_j}\xi^j = 0,
%\end{equation}
%where the operators $A$ and $C$ are defined by
%
%\begin{align}
%\label{proj:Afull}
%A\indices{_i_j}\xi^j \equiv \rho \xi^j \\
%\label{proj:Cfull}
%C\indices{_i_j}\xi^j \equiv \nabla_j T\indices{_i^j} - \rho\nabla_i\Phi.
%\end{align}
%Here $T$ is the stress-energy tensor 
%
%\begin{equation}
%\label{T}
%T_{ij} = -\delta p \,\delta_{ij} + \mu(\nabla_i\xi_j+\nabla_j\xi_i).
%\end{equation}
%The boundary condition \eqref{modes:elastic bc repeat}
%at the surface $r=R$ can then be written in 
%terms of $T_{ij}$ as
%
%\begin{equation}
%\label{proj:bcs rewrite}
%(T_{ij} - \xi^k\nabla_k p \delta_{ij})\hat{r}^i = 0.
%\end{equation}
%The operator $A$ is just multiplication by a scalar function, so
%it is Hermitian.

%%%%%%%%%%%%%%%%%%%%%%%%%%%%%%%%%%%%%%%%%%%%%%%%%%%%%%%%%%%%%%%%%%
% BIBLIOGRAPHY
%%%%%%%%%%%%%%%%%%%%%%%%%%%%%%%%%%%%%%%%%%%%%%%%%%%%%%%%%%%%%%%%%%

\bibliographystyle{mn2e}
\bibliography{bibfile}

\end{document}